\pdfoutput=1
\newcommand*{\ATLASLATEXPATH}{latex/}

\documentclass[UKenglish,texlive=2016,atlasstyle,subfigure,cernpreprint]{\ATLASLATEXPATH atlasdoc}
\usepackage[block=space]{\ATLASLATEXPATH atlaspackage}

\usepackage{longtable}
\usepackage{\ATLASLATEXPATH atlasbiblatex}
\usepackage{\ATLASLATEXPATH atlascontribute}
\usepackage{\ATLASLATEXPATH atlasphysics}
\graphicspath{{logos/}{figures/}}

\usepackage{EXOT-2015-04-PAPER-defs}

\AtlasTitle{Search for heavy particles decaying into top-quark pairs using lepton-plus-jets events in proton--proton collisions at $\sqrt{s} = 13$~\TeV\ with the ATLAS detector}

\author{The ATLAS Collaboration}

\AtlasRefCode{EXOT-2015-04}

\PreprintIdNumber{CERN-EP-2018-48}

\AtlasJournal{Eur.\ Phys.\ J.\ C}
\AtlasJournalRef{\EPJC 78 (2018) 565}
\AtlasDOI{10.1140/epjc/s10052-018-5995-6}

\AtlasAbstract{%
  A search for new heavy particles that decay into top-quark pairs is performed using data collected from proton--proton collisions at a centre-of-mass energy of $13$~\TeV{} by the ATLAS detector at the Large Hadron Collider. The integrated luminosity of the data sample is 36.1~fb$^{-1}$. Events consistent with top-quark pair production are selected by requiring a single isolated charged lepton, missing transverse momentum and jet activity compatible with a hadronic top-quark decay. Jets identified as likely to contain $b$-hadrons are required to reduce the background from other Standard Model processes.
The invariant mass spectrum of the candidate top-quark pairs is examined for local excesses above the background expectation.
No significant deviations from the Standard Model predictions are found. Exclusion limits are set on the production cross-section times branching ratio for hypothetical $Z'$ bosons, Kaluza--Kein gluons and Kaluza--Klein gravitons that decay into top-quark pairs.
}

\hypersetup{pdftitle={ATLAS},pdfauthor={The ATLAS Collaboration}}

\begin{document}

\maketitle

\section{Introduction}
\label{sec:intro}

This paper presents a search for new
 particles in the top-quark pair ($t\bar{t}$) final state. The 
signature is a deviation from the 
$t\bar{t}$ invariant mass ($m^{\text{reco}}_{t\bar{t}}$) spectrum predicted by the Standard Model (SM). The search uses
 a data sample with  an integrated luminosity of 36.1~fb$^{-1}$ collected by the ATLAS detector from
the Large Hadron Collider (LHC) proton--proton collisions at $\sqrt{s}=13$~\TeV{} in 2015 and 2016. Previous searches for this signature with 8~\TeV{} data at the LHC were performed by the ATLAS~\cite{Aad:2015fna} and CMS~\cite{Khachatryan:2015sma} collaborations. The CMS Collaboration also searched in 13~\TeV{} LHC data using a smaller sample of 2.6~fb$^{-1}$~\cite{Sirunyan:2017uhk}.

The analysis selects events consistent with $t\bar{t}$ production followed by 
subsequent decay into the {\em{lepton-plus-jets}} topology.
In this topology, most of the top quarks decay into a bottom quark plus a
$W$ boson, $t\rightarrow Wb$, and one of the $W$ bosons decays into an electron or muon plus a neutrino while the other decays
into quarks.
If the $W$ boson decays into a $\tau$-lepton and a neutrino, and the $\tau$-lepton subsequently decays into an electron or a muon, and neutrinos, these decays are included in the search. No attempt is made to identify hadronically decaying $\tau$-leptons. 
Approximately 30\% of $t\bar{t}$ pairs decay this 
way, and the non-$t\bar{t}$ background is much smaller than in the all-hadronic topology.
 The selection requires a single isolated electron or muon, large missing 
transverse momentum, and hadronic jets. At least one of the jets must be 
identified as likely to contain a $b$-hadron ($b$-jet).

The $m^{\text{reco}}_{t\bar{t}}$  variable is reconstructed using the jets, 
charged leptons and missing transverse momentum in the events. The $m^{\text{reco}}_{t\bar{t}}$ distribution is
 then examined for deviations from the SM predictions. In the 
absence of significant deviations, upper 
limits are set on the cross-section for the possible production of new heavy particles
that decay into $t\bar{t}$. For comparison with other searches, these limits
 are transformed to lower limits on the allowed mass within particular
 {\em{benchmark}} models. The sensitivity of the  search is tested for new colour-singlet and colour-octet bosons with spin 1 or spin 2 and masses from 0.4 to 5~\TeV{}. The resonance widths for the specific models vary from very narrow (1\% of the heavy particle mass) to a value (30\% of the heavy particle mass) larger than that of the experimental resolution.

The paper is organised as follows. Details of the potential signals tested in this search are given in Section~\ref{sec:signals}. The ATLAS detector is introduced in Section~\ref{sec:detector} and the data samples used for the analysis are described in Section~\ref{sec:samples}. The event selection and reconstruction of the $t\bar{t}$ system are described in Section~\ref{sec:selrec} and the estimation of background contributions using data is described in Section~\ref{sec:estimate}. The systematic uncertainties affecting the analysis are detailed in Section~\ref{sec:systematics} and the expected background contributions are compared with data in Section~\ref{sec:expectations}. The results are presented in Section~\ref{sec:results} and the paper is summarised in Section~\ref{sec:summary}.

\section{Signal models tested}
\label{sec:signals}
The details of potential signals considered in this search are reviewed below. Interference between the signal processes and SM $t\bar{t}$ production is not considered here since these signals are not expected to interfere strongly with the dominant component of the SM $t\bar{t}$ background. The effect of interference is particularly important for new heavy scalar particles produced via gluon--gluon fusion, and was studied by ATLAS using  8~\TeV{} data~\cite{Aaboud:2017hnm}; such signals are not considered in this search.

\begin{figure}[htbp]

\centering{
\subfigure[$Z' \rightarrow t\bar{t}$]{\includegraphics[width=0.30\textwidth]{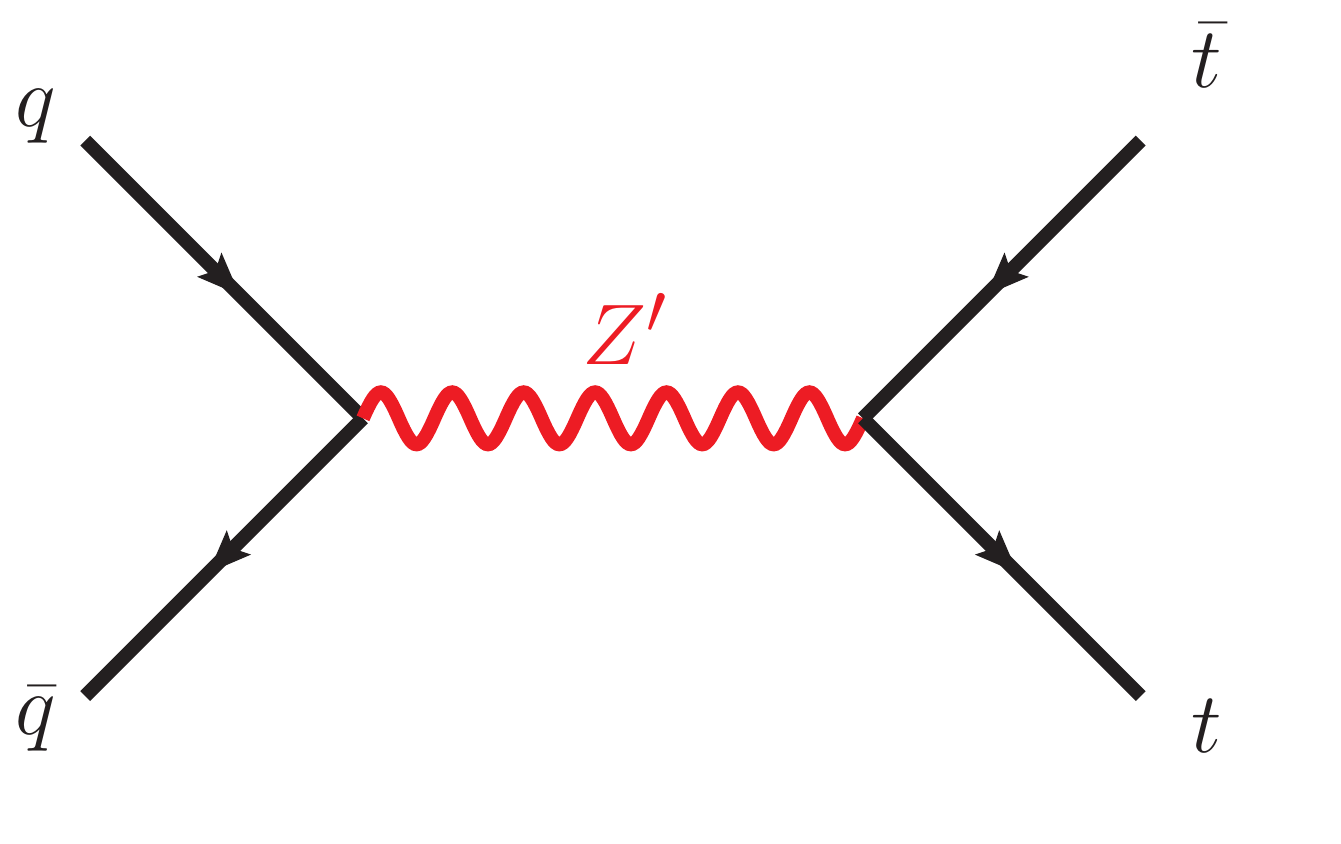}\label{fig:BSMFeynZp}}
\subfigure[$G_{KK} \rightarrow t\bar{t}$]{\includegraphics[width=0.30\textwidth]{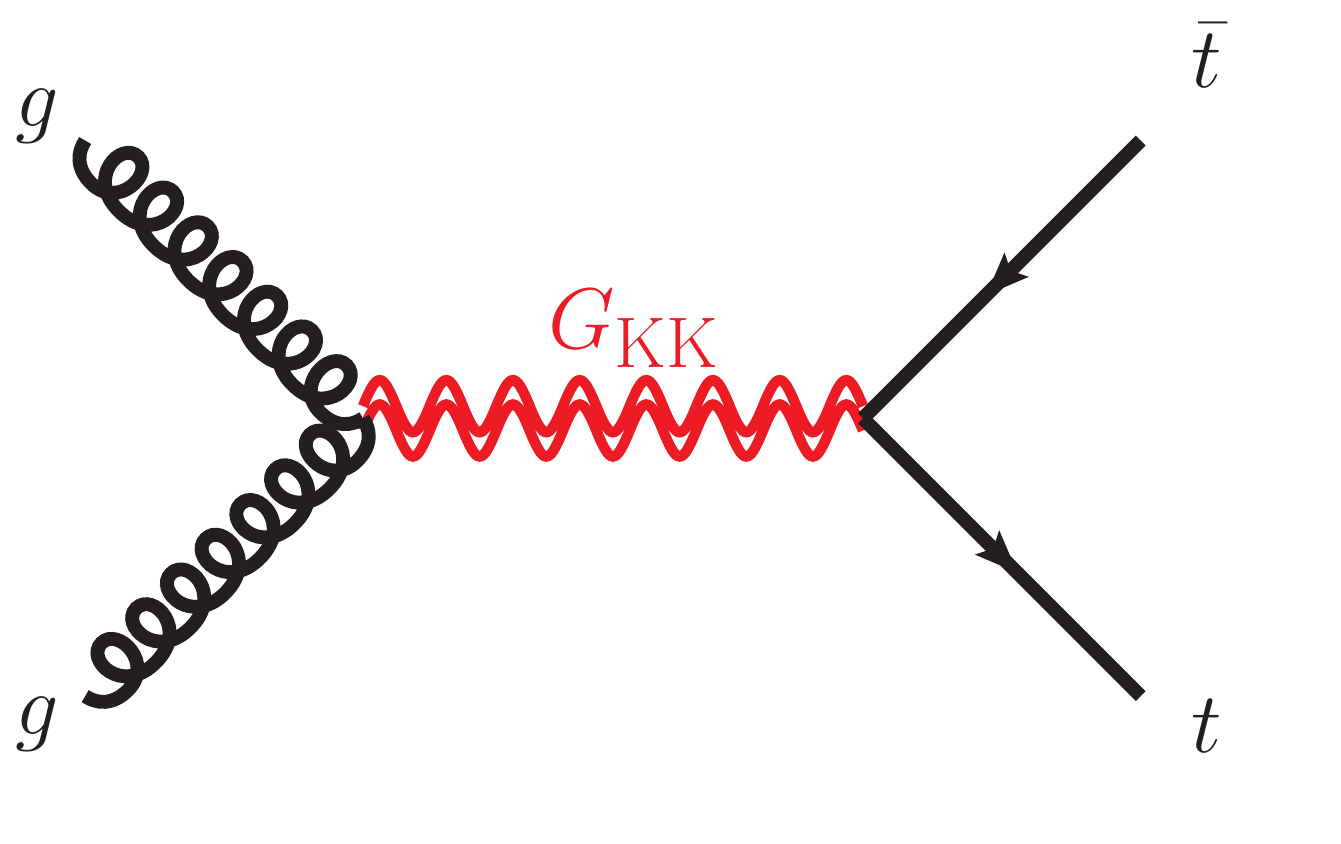}\label{fig:BSMFeynGrav}}
\subfigure[$g_{KK} \rightarrow t\bar{t}$]{\includegraphics[width=0.30\textwidth]{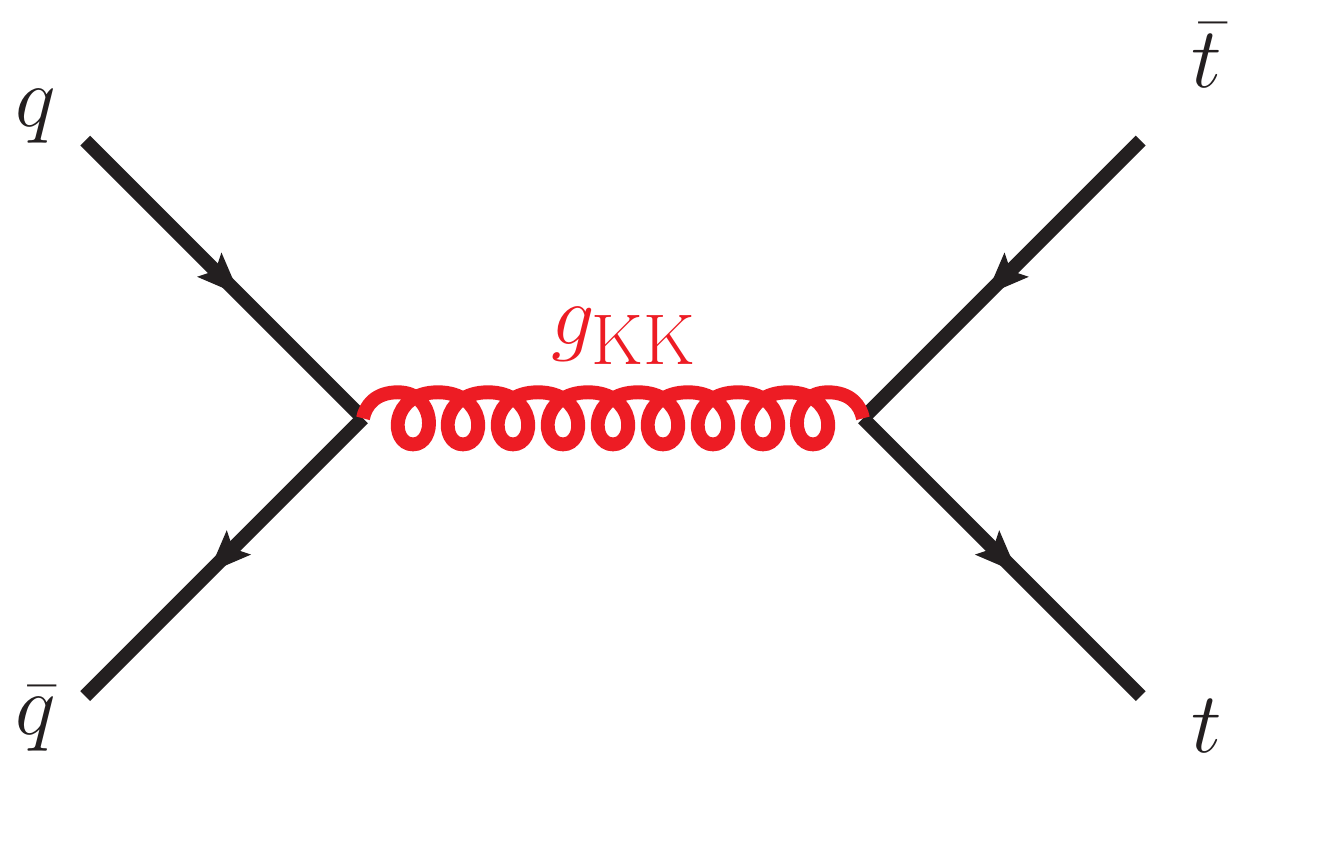}\label{fig:BSMFeynGlue}}}
\caption{Leading-order Feynman diagrams for the signal processes studied in this search. The $Z'$ (a) and Kaluza--Klein gluons ($g_{KK}$) have spin 1 (b), while the Kaluza--Klein graviton ($G_{KK}$) has spin 2 (c).
\label{fig:BSMFeyn}}
\end{figure}

\subsection{Spin-1 colour singlet}

Spin-1 colour singlets that decay into $t\bar{t}$ are predicted in many SM extensions. Three different types of $Z'$ bosons are explored in this study: one arising in topcolor-assisted-technicolor (TC2) models~\cite{Hill:1993hs,Hill:1994hp} and two others arising in simplified models of dark matter~\cite{Albert:2017onk}. The primary production mode is $q\bar{q}$ annihilation as shown in Figure~\ref{fig:BSMFeynZp}.

 The TC2 benchmark model chosen for this search  produces a $Z'$ boson, denoted $Z'_{\text{TC2}}$. This is a leptophobic boson, with couplings only to first- and third-generation quarks, referred to as Model IV~\cite{topcolor2}. The properties of the boson are controlled by three parameters:  the topcolour tilting parameter, $\cot \theta_\mathrm{H}$, which controls the width and the production cross-section, and $f_1$ and $f_2$, which are related to the coupling to up-type and down-type quarks, 
respectively. Here $f_1=1$ and $f_2=0$, which maximises the fraction of $Z'_\mathrm{TC2}$ bosons that decay into $t\bar{t}$.  The parameter $\cot \theta_\mathrm{H}$ is tuned\footnote{There is a one-to-one mapping between $\cot \theta_\mathrm{H}$ and the width, given a fixed mass, as shown in Eq.~(6) of Ref.~\cite{Harris:2011ez}.} for each mass point such that the resonance has a width of 1\%  of its mass~\cite{Harris:2011ez}. Previous searches by the ATLAS~\cite{Aad:2015fna} and CMS~\cite{Khachatryan:2015sma, Sirunyan:2017uhk} collaborations set lower limits of $ m(Z'_{\text{TC2}}) > 1.8 \,\TeV$ and $ m(Z'_{\text{TC2}}) > 2.5 \,\TeV$, respectively, on the allowed mass for such bosons.
As the detector resolution is not sufficient to resolve the resonance width for the $Z'_{\text{TC2}}$ model, limits are also quoted assuming a 3\% width.
A previous search by the ATLAS Collaboration~\cite{Aad:2015fna} set a lower limit of $ m(Z'_{\text{TC2}}) > 2.3 \,\TeV$ on the mass for such bosons. 

Interactions between dark matter and normal matter may be mediated by weakly coupled \TeV{}-scale particles. This search considers an axial-vector mediator, $Z'_{\text{DM,ax}}$ and a vector mediator, $Z'_{\text{DM,vec}}$, within a framework of simplified models proposed by the LHC Dark Matter Working group~\cite{Albert:2017onk}. There are five free parameters for these mediators: the coupling to quarks ($g_q$), the coupling to leptons ($g_\ell$), the coupling to dark matter ($g_{\text{DM}}$), the dark-matter mass ($m_{\text{DM}}$) and the mediator mass. The mediator mass is varied between 0.5~\TeV{} and 5~\TeV{} with the other parameters set to $g_q=0.25$, $g_\ell=0$, $g_{\text{DM}}=1$, and $m_{\text{DM}}=10$~\GeV{} following the benchmarks A1 and V1 defined in Ref.~\cite{Albert:2017onk}.
The width of $Z'_{\text{DM,ax}}$ and $Z'_{\text{DM,vec}}$ are 5.6\% of their masses, with the $Z'_{\text{DM,ax}}$ width kinematically limited to 5.3\% at 0.5~\TeV{}.

\subsection{Spin-2 colour singlet}

Spin-2 colour-singlet bosons are produced in models that postulate extra dimensions of space leading to Kaluza--Klein excitations of the graviton. This search considers a Randall--Sundrum (RS) model with an extra dimension where the SM fields are in the warped bulk and the fermions are localised appropriately to explain the flavour structure of the SM~\cite{Randall:1999ee, Agashe:2007zd, Fitzpatrick:2007qr}. This kind of graviton ($G_\mathrm{KK}$) is commonly referred to as a `Bulk' RS graviton and is characterised by a dimensionless coupling constant $k/\bar{M}_{\mathrm{Pl}} \sim 1$, where $k$ is the curvature of the warped extra dimension and $ \bar{M}_{\mathrm{Pl}} = M_{\mathrm{Pl}}/\sqrt{8 \pi}$ is the reduced Planck mass. For these gravitons, decays into light fermions are suppressed and the branching ratio to photons is negligible.  The primary production mode is gluon--gluon fusion as shown in Figure~\ref{fig:BSMFeynGrav}.
The branching ratios to $t\bar{t}$, $WW$, $ZZ$ and $HH$ are significant. In this particular model, $k/\bar{M}_{\mathrm{Pl}}$ is chosen to be 1, and the $G_\mathrm{KK}$ width varies from 3\% to 6\% in the  mass range 0.4--3 \TeV{}.
The branching ratio of the $G_\mathrm{KK}$ decay into a $t\bar{t}$ pair increases rapidly from 18\% to 50\% for masses between 400 and 600~\GeV, plateauing at 68\% for masses larger than 1~\TeV.
The ATLAS Collaboration's search for such gravitons in $\sqrt{s}=8$~\TeV\ data in the \ttbar decay channel set cross-section limits but did not exclude any graviton masses~\cite{Aad:2015fna}, while the search for the same model in the $G_\mathrm{KK} \rightarrow ZZ$ channel~\cite{Aad:2014xka} excluded a Bulk RS $G_\mathrm{KK}$ with mass less than 740~\GeV. The CMS Collaboration  performed
searches in the $G_\mathrm{KK} \rightarrow ZZ$ and $G_\mathrm{KK} \rightarrow WW$ decay channels~\cite{Khachatryan:2014gha, Sirunyan:2017jtu} excluding such RS gravitons with masses less than 1.3~\TeV{}.

\subsection{Spin-1 colour octet}
Spin-1 colour-octet bosons are produced in models that postulate extra dimensions of space leading to Kaluza--Klein excitations of the gluon.
This search considers heavy Kaluza--Klein gluons, $g_\mathrm{KK}$, as produced in RS  models with a single warped extra dimension~\cite{Lillie:2007yh, Ask:2011zs}, with widths varying between 10\% and 40\% of the $g_{\text{KK}}$ mass.  The primary production mode in both cases is $q\bar{q}$ annhilation as shown in Figure~\ref{fig:BSMFeynGlue}. The strong coupling of these gluon excitations to light quarks is set to $g_q=-0.2 g_{\mathrm{s}}$, where $g_{\mathrm{s}}$ is the SM gluon coupling.\footnote{The couplings used here correspond to the configuration mentioned in Eq.~(2.3) of Ref.~\cite{Lillie:2007yh}.} The left-handed coupling to the top quark is fixed at $g_{\mathrm{L}^{(t)}} = g_{\mathrm{s}}$, and the right-handed coupling to the top quark, $g_{\mathrm{R^{(t)}}}$, is varied to obtain the desired width.  A previous search using $\sqrt{s}=8$~\TeV\ ATLAS data~\cite{Aad:2013nca} excludes a similar $g_\mathrm{KK}$ (15\% width) with a mass less than  2.2~\TeV. The CMS Collaboration searched for similar resonances~\cite{Sirunyan:2017uhk}, using a slightly different benchmark model~\cite{Agashe:2006hk}. The CMS choice leads to a natural width of 20\% and a larger production cross-section, and, for such a scenario, CMS excludes the existence of $g_\mathrm{KK}$ with masses less than 3.3~\TeV.

\section{ATLAS detector}
\label{sec:detector}

The ATLAS detector~\cite{PERF-2007-01} at the LHC covers nearly the entire solid angle around the collision point.
It consists of an inner tracking detector surrounded by a thin superconducting solenoid, electromagnetic and hadronic calorimeters and a muon spectrometer incorporating three large superconducting toroid magnets.

A high-granularity silicon pixel detector covers the vertex region and typically provides four measurements per track. The innermost layer, known as the insertable B-Layer~\cite{ATLAS-TDR-19}, was added in 2014 and provides high-resolution hits at small radius to improve the tracking performance. The silicon pixel detector is followed by a silicon
microstrip tracker that
typically provides four measurements from four strip double layers.
These silicon detectors are complemented by a transition radiation tracker (TRT),
which enables radially extended track reconstruction up to $|\eta| = 2.0$.\footnote{ATLAS uses a right-handed coordinate system with its origin at the nominal interaction point (IP)
in the centre of the detector and the $z$-axis along the beam pipe.
The $x$-axis points from the IP to the centre of the LHC ring,
and the $y$-axis points upwards.
Cylindrical coordinates $(r,\phi)$ are used in the transverse plane, 
$\phi$ being the azimuthal angle around the $z$-axis.
The pseudorapidity is defined in terms of the polar angle $\theta$ as $\eta = -\ln \tan(\theta/2)$.
Angular distance is measured in units of $\Delta R \equiv \sqrt{(\Delta\eta)^{2} + (\Delta\phi)^{2}}$.}
The TRT also provides electron identification information 
based on the fraction of hits (typically 30 in total) above a higher energy-deposit threshold corresponding to transition radiation.
The inner-detector system (ID) is immersed in a \SI{2}{\tesla} axial magnetic field 
and provides charged-particle tracking in the pseudorapidity range $|\eta| < 2.5$.

The calorimeter system covers the pseudorapidity range $|\eta| < 4.9$.
Within the region $|\eta|< 3.2$, electromagnetic calorimetry is provided by barrel and 
endcap high-granularity lead/liquid-argon (LAr) electromagnetic calorimeters,
with an additional thin LAr presampler covering $|\eta| < 1.8$
to correct for energy loss in material upstream of the calorimeters.
Hadronic calorimetry is provided by a steel/scintillator-tile calorimeter,
segmented into three barrel structures within $|\eta| < 1.7$, and two copper/LAr hadronic endcap calorimeters.
The solid angle coverage is completed with forward copper/LAr and tungsten/LAr calorimeter modules
optimised for electromagnetic and hadronic measurements, respectively.

The muon spectrometer  comprises separate trigger and
high-precision tracking chambers measuring the deflection of muons in a magnetic field generated by superconducting air-core toroids.
The precision chamber system covers the region $|\eta| < 2.7$ with three layers of monitored drift tubes,
complemented by cathode strip chambers in the forward region, where the background is highest.
The muon trigger system covers the range $|\eta| < 2.4$ with resistive plate chambers in the barrel and thin gap chambers in the endcap regions.

A two-level trigger system~\cite{Artz:2015pca,Aaboud:2016leb} is used to select interesting events.
The first level of the trigger is implemented in hardware and uses a subset of detector information
to reduce the event rate to a design value of at most \SI{100}{\kHz}.
This is followed by a software-based trigger that reduces the event rate to a maximum of around \SI{1}{\kHz} for offline storage.

\section{Data and Monte Carlo samples}
\label{sec:samples}

This search is performed using data from $\sqrt{s}=13$~\TeV{} proton--proton collisions recorded by the ATLAS detector in 2015 and 2016. Only data recorded during stable beam conditions and with all relevant subdetector systems operational are used.  The integrated luminosity of the data sample is  $36.1$~fb$^{-1}$. Lepton-plus-jets events were collected using single-electron and single-muon triggers.

The SM background processes are, in order of decreasing importance: the production of $t\bar{t}$, a $W$ or $Z$ boson in association with additional jets ($W/Z$ + jets), a single top quark, multi-jets and dibosons. Simulated Monte Carlo (MC) data samples are used for signal processes, as well as for background processes that produce jets and prompt leptons. The MC samples are used to optimise the event selection, provide SM background estimates, and evaluate signal efficiencies. The multi-jet background is evaluated directly from data as described in Section~\ref{sec:estimate}.

For the generation of SM $t\bar{t}$ events~\cite{Frixione:2007nw} and 
single-top-quark events in the $Wt$-~\cite{Re:2010bp} and 
s-channels~\cite{Alioli:2009je}, the \POWHEGV{v2}~\cite{Nason:2004rx,Frixione:2007vw,Alioli:2010xd} generator with the CT10~\cite{Lai:2010vv,Gao:2013xoa} parton distribution function (PDF) set was used. The overlap between $t\bar{t}$ and $Wt$ production 
was treated within the diagram removal (DR) scheme~\cite{Frixione:2008yi}. 
Electroweak t-channel single-top-quark events 
were generated using \POWHEGBOXV{v1}~\cite{Frederix:2012dh}. This 
generator uses the four-flavour scheme for the next-to-leading-order (NLO) matrix element calculations 
together with the four-flavour PDF set CT10f4. For this process, the top-quark
decays were simulated using \MADSPIN~\cite{Artoisenet:2012st}, preserving all spin 
correlations. For all SM top-quark processes the parton shower, fragmentation
 and the  underlying event were simulated using \PYTHIAV{v6.428}~\cite{Sjostrand:2006za} with  the CTEQ6L1~\cite{Pumplin:2002vw} PDF set and the corresponding 
Perugia 2012 (P2012) set of tuned parameters~\cite{Skands:2010ak}. The top quark's mass was set to 
172.5 \GeV{}. The {\textsc{EvtGen}}~v1.2.0 program~\cite{Lange:2001uf} was used to model the decays of heavy-flavour hadrons. For the generation of $t\bar{t}$ events, the $h_{\mathrm{damp}}$ parameter, which controls the transverse momentum of the first additional emission beyond the Born configuration, was set to the mass of the top quark. The main effect of this parameter is to regulate the high transverse momentum emission against which the $t\bar{t}$ system recoils. The top-quark kinematics in all SM $t\bar{t}$ samples were corrected to account for  higher-order electroweak (EW) effects~\cite{Kuhn:2013zoa}.
This correction to the generated samples was made by applying a weight 
that depends on the flavour and energy of the initial partons in the 
centre-of-mass frame, and on the decay angle of the top quarks in the 
same frame.
The value of the correction factor decreases with the invariant mass of the $t\bar{t}$ system from 0.98 at a mass of 0.4~\TeV{} to 0.87 at a mass of 3.5~\TeV{}.

Samples of $W/Z$ + jets events were simulated using the \SHERPAV{2.2.1}~\cite{Gleisberg:2008ta} generator. Matrix elements were calculated for up to two partons at NLO in QCD and four partons at leading order (LO) using the {\textsc{Comix}}~\cite{Gleisberg:2008fv} and {\textsc{OpenLoops}}~\cite{Cascioli:2011va} matrix element generators and merged with the \SHERPA\ parton shower~\cite{Schumann:2007mg} using the ME+PS@NLO prescription~\cite{Hoeche:2012yf}. The NNPDF3.0 NLO PDF set~\cite{Ball:2014uwa} was used in conjunction with dedicated parton shower tuning developed by the authors of \SHERPA. The $W/Z$ + jets events were normalised to the next-to-next-to-leading-order (NNLO) cross-sections~\cite{Gavin:2012sy}.

Diboson ($WW, WZ, ZZ$) production  processes with four charged leptons ($4\ell$), three charged leptons and one neutrino ($3\ell+\nu$), two charged leptons and two neutrinos ($2\ell+2\nu$), or one charged lepton and one neutrino plus jets ($\ell \nu q\bar{q}'$) were simulated using the \SHERPAV{2.1.1} generator. The matrix elements contain all diagrams with four EW vertices. They were calculated for zero ($3\ell+\nu$, $\ell \nu q\bar{q}'$) or up to one  ($4\ell$, $2\ell+2\nu$) additional partons at NLO in QCD and up to three partons at LO using the {\textsc{Comix}} and {\textsc{OpenLoops}} matrix element generators and were merged with the \SHERPA\ parton shower using the ME+PS@NLO prescription. The CT10 PDF set was used  with the dedicated parton shower tuning developed by the \SHERPA\ authors. The cross-sections from the generator were used for sample normalisation.

Production of a new spin-1 colour-singlet particle that decays into $t\bar{t}$ was modelled using the $Z' \rightarrow t\bar{t}$ process from \PYTHIAV{v8.165}~\cite{Sjostrand:2007gs} with the NNPDF2.3 LO~\cite{Ball:2012cx} PDF set and the A14~\cite{ATL-PHYS-PUB-2014-021} set of tuned parameters. This search uses topcolour-assisted technicolour $Z'_{\mathrm{TC2}}$~\cite{Hill:1994hp,topcolor2,Harris:2011ez} as a benchmark.  To account for higher-order contributions to the cross-section, the samples were normalised to cross-section calculations performed at NLO in QCD~\cite{Bonciani:2015hgv} using the PDF4LHC2015 PDF set~\cite{Butterworth:2015oua}. The same sample, reweighted to have the appropriate resonance width as simulated in \MGMCatNLO~\cite{Alwall:2014hca}, was used to model $Z'_{\text{DM,ax}}$ and $Z'_{\text{DM,vec}}$ with the cross-sections normalised to LO QCD calculations using the NNPDF2.3 LO PDF set.
No corrections for higher-order EW effects were applied to these signal samples.

Production of a spin-1 colour-octet particle that decays into $t\bar{t}$ was modelled using the $g_{\mathrm{KK}} \rightarrow t\bar{t}$ process from \PYTHIAV{8.165} at leading order with the NNPDF2.3 LO PDF set and the A14 set of tuned parameters. 

The case of a spin-2 colour-singlet signal 
was modelled using \MGMCatNLO\ with the NNPDF2.3 LO PDF set, 
with parton showering performed by \PYTHIAV{v8.165} with the A14 set of tuned parameters.

The MC samples were processed through the full ATLAS detector 
simulation~\cite{Aad:2010ah} based on \GEANT~4~\cite{Agostinelli:2002hh} or through a faster simulation making use of parameterised showers in the 
calorimeters~\cite{ATLAS:2010bfa}.
The $t\bar{t}$ parton shower uncertainty is estimated using samples passed through the ATLAS fast simulation.
Additional  simulated proton--proton 
collisions generated using \PYTHIAV{v8.165} with the A2 set of tuned parameters~\cite{ATL-PHYS-PUB-2012-003} and the MSTW2008LO PDF  set~\cite{Martin:2009iq} were overlaid to simulate the effects of additional 
collisions from the same and nearby bunch crossings (pile-up). All simulated events were then processed using the same reconstruction algorithms and analysis chain as used for real data.

\section{Event selection and $t\bar{t}$ reconstruction}
\label{sec:selrec}

This section describes the selection of events containing a single charged lepton, hadronic jets, and large missing transverse momentum.
The construction of an observable that approximates the mass of the 
$t\bar{t}$ system and the categorisation of the events are also described.

\subsection{Event selection}
\label{sec:sel}
The event selection criteria are applied to the following physics objects:
\begin{description}
\item {\textbf{Hadronic jets}} defined in three different ways are used in this analysis.
\begin{description}
\item {\em{Small-$R$} jets} are built from three-dimensional topological clusters~\cite{Aad:2016upy} of energy in the calorimeters, calibrated at the electromagnetic (EM) energy scale, using the anti-$k_t$ algorithm~\cite{Cacciari:2008gp} with a radius parameter $R=0.4$. The jet energy is calibrated using a correction that relates the reconstructed jet energy to the true jet energy when reconstructed from stable particles with a lifetime of at least 30~ps (excluding muons and neutrinos)~\cite{Aad:2011he}.
The correction depends on the transverse momentum and pseudorapidity of each jet, and accounts for pile-up effects~\cite{ATL-PHYS-PUB-2015-015}.
They are required to have transverse momentum, $p_{\text{T}}$, greater than 25~\GeV{} and $|\eta| < 2.5$.
For jets with $p_{\text{T}} < 60$~\GeV{} and $|\eta| < 2.4$, a jet-vertex-tagger requirement corresponding to a 92\% efficiency while rejecting 98\% of jets from pile-up and noise is imposed~\cite{ATLAS-CONF-2014-018}.

\item {\em{Large-$R$ jets}} are built from three-dimensional topological clusters of energy in the calorimeters, calibrated with the local cluster weighting (LCW) procedure~\cite{PERF-2014-07},  using the anti-$k_t$ algorithm with a radius parameter $R=1.0$. In the LCW calibration procedure, corrections for the non-compensating response of the calorimeter and for the energy lost in dead material and from out-of-cluster leakage are applied to the cluster energy  before applying the jet algorithm. These corrections are obtained from simulations of charged and neutral particles. These jets are further {\em{trimmed}}~\cite{Krohn:2009th}, which mitigates the effects of pile-up~\cite{Aad:2013gja}. In trimming, the constituents of a jet are reclustered into subjets according to the $k_t$ algorithm~\cite{Catani:1991hj,Ellis:1993tq,Catani:1993hr} with a radius parameter $R_{\text{sub}}$. Subjets with a transverse momentum smaller than a fraction $f_{\text{cut}}$ of the parent jet's transverse momentum are then discarded. The surviving subjets are recombined to produce the final trimmed jet. Based on a study of sensitivity to pile-up, the trimming parameters used are $R_{\text{sub}}= 0.2 $ and $f_{\text{cut}}=0.05$~\cite{ATLAS-CONF-2015-035}. The jets are calibrated using corrections that relate the reconstructed jet to its true jet when clustered from stable particles with a lifetime of at least 30~ps (excluding muons and neutrinos)~\cite{Aad:2011he,ATLAS-CONF-2016-035}. The resultant jets are required to have $p_{\text{T}} > 300$~\GeV{} and $|\eta| < 2.0$.  Large-$R$ jets consistent with the decay products of a hadronically decaying top quark are identified ({\em{top-tagged}}) using an algorithm~\cite{ATL-PHYS-PUB-2015-053} based on the invariant mass of the jet and the N-subjettiness ratio $\tau_{32}$~\cite{Thaler:2010tr,Thaler:2011gf}. This algorithm has an efficiency of approximately 80\% for selecting top-quark jets  with $p_{\text{T}}> 300$~\GeV{} and $|\eta| < 2.0$ in simulated SM $t\bar{t}$ events. 
\item {\em{Track-jets}} are built from charged-particle tracks using the anti-$k_t$ algorithm with a radius parameter $R=0.2$. These jets are required to have $p_{\text{T}} > 10$~\GeV{} and $|\eta| < 2.5$ and at least two constituent charged-particle tracks. The charged-particle tracks used to build the jets must themselves have $p_{\text{T}} > 0.4$~\GeV{} and $|\eta| < 2.5$, and pass quality requirements that test the number of hits used to reconstruct the track and the matching to the primary vertex~\cite{ATL-PHYS-PUB-2015-035}.
Track-jets consistent with including the decay products of a $b$-hadron are identified ({\em{$b$-tagged}}) using the MV2c20 algorithm~\cite{ATL-PHYS-PUB-2015-022}. The $b$-tagging working point chosen has approximately 70\% efficiency for such jets to contain a $b$-hadron in simulated SM $t\bar{t}$ events.
The track-jets are used in this analysis for the identification of the $b$-tagged small-$R$ calorimeter-measured jets.
Small-$R$ calorimeter-measured jets, $j_{\text{calo}}$, are identified as $b$-jets if a track-jet that passes the $b$-tagging selection, $j_{\text{track}}$, satisfies the $\Delta R(j_{\text{calo}}, j_{\text{track}}) < 0.4$ requirement.
\end{description}
The anti-$k_t$ and $k_t$ algorithms are applied through their implementation in FastJet~\cite{Cacciari:2005hq,Cacciari:2011ma}.
\item{\textbf{Muon candidates}} are reconstructed by combining tracks found in the ID with tracks found in the muon spectrometer that satisfy $p_{\text{T}} > 25$~\GeV{} and $|\eta| < 2.5$.
Muons are required to be isolated using the requirement that the sum of the $p_{\text{T}}$ of the tracks in a variable-size cone around the muon direction (excluding the track identified as the muon) be less than 6\% of the transverse momentum of the muon. The track isolation cone size is given by the minimum of $\Delta R = 10\ \GeV{}/p_{\text{T}}^\mu$ and $\Delta R = 0.3$, where $p_{\text{T}}^{\mu}$ is the muon $p_{\text{T}}$. Thus, the cone radius increases with decreasing $p_{\text{T}}$ up to a maximum of 0.3. 
To reduce the background contributions due to muons from heavy-flavour decays inside jets, muons are removed if they are separated from the nearest jet by $ \Delta R < 0.04+10\ \GeV{}/p^{\mu}_{\text{T}}$. However, if the jet has fewer than three associated tracks, the muon is kept and the jet is removed instead; this avoids an inefficiency for high-energy muons undergoing significant energy loss in the calorimeter.
\item {\textbf{Electron candidates}} are reconstructed from an isolated energy deposit in the electromagnetic calorimeter  matched to an ID track, within the fiducial region of transverse energy $E_{\text{T}} > 25$~\GeV{} and $|\eta| < 2.47$. Candidates within the transition region between the barrel and endcap electromagnetic calorimeters, $1.37 < |\eta| < 1.52 $, are removed. A tight likelihood-based requirement~\cite{ATLAS-CONF-2014-032} is used to further suppress the background from multi-jet production.
Electrons are also required to be isolated, using the same track-based variable as for muons, except that the maximum $\Delta R$ in this case is 0.2.
Electrons sharing the same track with a muon candidate are assumed to be bremsstrahlung photon and are rejected as electron candidates.
To prevent double-counting of electron energy deposits as jets, the closest small-$R$ jet within $\Delta R = 0.2$ of a reconstructed electron is removed. Finally, if the nearest small-$R$ jet surviving this selection is within $\Delta R = 0.4$ of the electron, the electron is discarded, to ensure it is sufficiently separated from nearby jet activity. This procedure is referred to as ``overlap removal''.
\item {The \textbf{Missing transverse momentum}}, $\met{}$,  is defined as the magnitude of $\vec{E}^{\text{miss}}_{\text{T}}$, which is the negative of the total vector sum $p_{\text{T}}$ of all selected physics objects (electrons, muons, small-$R$ jets) as well as specific `soft terms' considering tracks that do not match the selected physics objects. In this way, the missing transverse momentum is adjusted to take into account the best calibration of the identified physics objects~\cite{ATL-PHYS-PUB-2015-027}.
\end{description}
In addition:
\begin{description}
\item {The \textbf{primary vertex}} is defined as the vertex with the highest sum of squared transverse momentum of the tracks associated with it.
\end{description}

Following the initial selection by the triggers described in Section~\ref{sec:samples}, the event selection proceeds with the following steps:
\begin{enumerate}
\item {\textbf{Event cleaning requirement:}} Events are required to have been recorded when all subsystems of the ATLAS detector were working acceptably. Events are also required to have at least two tracks associated with the primary vertex.
\item {\textbf{Charged-lepton selection:}} Exactly one charged-lepton candidate (electron or muon) is required with a minimum $p_{\text{T}}$ of 30~\GeV{}. The lepton candidates must geometrically match the candidate that triggered the event.  Events containing a second charged lepton with a transverse momentum larger than 25~\GeV{} are rejected.
\item {\textbf{Leptonic-$W$ selection:}} The event is required to have a charged lepton and missing transverse momentum consistent with the leptonic decay of a $W$ boson. This is achieved by requiring that the event satisfies two criteria. Firstly, the \met is required to be greater than 20~\GeV. Secondly, the transverse mass of the selected lepton, $\ell$, and $\vec{E}^{\text{miss}}_{\text{T}}$, $m_{\text{T}}^{W}=\sqrt{2 {\ensuremath{p_{\text{T}}^{\ell}}\xspace}\met(1-\cos \Delta \phi (\ell,\vec{E}^{\text{miss}}_{\text{T}}))}$, is required to satisfy $\met + m_{\text{T}}^{W}> 60$ \GeV{}.
\item {\textbf{$b$-tagging:}} The event is required to contain at least one $b$-tagged track-jet.
The $b$-tagged track-jets are used to categorise the accepted events into several channels. More information about this is given at the end of this section.
\item{\textbf{Classification into Boosted or Resolved selection:}} Based on the hadronic activity, the event is classified as {\textit{Boosted}} or {\textit{Resolved}} as described below.
\end{enumerate}

An event passes the boosted selection if it meets the following criteria:
\begin{enumerate}
\item {\textbf{Leptonic-top $b$-jet:}} Events are required to contain at least one small-$R$ jet with $\Delta R(\text{jet},\text{lepton}) < 1.5$. If multiple jets satisfy this condition, the one with the highest $p_{\text{T}}$ is chosen and subsequently referred to as the {\em{selected jet}}, $j_{\text{sel}}$. This is identified with the expected $b$-jet from the leptonic top-quark decay, although no $b$-tagging requirement is enforced on it. 
This definition is found to yield better resolution for the invariant mass of the \ttbar\ system than others based on $b$-tagging or information about the top-quark candidate's mass.
\item {\textbf{Hadronic-top jet:}} Events are required to contain at least one large-$R$ jet, $j_{\text{top}}$, passing the top-tagging requirements. The jet is further required to be well separated from the leptonically decaying top quark by requiring differences in azimuthal angle between it and the charged lepton $\Delta \phi(j_{\text{top}},\text{lepton}) > 2.3$ and   $\Delta R(j_{\text{top}},j_{\text{sel}})> 1.5$. The highest-$p_{\text{T}}$ jet passing all of these requirements is referred to as the hadronic-top jet.
\end{enumerate}
Events that fail any of these boosted selection requirements are classified as passing the resolved selection if there are at least four small-$R$ jets with $p_{\text{T}} > 25$~\GeV{} and if the $\chi^2$ algorithm for reconstructing the $t\bar{t}$ system (described in Section~\ref{sec:rec}) yields a value of $\log_{10}(\chi^2) < 0.9$. 
This selection requirement has been found to effectively reject $t\bar{t}$ events not correctly reconstructed and a fair fraction of the other background, while improving the actual resolution on the ttbar mass system.

\begin{figure}[htbp]

\centering{
\subfigure[$Z' \rightarrow t\bar{t}$]{\includegraphics[width=0.45\textwidth]{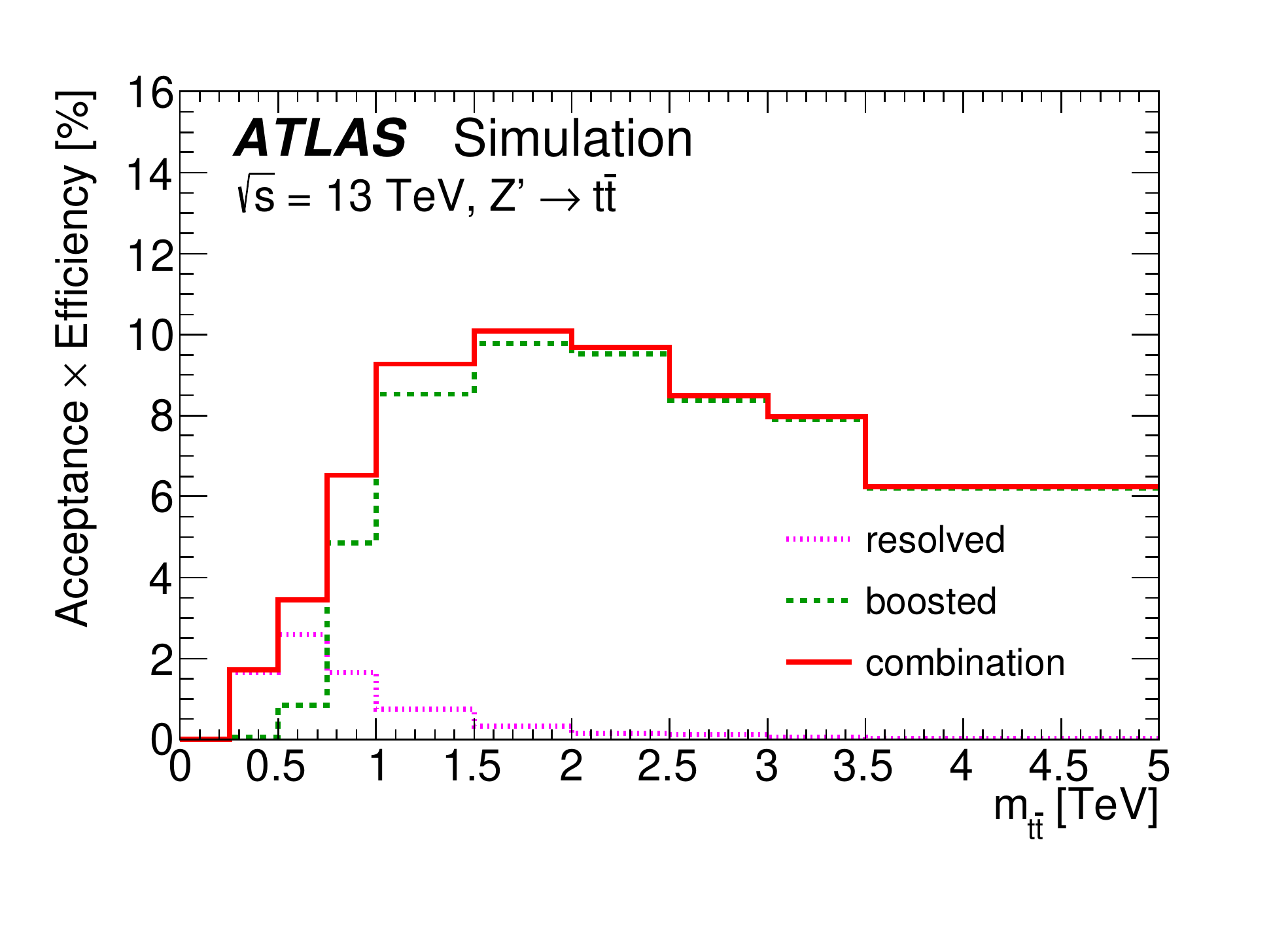}}
\subfigure[$g_{\mathrm{KK}} \rightarrow t\bar{t}$]{\includegraphics[width=0.45\textwidth]{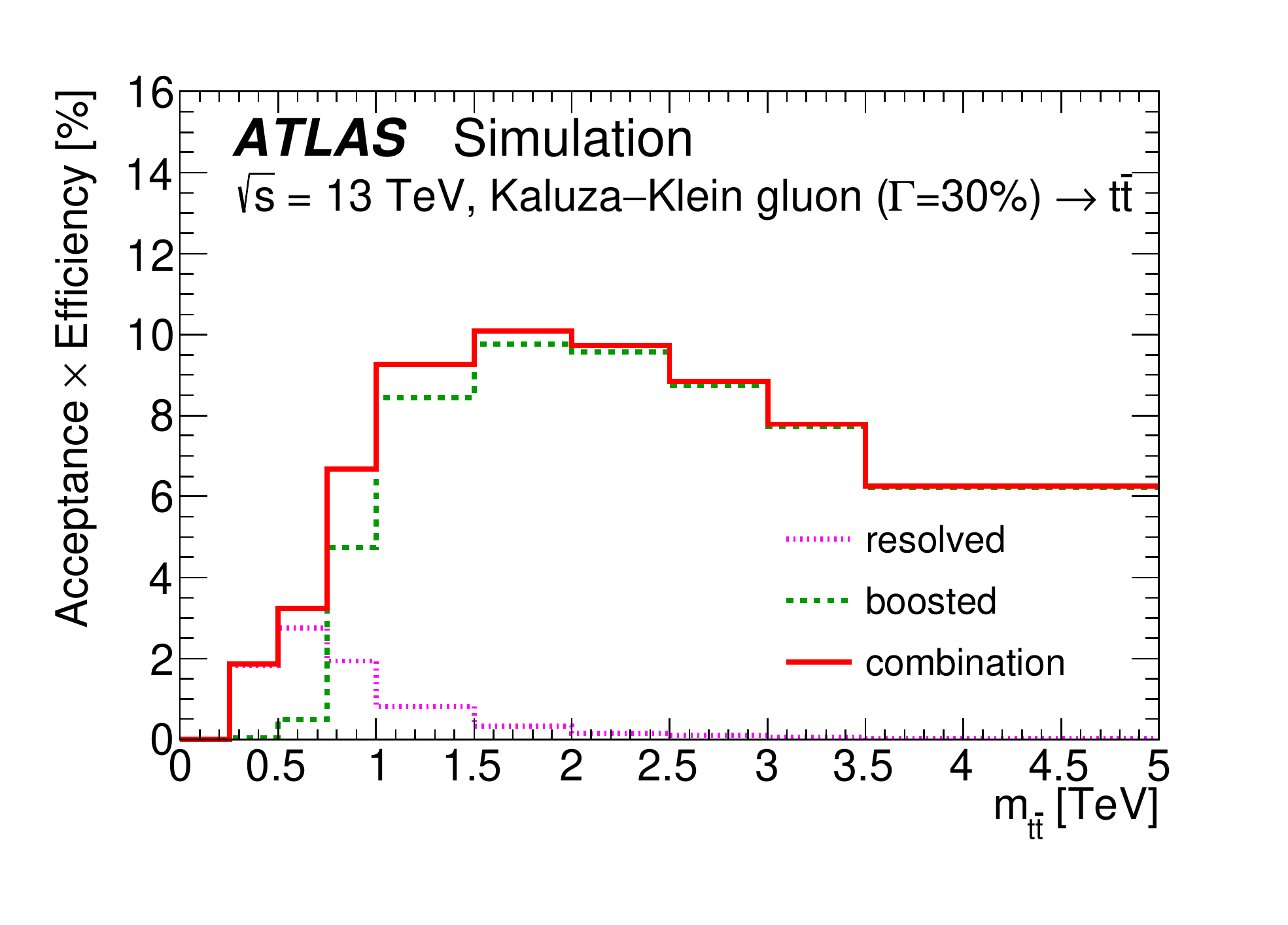}}
\subfigure[$G_{\mathrm{KK}} \rightarrow t\bar{t}$]{\includegraphics[width=0.45\textwidth]{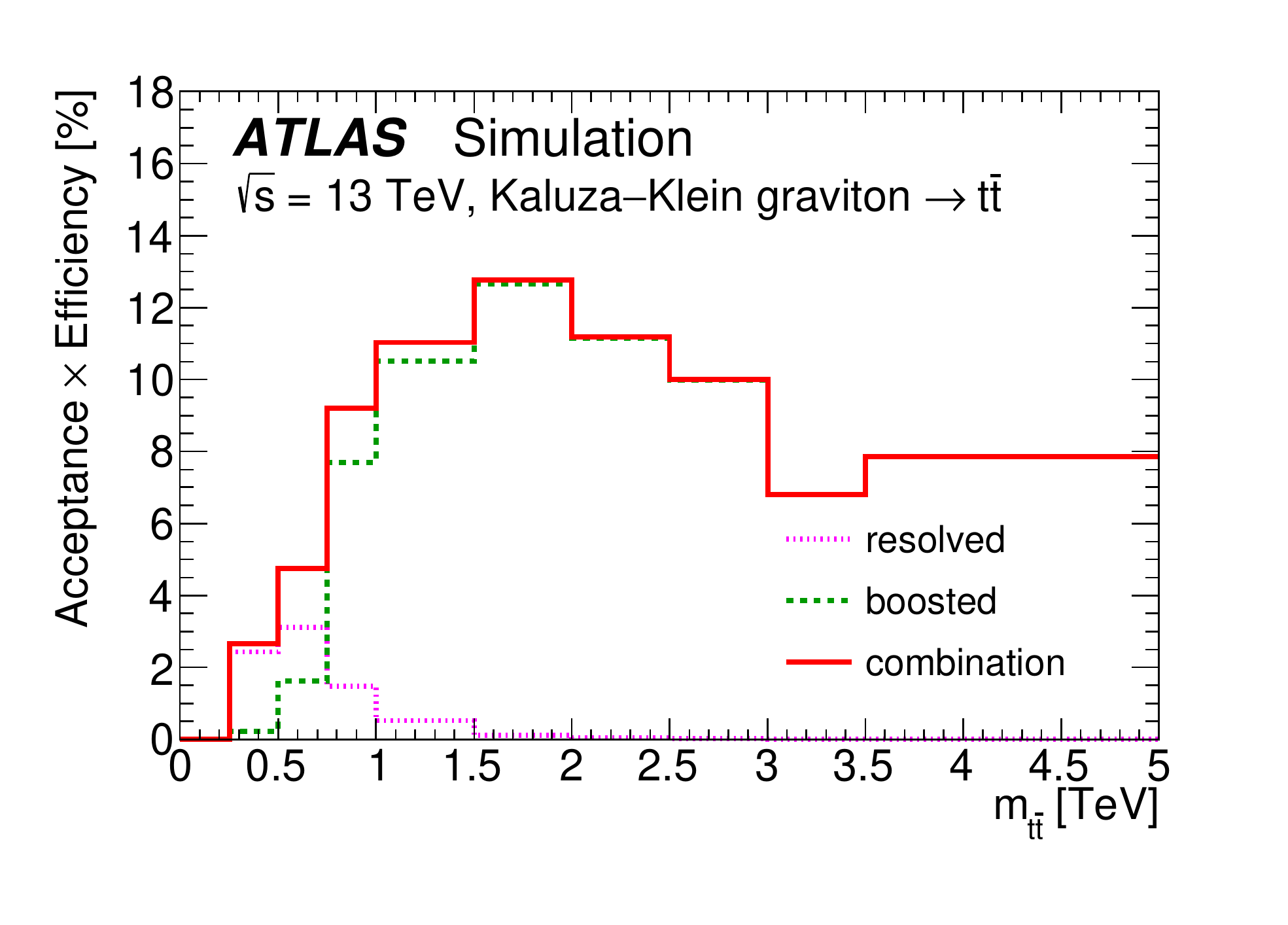}}

}
\caption{Acceptance times efficiency ($A \times \epsilon$), including the branching ratio for MC simulated BSM particles decaying into $t\bar{t}$, 
as a function of the \ttbar invariant mass $m_{t\bar{t}}$ (computed before parton radiation) for simulated signal events.
The signal samples shown here include events from generated masses ranging from 0.4~\TeV{} to 5~\TeV{}.
All $t\bar{t}$ decay modes are simulated. The $e$ and $\mu$ channel efficiencies are combined. 
\label{fig:BSMSelectionEffSum}}
\end{figure}

The acceptance times efficiency ($A \times \epsilon$) including the branching ratio for simulated beyond-the-SM (BSM) particles decaying into $t\bar{t}$ is given in Figure~\ref{fig:BSMSelectionEffSum}. For reference, the branching ratio for $t\bar{t}$ to electron- or muon-plus-jets is about 17\% for each lepton flavour, taking into account leptonic $\tau$-lepton decays~\cite{Agashe:2014kda}.
There are efficiency losses from the large-$R$ jet requirements and the $b$-tagging requirement, as well as the four-jet and $\chi^2$ kinematic fit requirement in the resolved channel. The value of $A \times \epsilon$ is smaller for $e$+jets events than $\mu$+jets for resonance masses above 1.5~\TeV{}, due to the inefficiency of the electron identification and overlap removal in an environment with highly boosted top quarks. For the $Z'$ and $g_{\text{KK}}$ signals, the $A \times \epsilon$ values are very similar to each other, whereas the total $G_{\text{KK}}$ $A \times \epsilon$ is about two percentage points higher than the other signals for masses greater than 0.8~\TeV{},
because the $G_{\text{\text{KK}}}$ produces top quarks that are more central than those produced by $g_{\mathrm{KK}}$.

\subsection{Mass reconstruction and event categorisation}
\label{sec:rec}

Following the event selection, an observable $m_{t\bar{t}}^{\text{reco}}$ is constructed from the physics objects described above to approximate the invariant mass of the $t\bar{t}$ system.  The construction of the variable in the boosted and resolved selections uses different physics objects.

For events passing the boosted selection, the four-momentum of the hadronic-top jet is used for the {\em{hadronic-top candidate}}. The {\em{leptonic-top candidate}}  is constructed by summing the four-momenta of the charged lepton, the neutrino candidate, and $j_{\text{sel}}$. The neutrino candidate's transverse momentum is taken equal to $\vec{E}^{\text{miss}}_{\text{T}}$. The $z$ component of its momentum, $p_z$, is estimated by assuming that the neutrino and the lepton come from an on-shell $W$ boson decay and imposing a $W$ mass constraint on the neutrino--lepton system~\cite{Aad:2015fna}.
If no real solution is found for the neutrino's $p_z$, it is assumed that a mismeasurement of the $\vec{E}^{\text{miss}}_{\text{T}}$ leads to this effect, in which
case the $\vec{E}^{\text{miss}}_{\text{T}}$ is rescaled and rotated by the minimal amount until a real solution is found. If more than one solution is available, the solution with smallest absolute value of the neutrino's $p_z$ is taken.
The value of $m_{t\bar{t}}^{\text{reco}}$ is then the mass of the summed four-momenta of the leptonic- and hadronic-top candidates. 

For events passing the resolved selection, following the approach of previous ATLAS searches~\cite{Aad:2015fna}, a $\chi^2$ algorithm is employed 
to find the best assignment of jets to the leptonic-top candidate and hadronic-top  candidate.  Using the four-momenta of the neutrino, lepton, and all small-$R$ jets in the event, 
a $\chi^2$ is defined using the expected top-quark and $W$ boson masses:
\begin{eqnarray}
\chi^2 &=& \left[ \frac{m_{jj}-m_{W_\mathrm{h}}}{\sigma_{W_\mathrm{h}}} \right]^2 + 
         \left[ \frac{m_{jjb}-m_{jj}-m_{t_\mathrm{h}-W_\mathrm{h}}}{\sigma_{t_\mathrm{h}-W_\mathrm{h}}} \right]^2 +
         \left[ \frac{m_{b\ell\nu}-m_{t_\ell}}{\sigma_{t_\ell}} \right]^2 \nonumber \\ \nonumber
        & &+ \left[ \frac{(p_{\mathrm T,jjb}-p_{\mathrm T,b\ell\nu}) - 
                (p_{\mathrm T,t_\mathrm{h}}-p_{\mathrm T,t_\ell})}{\sigma_{p_{\mathrm T,t_\mathrm{h}}-p_{\mathrm T,t_\ell}}}  \right]^2.
\end{eqnarray}
The first term is a constraint using the mass of the hadronically decaying $W$ boson.
The second term is a constraint using the mass difference between the hadronically decaying top quark and the hadronically decaying $W$ boson.   
Since the mass of the hadronically decaying $W$ boson, $m_{jj}$,  
and the mass of the hadronically decaying top quark, $m_{jjb}$,
are highly correlated, the mass of the hadronically decaying $W$ boson
is subtracted from the second term to decouple it from the first term. 
The third term is a constraint using the mass of the semileptonically decaying top quark. 
The last term arises as a constraint on the expected transverse momentum balance between the two decaying top quarks. 
In the $\chi^2$ definition above, $t_\mathrm{h}$ and $t_\ell$
refer to the hadronically and semileptonically decaying top quarks. 
Only arrangements in which $b$-quarks are assigned to $b$-tagged jets are considered.\footnote{If there is only one $b$-tagged jet in the event, then only arrangements in which
it is assigned to a $b$-quark in the $\chi^2$ kinematic fit are considered and one of the top quark candidates is allowed not to have a $b$-quark candidate associated with it.}
The values of the $\chi^2$ central-value parameters $m_{W_\mathrm{h}}$, $m_{t_\mathrm{h}-W_\mathrm{h}}$, $m_{t_\ell}$, and $p_{\mathrm T,t_\mathrm{h}}-p_{\mathrm T,t_\ell}$, and the values of the width parameters $\sigma_{W_\mathrm{h}}$,  $\sigma_{t_\mathrm{h}-W_\mathrm{h}}$, $\sigma_{t_\ell}$, and $\sigma_{p_{\mathrm T,t_\mathrm{h}}-p_{\mathrm T,t_\ell}}$ are obtained from Gaussian fits to the distributions of relevant reconstructed variables, using MC events for which the reconstructed objects are matched to partons,
from $Z'$ samples with masses from 0.5 to 2.0 \TeV{}.
As in the case of the boosted reconstruction,
the neutrino candidate's transverse momentum is taken to be the $\vec{E}^{\text{miss}}_{\text{T}}$ and the neutrino $z$
component is estimated by assuming that the neutrino and the lepton come from an on-shell $W$ boson decay.
All possible neutrino $p_z$ solutions and jet permutations 
are considered, and the one with the lowest $\chi^2$ value is selected.
The $m_{t\bar{t}}^{\text{reco}}$ observable is estimated as the mass of the
four-momentum obtained by summing the four-momenta of the objects that minimise the $\chi^2$ value.

\begin{figure}[htbp]
\centering
{
\subfigure[$Z' \rightarrow t\bar{t}$, resolved selection]{\label{fig:mtt_resolved_zprime}\includegraphics[width=0.45\textwidth]{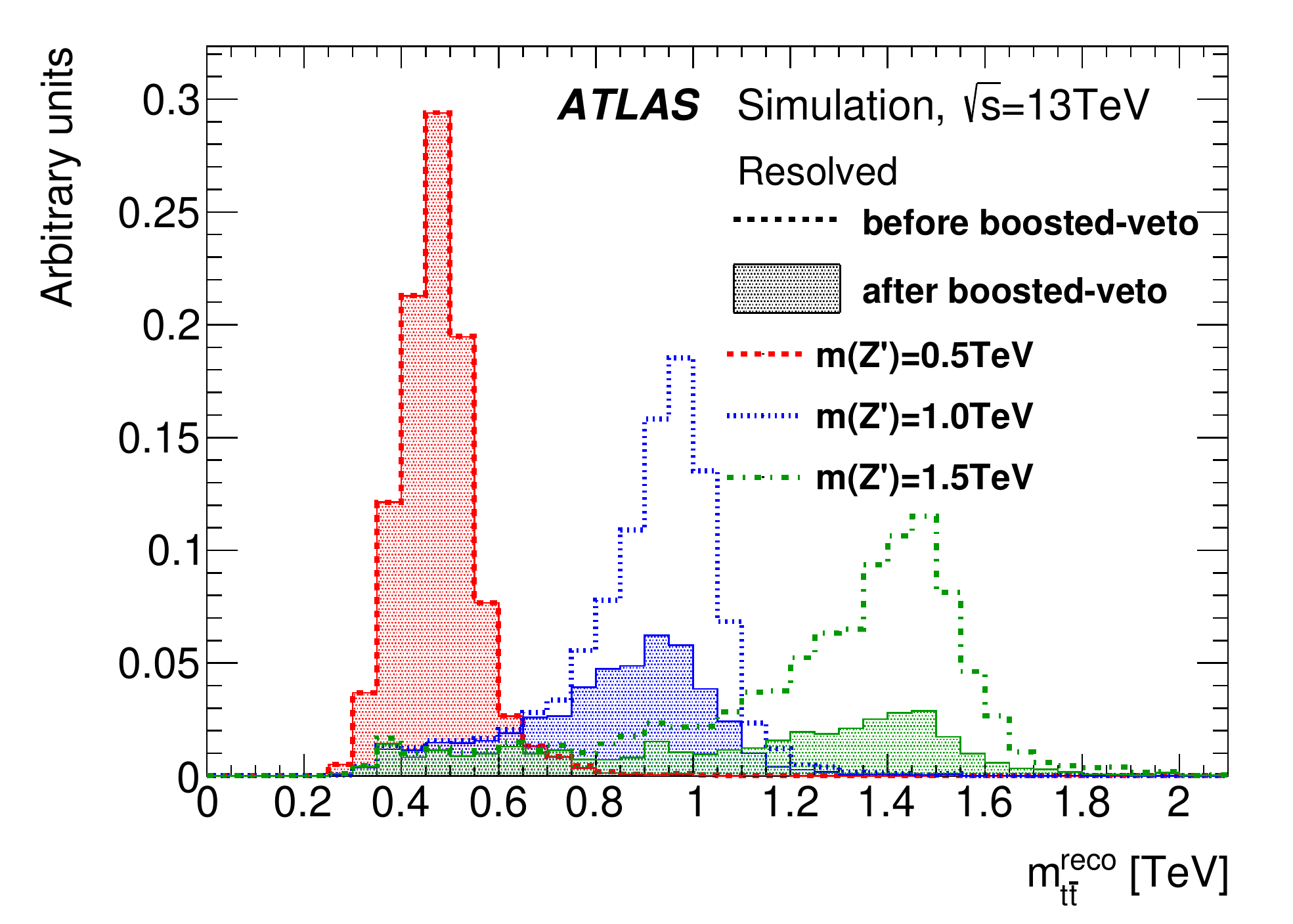}}
\subfigure[$Z' \rightarrow t\bar{t}$, boosted selection]{\includegraphics[width=0.45\textwidth]{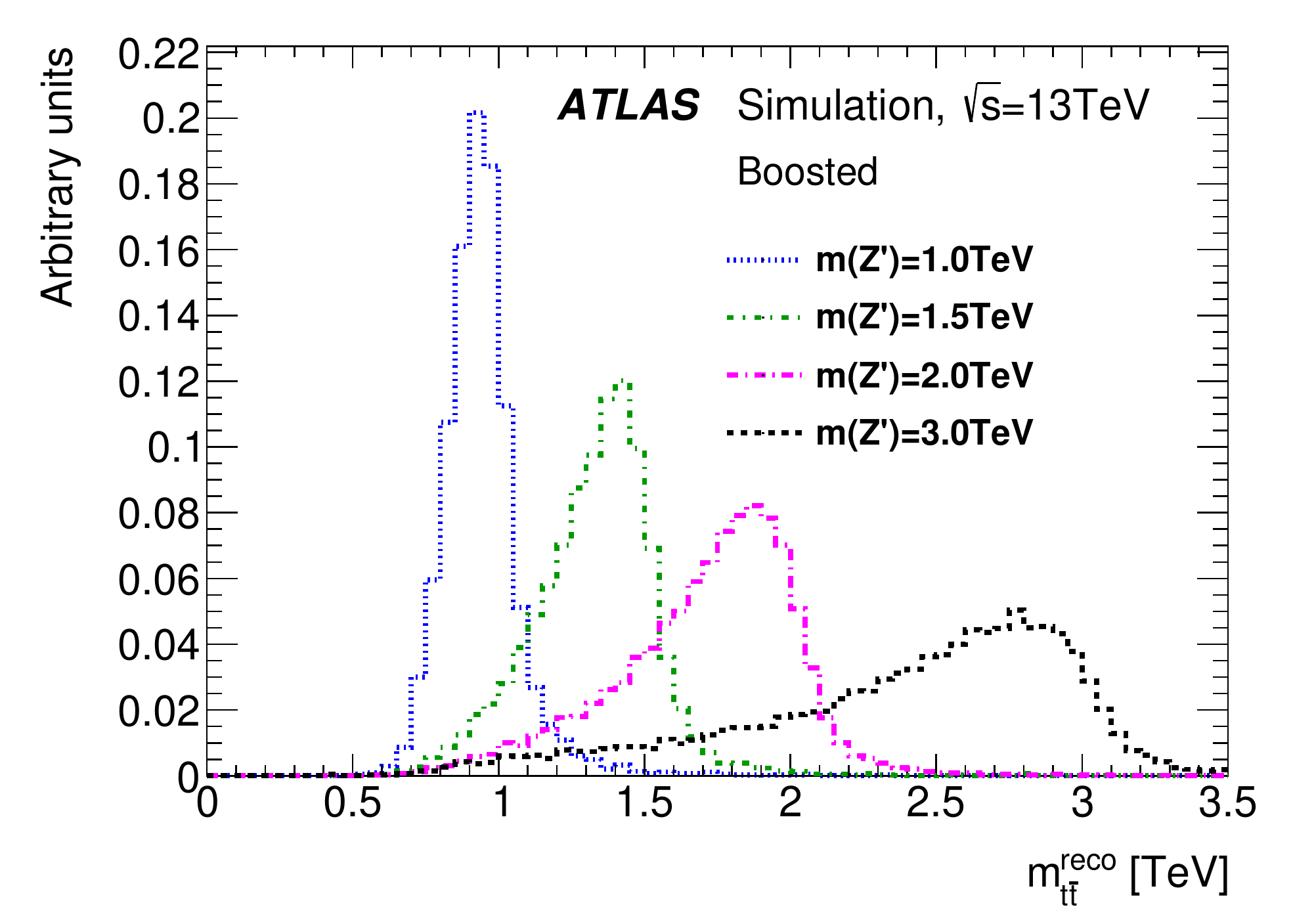}}
\subfigure[$g_{\mathrm{KK}}$, resolved selection]{\label{fig:mtt_resolved_kkgluon}\includegraphics[width=0.45\textwidth]{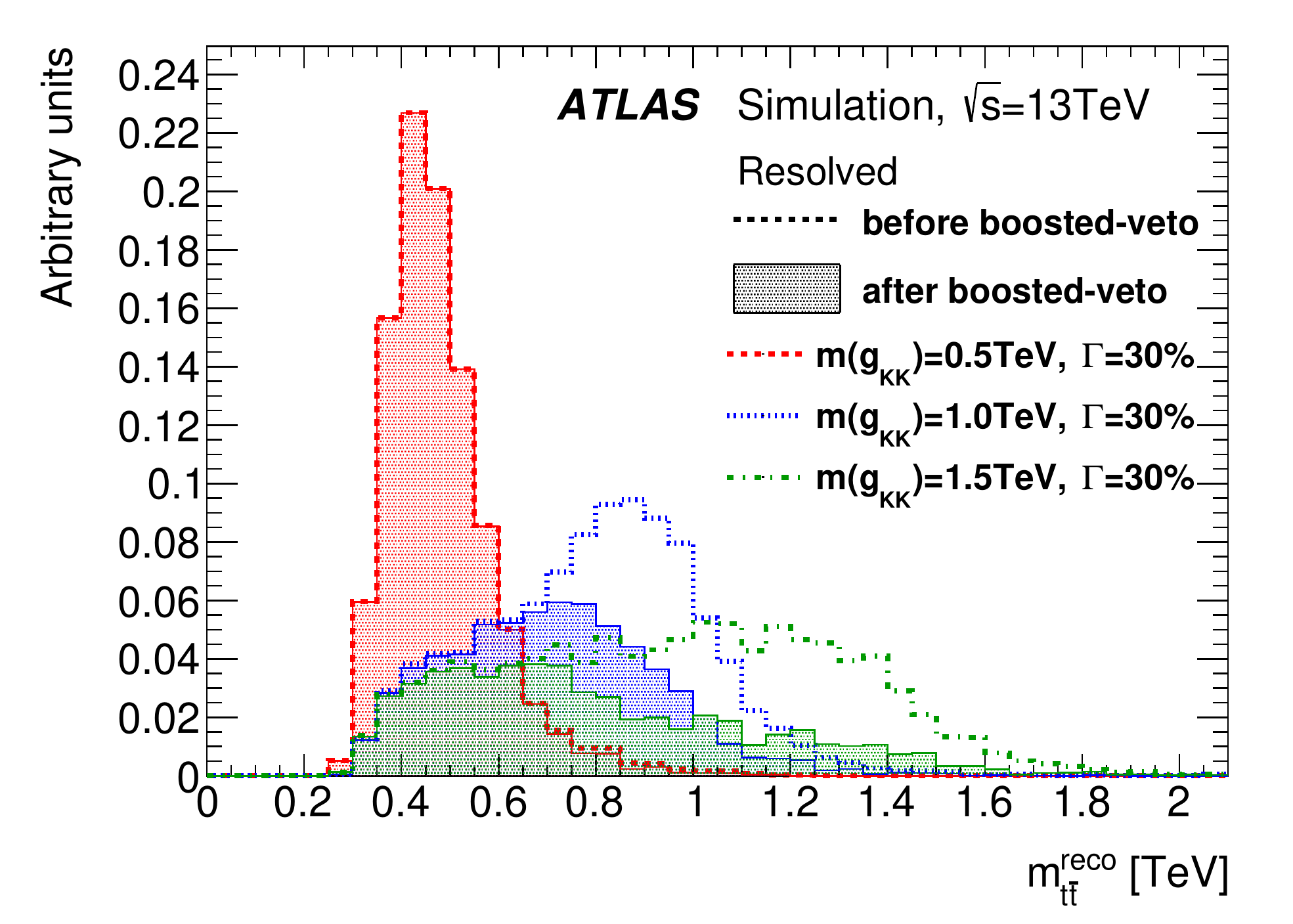}}
\subfigure[$g_{\mathrm{KK}}$, boosted selection]{\includegraphics[width=0.45\textwidth]{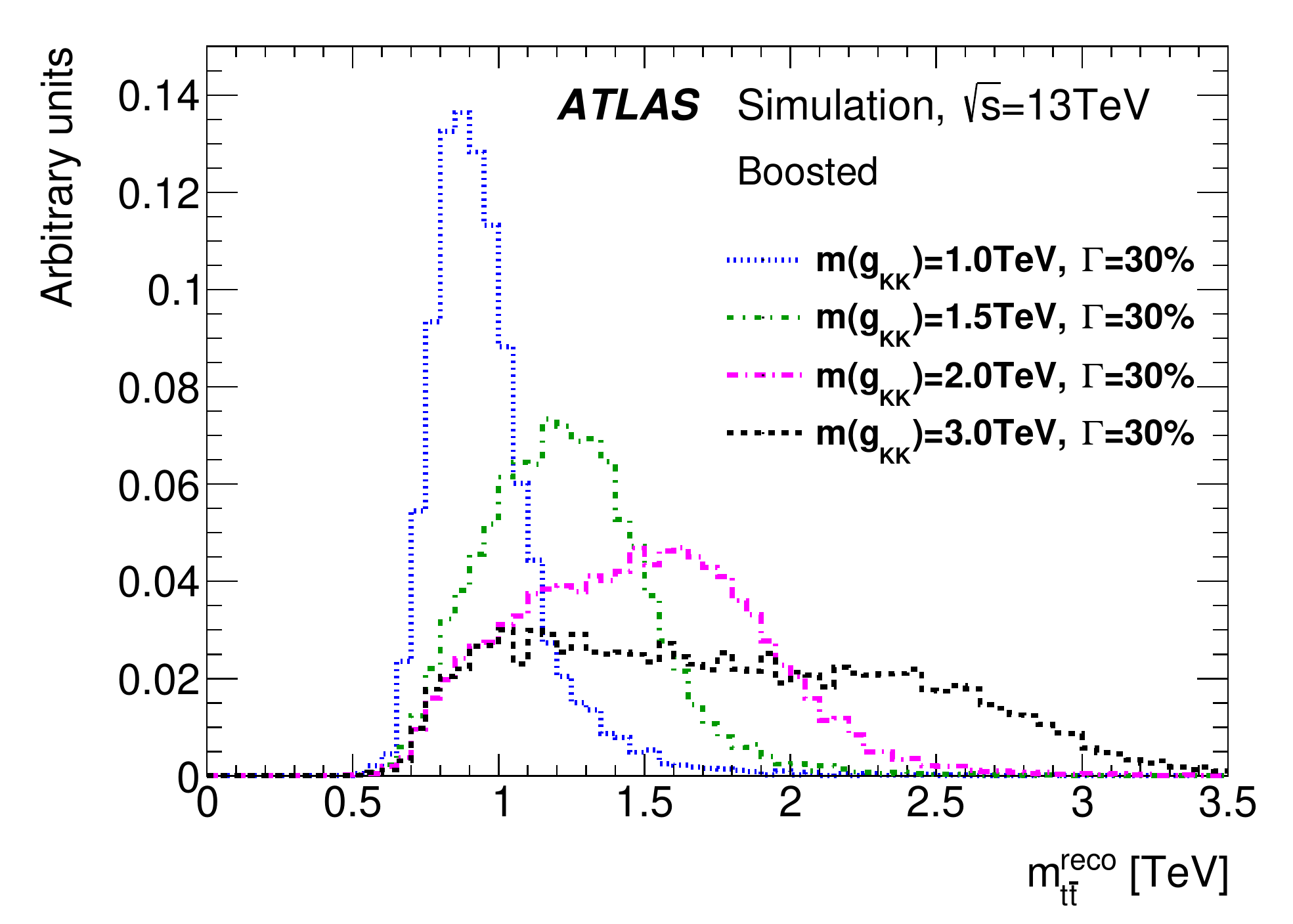}}
\subfigure[$G_{\mathrm{KK}}$, resolved selection]{\label{fig:mtt_resolved_kkgrav}\includegraphics[width=0.45\textwidth]{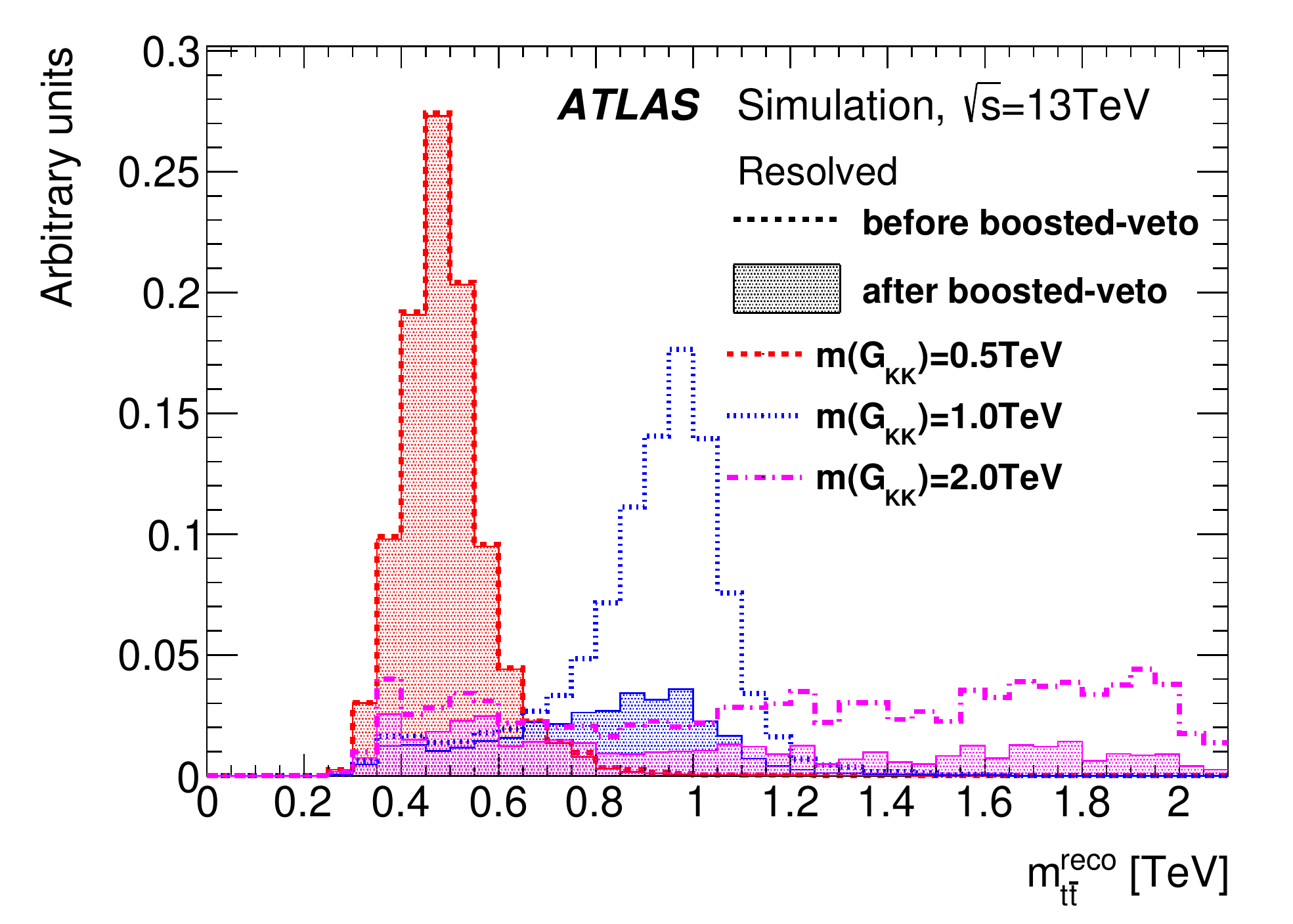}}
\subfigure[$G_{\mathrm{KK}}$, boosted selection]{\includegraphics[width=0.45\textwidth]{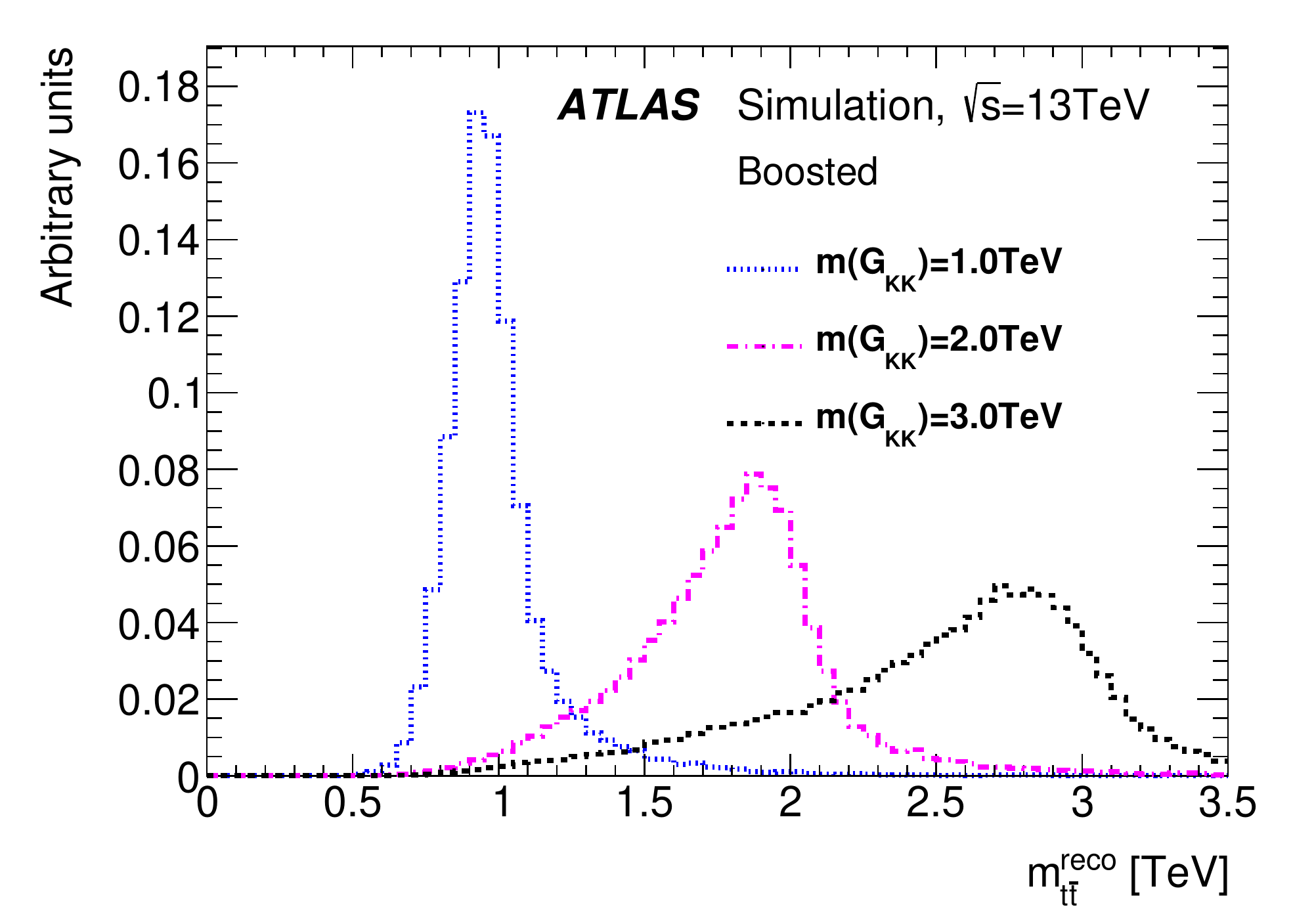}}
}
\caption{Reconstructed top-quark pairs invariant mass,  $m_{t\bar{t}}^{\text{reco}}$,  for simulated signal events satisfying the selection criteria. The $Z'$ in the simulated samples used here has a width of 3\% of its mass.
The $g_{\mathrm{KK}}$ shown here has a width of 30\% of its mass and the width of the $G_{\mathrm{KK}}$ width varies between 3\% to 6\% of its mass.
The figure shows the distribution including events that may satisfy both the boosted and resolved
selections in the line marked as ``before boosted-veto''. The line marked as ``after boosted-veto''
excludes events which satisfy both the boosted and resolved selections from the resolved selection.
}
\label{fig:mtt}
\end{figure}

The resulting  $m_{t\bar{t}}^{\text{reco}}$ distributions for several signal masses  are shown in Figure~\ref{fig:mtt}.
For this figure, all events satisfying the  selection
criteria are used. The low-mass tails arise from two effects: first, the $t\bar{t}$ system may emit radiation that is not included in the reconstruction, thus shifting $m_{t\bar{t}}^{\text{reco}}$ to lower values; second, before reconstruction the Breit--Wigner signal shape in $m_{t\bar{t}}$ has a tail at lower values due to the steep fall in parton luminosity with increasing partonic centre-of-mass energy. 
The former is particularly true for high-mass resonances, such as the benchmark processes used in this analysis, while the latter has a larger effect on broad resonances.
Figures~\ref{fig:mtt_resolved_zprime},~\ref{fig:mtt_resolved_kkgluon} and~\ref{fig:mtt_resolved_kkgrav} show the $m_{t\bar{t}}^{\text{reco}}$ distributions in the resolved
channel before and after the requirement that the events fail the boosted selection (`boosted channel-veto') is imposed.

\vspace{1em}

Following this reconstruction, events are placed into one of  four $b$-tagging categories:
\begin{description}
\item \textbf{Category 0:} there is no $b$-tagged jet matching the hadronic- nor leptonic-top candidates 
\item \textbf{Category 1:} only the leptonic-top candidate has a matching $b$-tagged jet
\item \textbf{Category 2:} only the hadronic-top candidate has a matching $b$-tagged jet
\item \textbf{Category 3:} the hadronic-top candidate and the leptonic-top candidate both have a matching $b$-tagged jet.
\end{description}

The matching requirement for the leptonic top candidate in the boosted channel is that at least one $b$-tagged track-jet must be within $\Delta R=0.4$ of the small-$R$ jet used for the leptonic top candidate reconstruction.
The criterion used to reconstruct the hadronic top candidate is that at least one $b$-tagged track-jet is within $\Delta R = 1.0$ of the large-$R$ jet used to reconstruct the hadronic top candidate.
In the resolved channel, this matching must be to one small-$R$ jet assigned as a $b$-quark jet by the $\chi^2$ algorithm. Events in Category 0 are rejected.

\section{Estimation of background contributions using data}
\label{sec:estimate}

SM $t\bar{t}$ production is the dominant source of background, followed by $W$+jets and multi-jet production. The SM $t\bar{t}$ background is estimated using MC samples and fixed-order theory calculations as described in Section~\ref{sec:samples}.
The background contributions from multi-jet and $W$+jets
production are estimated using data, as described in this section.  

\subsection{Multi-jet background}
\label{sec:qcd}
The multi-jet background consists mainly of events that have a
jet that is misreconstructed 
as a lepton.
The normalisation, kinematic distributions,  and statistical and systematic uncertainties 
associated with the multi-jet background are estimated 
from data using a technique known as a {\em{matrix method}}. The particular matrix method used in this search is a variation of the one used in the previous ATLAS $t\bar{t}$ resonance searches analyses described in detail in Ref.~\cite{Aad:2012qf}.

The matrix method uses lepton misidentification probabilities  and lepton identification efficiencies to estimate the multi-jet background.
The efficiency $f$, which is also referred to as the `fake rate', is defined as the probability that a jet from multi-jet production that satisfies a looser set of lepton identification criteria, in particular without an isolation requirement, also satisfies the tight lepton identification criteria. It is estimated from a control region with the same selection as the resolved signal, but with the missing transverse momentum and transverse mass requirements inverted. In this control region, which is enriched in multi-jet events, the subtraction of prompt-lepton contributions is based on MC simulation.
The efficiency $\epsilon$ is defined as the probability that a prompt lepton (from a $W$ or $Z$ boson) that satisfies the loose lepton identification criteria also satisfies the tight identification criteria.
It is determined using SM $t\bar{t}$ MC samples, 
corrected using comparisons of MC and data $Z \rightarrow \ell\ell$ events.

The number of multi-jet background events satisfying the selection criteria is estimated using
data events that satisfy all criteria, except that the loose lepton identification criteria are used.

The number of events with leptons satisfying the loose identification criteria, $N_\mathrm{L}$, is defined as
\begin{equation} \notag
N_\mathrm{L}=N_\text{prompt}+N_\text{multi-jet} 
\end{equation} 
where $N_\text{prompt}$  and $N_\text{multi-jet}$ are the numbers of events  satisfying those criteria with prompt leptons and  with leptons from other sources, respectively. The number of events satisfying the tight identification criteria, $N_\mathrm{T}$, is then
\begin{equation} \notag
N_\mathrm{T}=\epsilon{}\times{}N_\text{prompt}+f\times{}N_\text{multi-jet}. 
\end{equation} 
Solving these two equations for $N_\text{prompt}$ and $N_\text{multi-jet}$ gives the multi-jet
contribution from events satisfying all the selection criteria.
A large uncertainty 
is associated with this background, which was obtained by testing its modelling in a validation region, as described below.

Good modelling of the shape of kinematic distributions is achieved by parameterising the efficiencies as functions of relevant kinematic variables.
For electrons, the efficiencies are parameterised as a two-dimensional function of the transverse momentum of the lepton and a calorimeter-based isolation variable.
For muons, in addition to the transverse momentum and the calorimeter-based isolation variable, the angular separation between the lepton and the closest jet is also used.
The modelling is validated in separate dedicated validation regions, where only one of the $E^{\text{miss}}_{\text{T}}$ cut or the
$\met + m_{\text{T}}^{W}$ cut is inverted. Such validation regions contain a more similar mixture of contributions to the signal region samples' contributions, but
still have an enhanced multi-jet contribution.

The fake rates for electrons vary from 18\% to 92\%, with the largest values occurring at high lepton $p_{\text{T}}$, with low nearby calorimeter activity. This behaviour is explained by the track-based lepton isolation criterion that uses a $p_{\text{T}}$-dependent cone and leads to a looser isolation requirement at higher $p_{\text{T}}$.
The fake rates for muons vary from 4\% to 94\%, with the largest values occurring in conditions similar to the electron case.
Such variations are parameterised, as mentioned previously, using the lepton transverse momentum, the $\Delta R$ between
the lepton and the closest jet, as well as a calorimeter-based isolation requirement around the lepton.

\subsection{$\mathbf{W}$+jets background}
\label{sec:wjet}
For the $W$+jets background, data are used to derive scale factors that are applied to correct the normalisation given by \SHERPA\ MC simulations of this background for possible mismodelling of the cross-section times acceptance. Furthermore, the data are used to correct the fractions of the different quark-flavour components of the $W$+jets background. 
The procedure used is implemented separately for the electron and muon channels, as the different lepton selections can lead to differences between the correction factors.

The scale factors that correct the normalisation are determined by comparing the measured 
$W$ boson charge asymmetry in data~\cite{ATLAS:2012an,Aad:2012hg}
with that predicted by the simulation. 
A relaxed set of selection criteria that does not include a $b$-tagging requirement is used, so that the $W$+jets purity of the control region is increased, while also reducing the statistical uncertainty in the scale factors used for this procedure.
Any bias induced by relaxing the selection criteria is found to be negligible compared to the statistical uncertainty in the scale factor determination.
The total number of $W$+jets events in data, $N_{W^+} + N_{W^-}$, is given by:
\begin{equation}
N_{W^+} + N_{W^-} = \left(\frac{r_\mathrm{MC} + 1}{r_\mathrm{MC} - 1}\right)(D_{\mathrm{corr}+} - D_{\mathrm{corr}-}), 
\label{eq:Wchargeasymm}
\end{equation}
where $r_\mathrm{MC}$ is the ratio given by MC simulation 
of the number of $W$+jets events with a positively charged lepton to that
with a negatively charged lepton
and $D_{\mathrm{corr}+(-)}$ is
the number of observed events with a positively (negatively) charged lepton.  
Contributions to $D_{\mathrm{corr}+(-)}$ from charge-asymmetric processes such as single 
top, $WZ$ and \ttbar{}+$W$ production are estimated from MC simulation and are subtracted. 
Contributions from charge-symmetric processes such as $t\bar{t}$ production cancel out in the 
difference on the right-hand side of Eq.~(\ref{eq:Wchargeasymm}). A scale factor, $C_{\mathrm{A}}$, applied to the 
MC simulated samples of $W$ + jets events, 
is then calculated as 
the ratio of $N_{W^+} + N_{W^-}$ evaluated from data to that predicted from MC simulation.  This evaluation is performed separately for four jet multiplicity bins; $n_{\text{jet}}=2$, $n_{\text{jet}}=3$, $n_{\text{jet}}=4$, and $n_{\text{jet}}\ge 5$.

The flavour fractions $f_{\text{flavour}} = N^{\text{flavour}}_{\text{MC},W}/N_{\text{MC},W}$ are extracted from a $W$+jets-dominated control region. This control region is selected using criteria identical to the signal event selection except for requirements on the hadronic jet activity: exactly two small-$R$ jets are required. Based on the lepton charge distribution of events with at least one $b$-tagged jet, scale factors are derived for the flavour components $W_{b\bar{b}}$, $W_{c\bar{c}}$, $W_{c}$, and $W_{\text{light}}$\footnote{The flavour components are: $W_{b\bar{b}}$ -- $W$ bosons produced in association with a $b\bar{b}$ pair;  $W_{c\bar{c}}$ -- $W$ bosons  produced in association with a $c\bar{c}$ pair; $W_c$ -- $W$ bosons produced in association with a single $c$- or $\bar{c}$-quark; and $W_\text{light}$ -- $W$ bosons produced in association with light quarks.} by solving a system of linear equations:

\begin{equation}\nonumber
  \left(
  \begin{matrix}
    C_A \cdot (N_{\text{MC},W^-}^{bb} + N_{\text{MC},W^-}^{cc}) & C_A \cdot N_{\text{MC},W^-}^{c} & C_A \cdot N_{\text{MC},W^-}^{\text{light}}  & N_{Q^{-}} \\
             (      f_{bb} +       f_{cc})       &          f_{c}          &       f_{\text{light}}              & 0                    \\
                              0                  &           1             &        0                  & 0                    \\
    C_A \cdot (N_{\text{MC},W^+}^{bb} + N_{\text{MC},W^+}^{cc}) & C_A \cdot N_{\text{MC},W^+}^{c} & C_A \cdot N_{\text{MC},W^+}^{\text{light}}  & N_{Q^{+}} \\
  \end{matrix}
  \right) 
  \cdot
  \left(\begin{matrix} K_{bb,cc} \\ K_c \\ K_{\text{light}} \\ K_{Q} \end{matrix}\right)  
  =
  \left(\begin{matrix} D_{W^-} + N_{Q^{-}} \\ 1.0 \\ 1.0 \\ D_{W^+} + N_{Q^{+}} \end{matrix}\right)
\end{equation} 

where $D_{W^\pm}$ is the expected number of $W$+jets events with a positively or negatively charged lepton in data after subtracting all non-$W$+jets MC background contributions and each $K_{\text{flavour}}$ is a correction factor extracted by this procedure. The $K_{bb,cc}$ factor refers to both the $W+bb$ and $W+cc$ contributions in the background. The variable $K_{Q}$, which is a normalisation factor for the multi-jet background, is also extracted by the procedure. The number of events in the MC simulation with positively charged (negatively charged) leptons for each flavour component is $N^{\text{flavour}}_{\text{MC},W^+}$ ($N^{\text{flavour}}_{\text{MC},W^-}$). The fraction of each flavour predicted by the MC simulation is $f_{\text{flavour}}$.  The contributions from multi-jet production in the different lepton charge regions, $N_{Q^+}$ and $N_{Q^-}$, are estimated using the same matrix method as described in Section~\ref{sec:qcd}. 

Solving this system of equations gives corrected heavy-flavour fractions for $W$+jets events with exactly two jets. 
Since the predicted charge asymmetry depends on the flavour fractions, the charge-asymmetry normalisation followed by flavour-fraction extraction is iterated until stable results for $C_A$ and $K_{\text{flavour}}$ are obtained.
The MC predictions of the flavour fractions for higher jet multiplicities are used together with these correction factors to obtain a corrected prediction for the flavour fractions at higher jet multiplicities.
The extracted correction factors depend on the selection and the jet multiplicity. The $K_{bb,cc}$ factors are between 1.19 and 1.27 (1.34 and 1.51) in the electron (muon) channel. The $K_{\text{light}}$ factor varies from 0.87 to 0.91 (0.78 to 0.88) in the electron (muon) channel. The $K_c$ factor is found to lie between 0.93 and 1 (0.86 and 1) in the electron (muon) channel. The normalisation factor $C_A$ extracted from the charge asymmetry varies from 0.78 to 1.05 (0.8 to 1.14) in the electron (muon) channel.

\section{Systematic uncertainties}
\label{sec:systematics}

In this section, the systematic uncertainties that affect this search are detailed. These are uncertainties in the normalisation and shape of predicted $m^{\text{reco}}_{t\bar{t}}$ distributions for signal and background. 

The uncertainty in the combined 2015+2016 integrated luminosity is 2.1\%. It is derived, following a methodology similar to that detailed in Ref.~\cite{Aaboud:2016hhf}, from a calibration of the luminosity scale using $x$--$y$ beam-separation scans performed in August 2015 and May 2016.
In addition, a `pile-up' uncertainty due to the observed disagreement between the instantaneous luminosities in data and simulation is estimated.

The modelling of the electron and muon trigger efficiencies, identification efficiencies, energy scales and resolutions are studied using leptonic $Z$ boson decays in data and simulation at $\sqrt{s} = 13$\,\TeV{}. Small corrections are applied to the simulation to better model the performance seen in data~\cite{ATL-PHYS-PUB-2015-043,Aad:2016jkr}. These corrections have associated uncertainties that are propagated to the estimated signal and background yields. The modelling of the isolation requirements on electrons and muons is studied in $13$\,\TeV{} data using $Z$ boson decays and parameterised as functions of the lepton $p_{\text{T}}$, $\eta$, and the hadronic activity near the lepton. The isolation efficiencies are found to be generally well modelled, and the measurements are extrapolated to the $t\bar{t}$ environment to give an uncertainty of 1\% for each electrons or muons.  

The small-$R$ jet energy scale (JES) uncertainty is derived using a combination of simulations, test-beam data, and in situ measurements. Additional contributions from jet flavour composition, punch-through, single-particle response, calorimeter response to different jet flavours and pile-up are taken into account, resulting in 19 eigenvector systematic uncertainty subcomponents, including the uncertainties in the jet energy resolution obtained with an in situ measurement of the jet response in di-jet events~\cite{PERF-2016-04}.

Correction factors are applied to the simulated event samples to compensate for differences between data and simulation~\cite{PERF-2012-04, ATL-PHYS-PUB-2016-012}
in the $b$-tagging efficiency for $b$-, $c$- and light-jets. The correction for $b$-jets is derived from $t\bar{t}$ events with
final states containing two leptons. The corrections are consistent with unity with uncertainties at the level
of a few percent over most of the jet $p_{\text{T}}$ range.
Uncertainties in the correction factors for the $b$-tagging identification response are estimated
by examining dedicated flavour-enriched samples in the data. An additional term is included to extrapolate the
measured uncertainties to the high-$p_{\text{T}}$ region of interest. This term is calculated from simulated events by considering
variations of quantities affecting the $b$-tagging performance such as the impact parameter resolution,
percentage of poorly measured tracks, description of the detector material and track multiplicity per jet.
The dominant effect on the uncertainty when extrapolating to high $p_{\text{T}}$ is related to the different tagging efficiency
when adjusting the track impact parameters according to the resolution measured in data and simulation.
 
The large-$R$ jet energy and mass scales and $\tau_{32}$ scale are varied in simulation according  to  the uncertainties derived from  $\sqrt{s} = 8$\,TeV{}~\cite{Aad:2016pux} simulation and in situ calibration, and the uncertainties are extrapolated to $\sqrt{s} = 13$\,\TeV{}~\cite{ ATL-PHYS-PUB-2015-053}. The uncertainties in the jet mass and $\tau_{32}$ are propagated into uncertainties in the top-tagging efficiency.  

Several uncertainties are specific to the dominant SM $t\bar{t}$ background process.   The $t\bar{t}$ cross-section for $pp$ collisions at a centre-of-mass energy of $\sqrt{s} = 13$~\TeV{} is $\sigma_{t\bar{t}}= 832^{+46}_{-52}$~pb for a top-quark mass of $172.5$~\GeV{}. It was calculated at next-to-next-to-leading order  in QCD including resummation of next-to-next-to-leading-logarithm (NNLL) soft gluon terms with Top++2.0~\cite{Cacciari:2011hy,Beneke:2011mq,Baernreuther:2012ws,Czakon:2012zr,Czakon:2012pz,Czakon:2013goa,Czakon:2011xx}. The uncertainties from the PDFs and $\alpha_{\mathrm{S}}$ were calculated using the PDF4LHC prescription~\cite{Botje:2011sn} with the MSTW2008 68\% CL NNLO~\cite{Martin:2009iq,Martin:2009bu}, CT10 NNLO~\cite{Lai:2010vv, Gao:2013xoa} and NNPDF2.3 5f FFN~\cite{Ball:2012cx} PDF sets and added in quadrature to the effect of the scale uncertainty.  The normalisation of the $t\bar{t}$ background is obtained from a fit to the data in the boosted channels, within the profile likelihood fit method described in Section~\ref{sec:results}. In addition to this normalisation uncertainty, the following top-modelling uncertainties that affect the shape of the $t\bar{t}$ kinematic distributions as well as the normalisation are considered:
\begin{description}
\item {\em{Choice of the event generator}}: this is evaluated by comparing the prediction from a \POWHEG{}+\HERWIG $t\bar{t}$ sample~\cite{Bahr:2008pv} with that from an \AMCatNLO{}+\HERWIG sample and symmetrising the difference. 
\item {\em{Choice of the parton shower model}}: this is evaluated by comparing the prediction from a \POWPYTHIA $t\bar{t}$ sample with that from a \POWHEG{}+\HERWIGV{7} sample~\cite{Bellm:2015jjp} and symmetrising the difference. 
\item {\em{Choice of the parton distribution functions}}: the uncertainties arising from the choice of the PDF set are evaluated using the PDF4LHC15 PDF set. The version that provides 30 separate uncertainty eigenvectors is used~\cite{Butterworth:2015oua}.
\item {\em{Modelling of extra QCD radiation}}: this is evaluated using  \POWPYTHIA samples in which the renormalisation and factorisation scales and the $h_{\text{damp}}$ parameter are varied within ranges consistent with measurements of $t\bar{t}$ production in association with jets~\cite{TOPQ-2011-21,TOPQ-2012-03,,ATL-PHYS-PUB-2015-002}.
\item {\em{EW corrections}}: the uncertainty in the EW corrections to $t\bar{t}$ production is  10\% of their deviation from unity.
\item {\em{NNLO QCD corrections}}: sensitivity of the $m_{t\bar{t}}$ distribution to higher-order QCD corrections relative to the MC generators used is accounted for by adding an uncertainty covering the difference between NLO and NNLO QCD calculations of $t\bar{t}$ production. Corrections are derived from recent calculations~\cite{Czakon:2016dgf} and applied as a function of top-quark $p_{\text{T}}$ and the transverse momentum of the $t\bar{t}$ system, following the recommended scales given in Ref.~\cite{Czakon:2016dgf}. The effect of this uncertainty in the $m_{t\bar{t}}$ distribution is very small at low mass, but increases to 7\% at masses of 2~\TeV{} in the resolved selection and 20\% above 3~\TeV{} in the boosted selection.
\end{description}

The normalisation of the single-top background is varied by $\pm 5.3\%$. This corresponds to the theoretical uncertainty in the dominant $Wt$-channel contribution at approximate NNLO in QCD~\cite{Kidonakis:2010ux,Kidonakis:2010tc,Kidonakis:2011wy}.
An additional shape and normalisation uncertainty is applied to account for differences between the predictions from diagram removal and diagram subtraction approaches~\cite{Frixione:2008yi} to the interference between $tW$ production and $t\bar{t}$. Such uncertainty has an effect of less than 1\% in the yields.
We have found that other single top modeling uncertainties are negligible.

Systematic uncertainties in the $W$+jets background are evaluated by varying the extracted correction factors for normalisation and flavour fractions by their associated uncertainties. The correction factors are also separately estimated for each of the systematic variations which affect the correction factor estimation described in this section.
A 30\% uncertainty is associated with the normalisation of the $W$+c component of the W+jets background.

Systematic uncertainties in the multi-jet background estimation are evaluated using various definitions of multi-jet control regions that result in slightly different estimates of $f$.
Systematic uncertainties associated with object reconstruction and MC simulation are also considered
and a total normalisation uncertainty of 50\% is assigned.

Table~\ref{tab:syst_signal} shows a summary of the systematic uncertainties in the yields for the total background and two signals.
The $t\bar{t}$ modelling and jet energy uncertainties provide the largest contributions to the overall uncertainties.

\begin{table}
\caption{The systematic uncertainties in the yields in the background, as well as in the 2~\TeV\ and 3~\TeV\ $Z^{\prime}_{\mathrm{TC2}}$ signal models, in percentages. Only rows with at least one column with an uncertainty larger than 2\% are shown individually. Systematic uncertainties associated with the muon and electron trigger, identification, energy scales and resolutions combined are smaller than 2\% for all signal regions and are not shown. JES and JER stand for jet energy scale and jet energy resolution.}
\label{tab:syst_signal}
\begin{center}
\begin{tabular}{|l|r|r|r|r|r|r|}
\hline
Systematic Uncertainty & \multicolumn{2}{c|}{Background [\%]} & \multicolumn{2}{c|}{$Z^{\prime}_{\mathrm{TC2}}$, 2~\TeV\ [\%]} & \multicolumn{2}{c|}{$Z^{\prime}_{\mathrm{TC2}}$, 3~\TeV\ [\%]} \\
 & resolved & boosted & resolved & boosted & resolved & boosted \\
\hline
$t\bar{t}$ extra QCD radiation  &        4.0 &        2.4 &        -- &        -- &        -- &        -- \\
$t\bar{t}$ QCD NNLO            &        0.8 &        7.4 &        -- &        -- &        -- &        -- \\
$t\bar{t}$ cross-section       &        5.2 &        -- &        -- &        -- &        -- &        -- \\
$t\bar{t}$ generator           &        1.7 &        3.8 &        -- &        -- &        -- &        -- \\
$t\bar{t}$ parton shower       &        0.6 &        3.2 &        -- &        -- &        -- &        -- \\
Multi-jet                 &        2.6 &        2.7 &        -- &        -- &        -- &        -- \\
Anti-$k_t$ $R=0.4$ JER           &        1.1 &        0.2 &        3.2 &        0.2 &        1.2 &        0.2 \\
Anti-$k_t$ $R=0.4$ JES           &        5.8 &        0.9 &        7.0 &        0.7 &        3.6 &        0.6 \\
Anti-$k_t$ $R=1.0$ JER           &        0.1 &        4.0 &        5.3 &        3.7 &        2.0 &        4.2 \\
Anti-$k_t$ $R=1.0$ JES           &        0.3 &        6.0 &        3.7 &        4.7 &        2.8 &        6.0 \\
$b$-tagging efficiency                 &        3.2 &        1.8 &        1.8 &        1.9 &        2.3 &        2.7 \\
$b$-tagging extrapolation              &        2.4 &        2.3 &        2.0 &        0.6 &        1.2 &        1.8 \\
Luminosity                     &        1.9 &        1.9 &        2.1 &        2.1 &        2.1 &        2.1 \\
Pile-up                        &        4.4 &        0.5 &        4.4 &        0.8 &        3.9 &        0.5 \\
\hline
Total                          &       11.6 &       12.8 &       11.7 &        7.1 &        7.6 &        8.7 \\
\hline
\end{tabular}
\end{center}
\end{table}

\section{Comparison of data with expected background contributions}
\label{sec:expectations}

After all event selection criteria are applied, $35\,612$ ($261\,554$) boosted (resolved)  events remain in the $e$+jets selection and $31\,188$ ($254\,277$) events remain in the $\mu$+jets selection. 
There is a deficit of data compared to expectation for the boosted selections; however, this deficit is consistent with the nominal prediction within the associated systematic uncertainties.
In the following figures, the legend `others' refers to single top, $Z$+jets, $t\bar{t}+W/Z$ and diboson production.

\begin{figure}[ht]
\centering
\subfigure[~$e$+jets, boosted selection.]{
 {\includegraphics[width=0.45\linewidth]{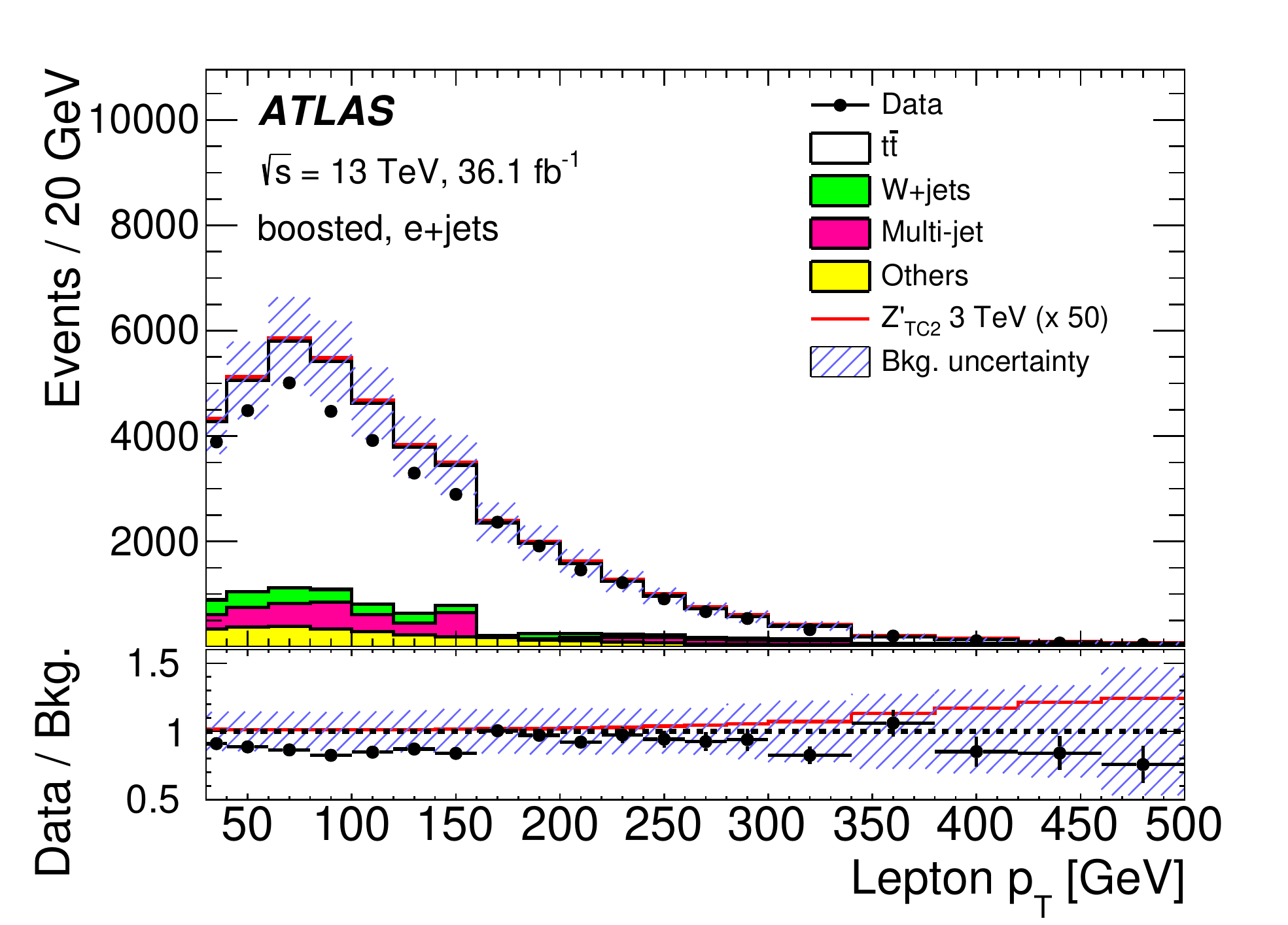}}
}
\subfigure[~$\mu$+jets, boosted selection.]{
 {\includegraphics[width=0.45\linewidth]{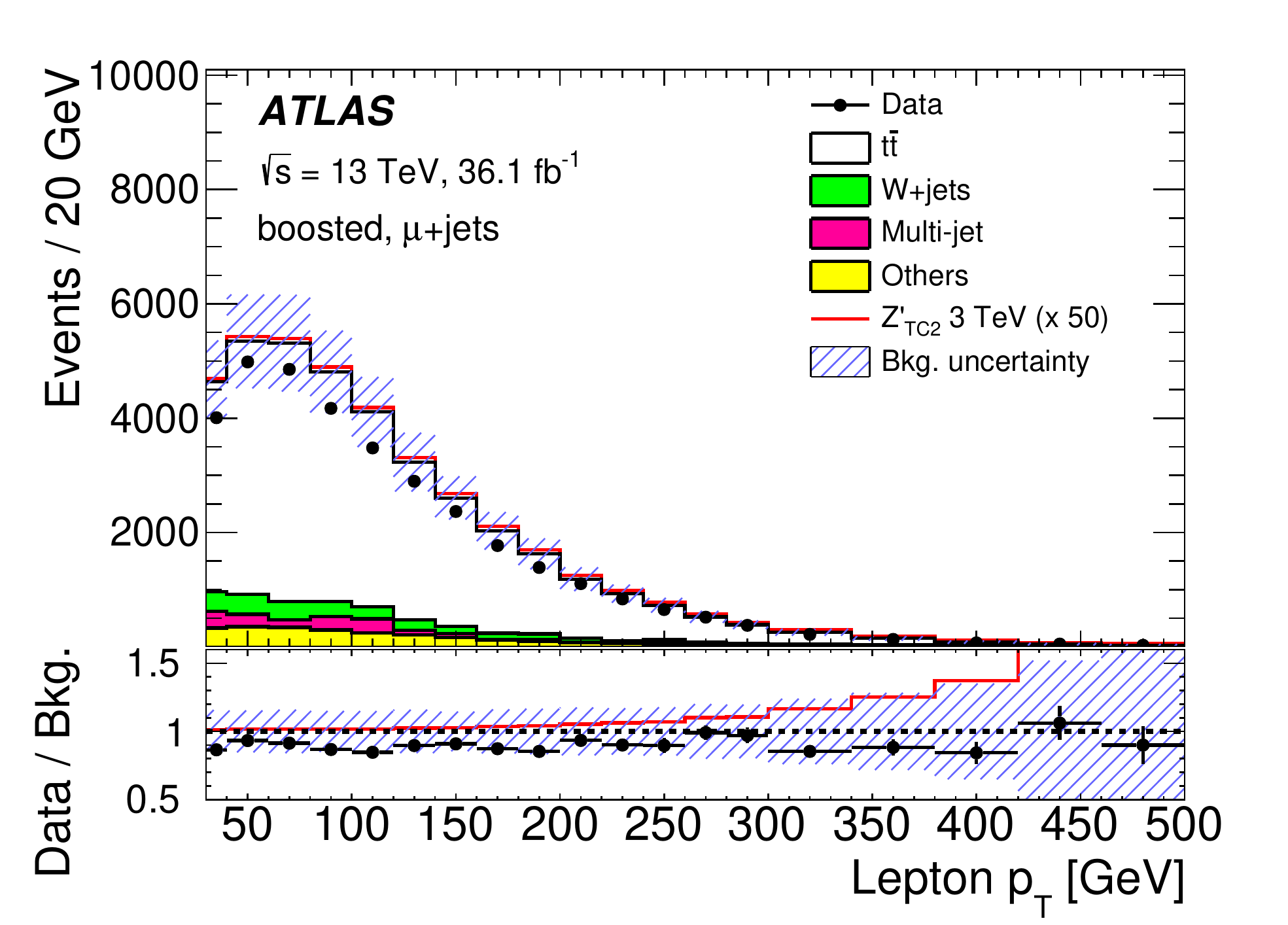}}
}
\subfigure[~$e$+jets, resolved selection.]{
{\includegraphics[width=0.45\linewidth]{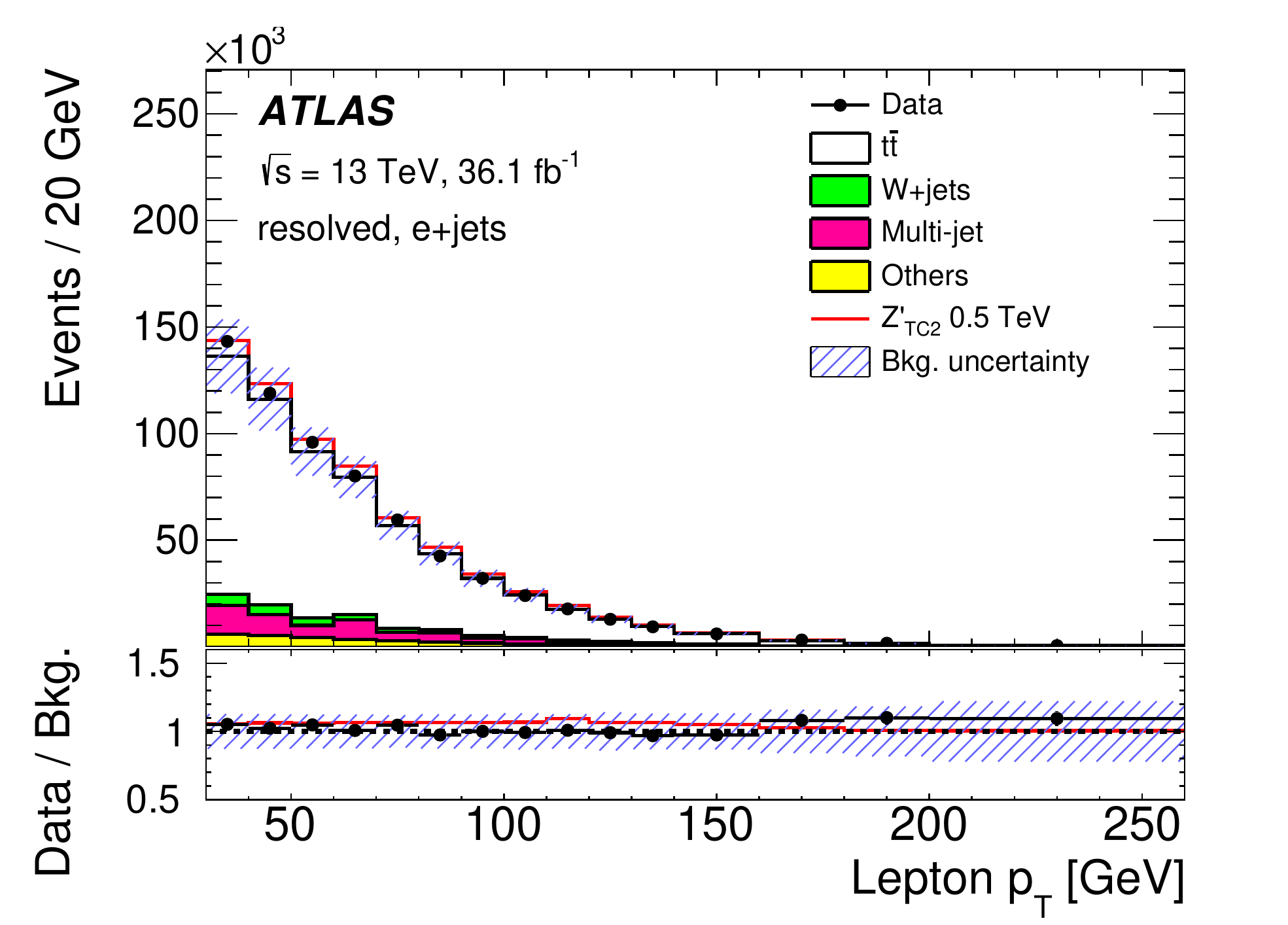}}
}
\subfigure[~$\mu$+jets, resolved selection.]{
 {\includegraphics[width=0.45\linewidth]{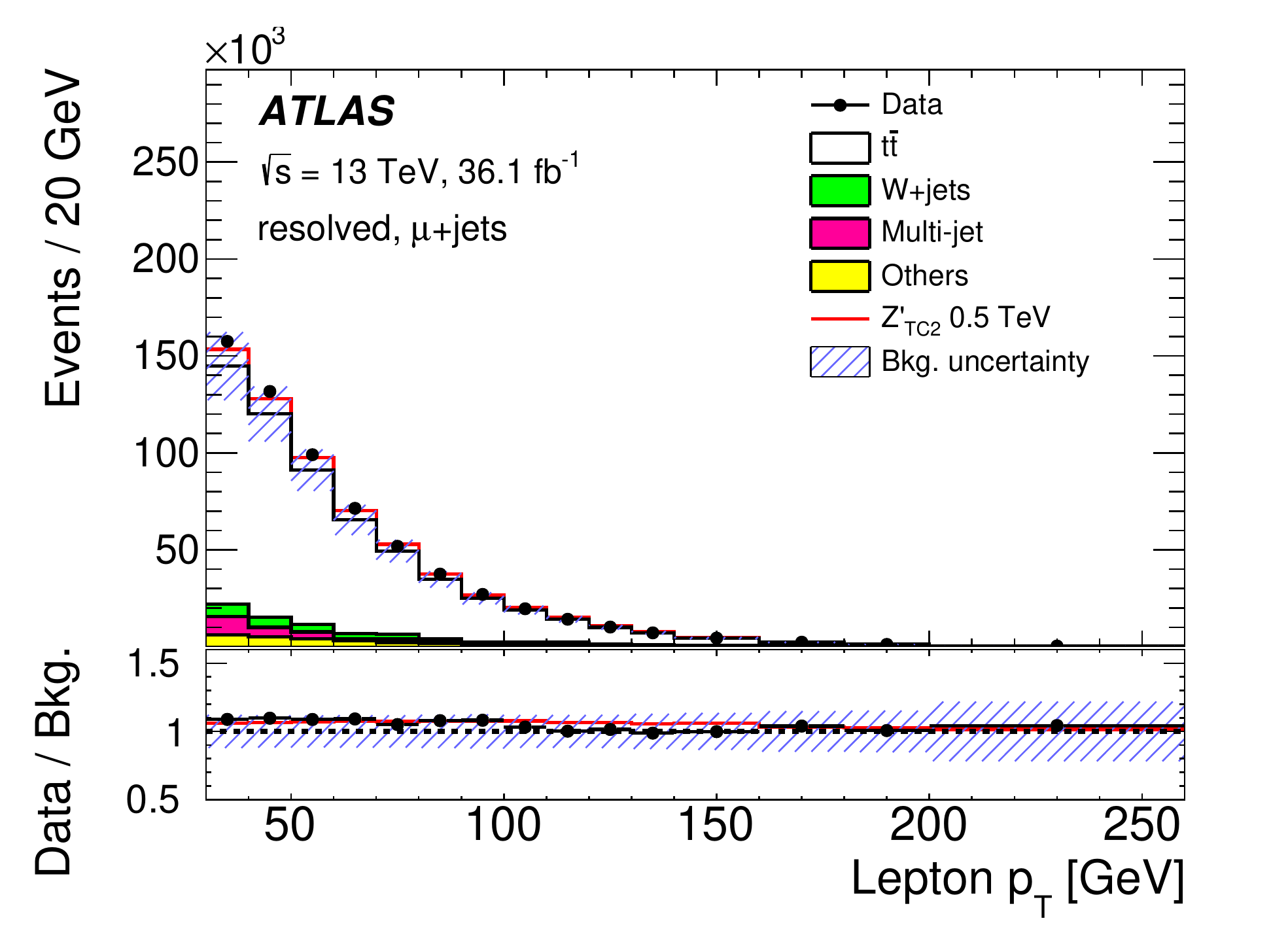}}
}
\caption{The distribution of the transverse momentum of the lepton in the (a) boosted $e$+jets, (b) boosted $\mu$+jets, (c) resolved $e$+jets, and (d) resolved $\mu$+jets selections. The SM background components are shown as stacked histograms. The shaded areas indicate the total systematic uncertainties.
The lower panels in each plot show the ratio of data (points) and a signal example (line) to the background expectation.}
\label{fig:lepPt_boosted}
\end{figure}

\begin{figure}[ht]
\centering
\subfigure[~$e$+jets, boosted selection.]{
  {\includegraphics[width=0.45\linewidth]{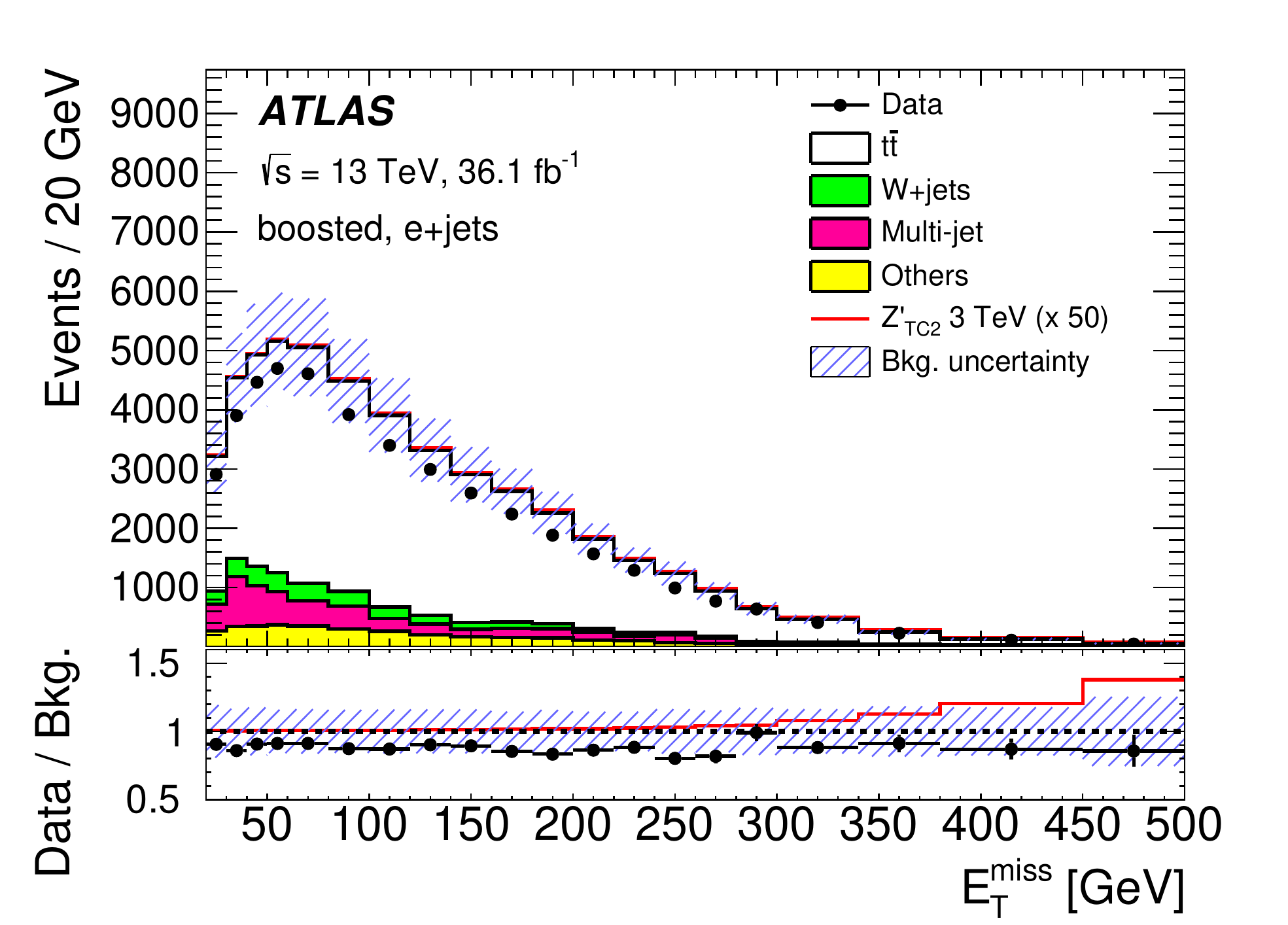}}
}
\subfigure[~$\mu$+jets, boosted selection.]{
  {\includegraphics[width=0.45\linewidth]{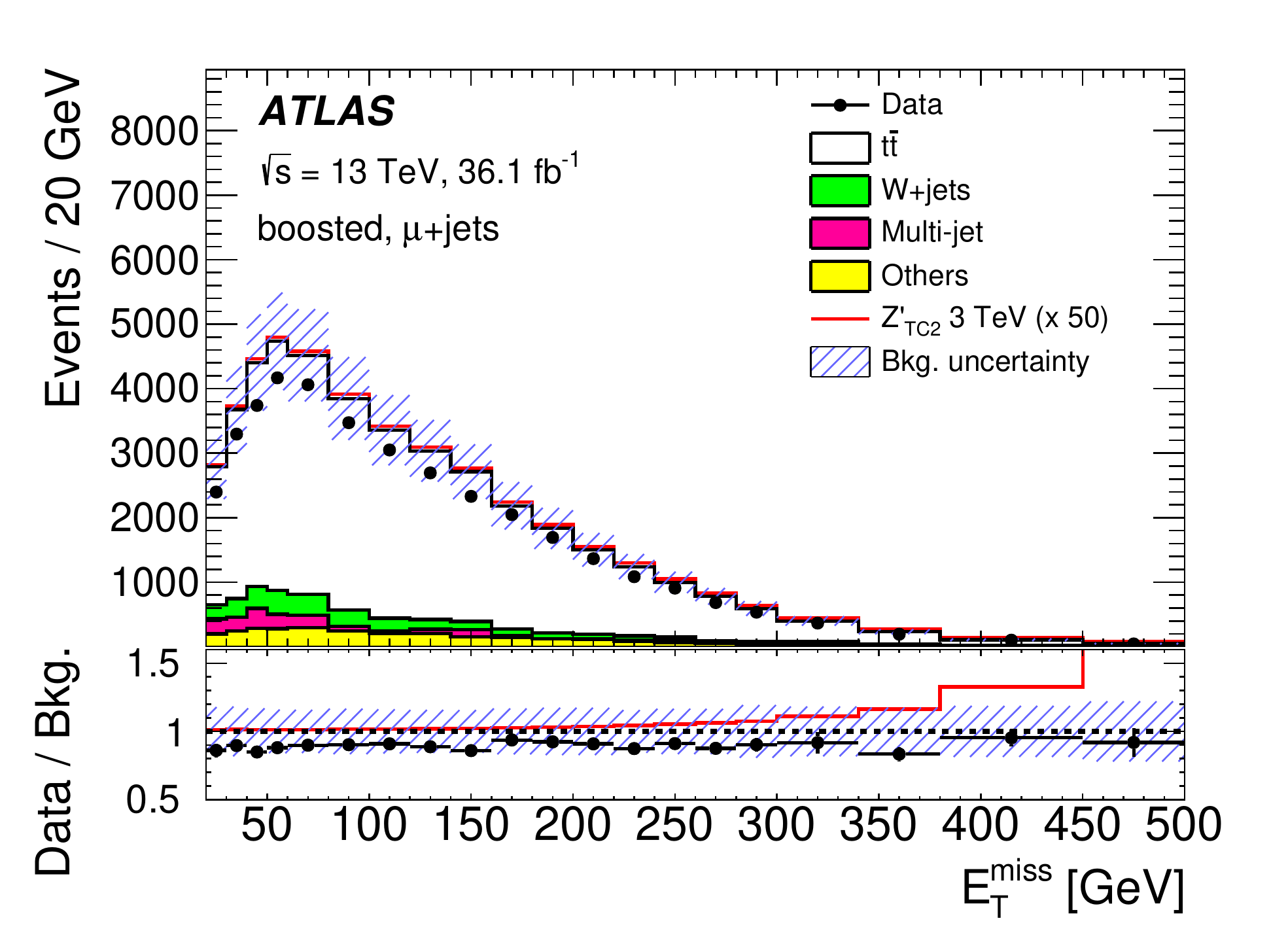}}
}

\subfigure[~$e$+jets, resolved selection.]{
  {\includegraphics[width=0.45\linewidth]{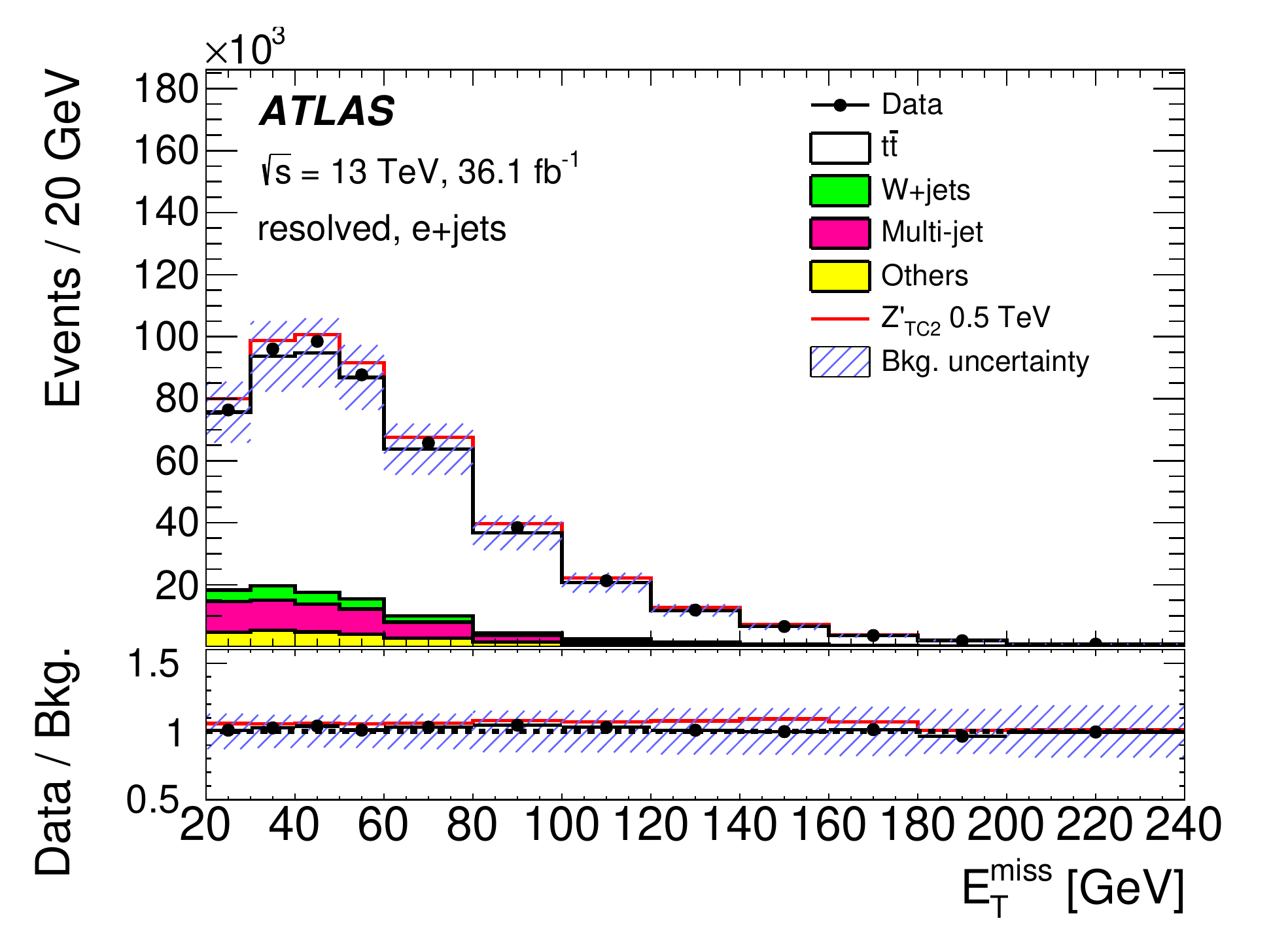}}
}
\subfigure[~$\mu$+jets, resolved selection.]{
  {\includegraphics[width=0.45\linewidth]{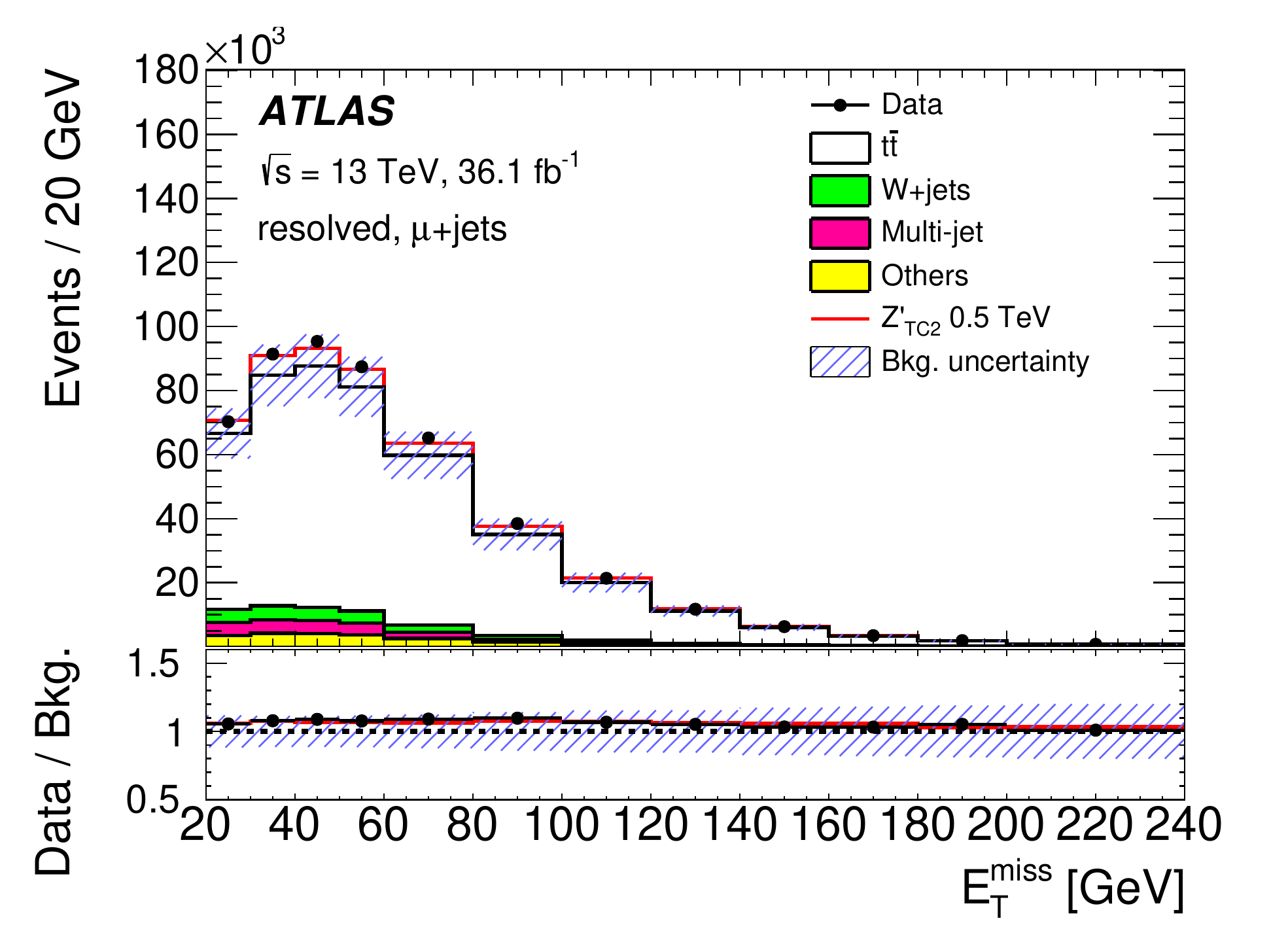}}
}
\caption{The distribution of the \met{} in the (a) boosted $e$+jets, (b) boosted $\mu$+jets, (c) resolved $e$+jets, and (d) resolved $\mu$+jets selections. The SM background components are shown as stacked histograms. The shaded areas indicate the total systematic uncertainties.
The lower panels in each plot show the ratio of data (points) and a signal example (line) to the background expectation.}
\label{fig:met_boosted}
\end{figure}

\begin{figure}[ht]
\centering
\subfigure[~$e$+jets selection.]{

\includegraphics[width=0.45\linewidth]{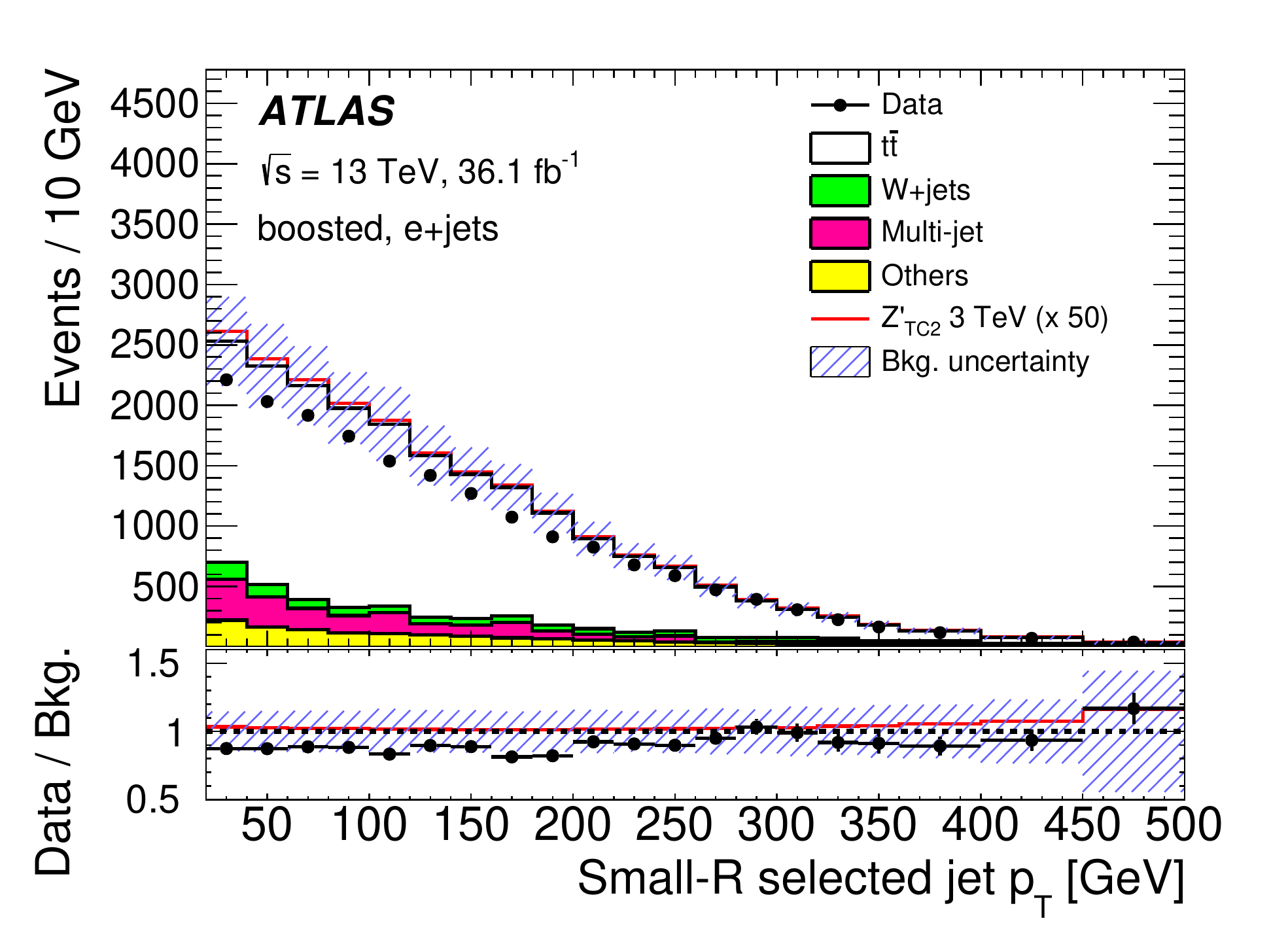}
}
\subfigure[~$\mu$+jets selection.]{
\includegraphics[width=0.45\linewidth]{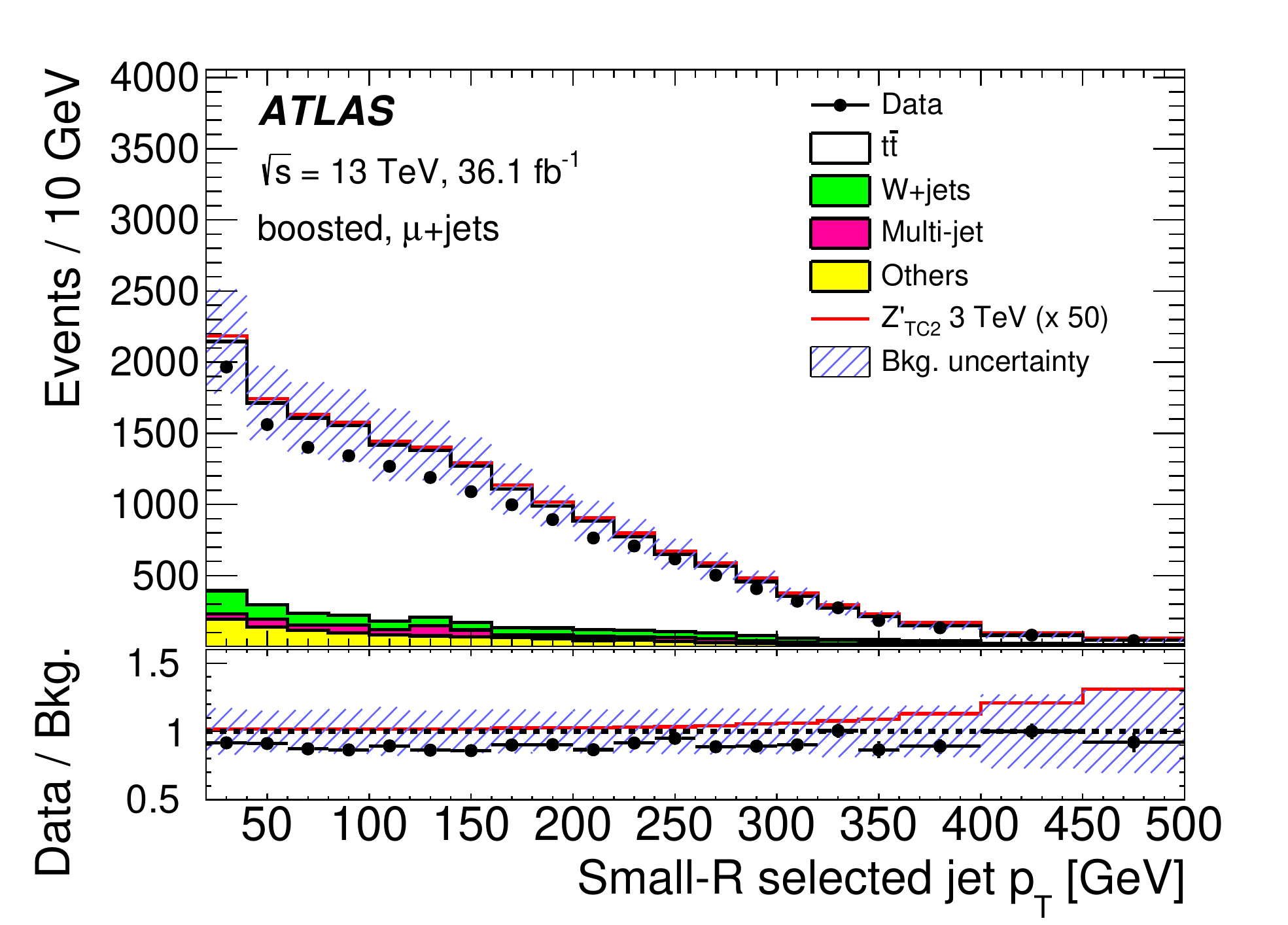}
}
\caption{The distribution of the transverse momentum of the hardest small-$R$ jet with $\Delta R(\ell, \textrm{jet}) < 1.5$ in the (a) boosted $e$+jets, and (b) boosted $\mu$+jets selections. The SM background components are shown as stacked histograms. The shaded areas indicate the total systematic uncertainties.
The lower panels in each plot show the ratio of data (points) and a signal example (line) to the background expectation.}
\label{fig:closeJetPt_boosted}
\end{figure}

\begin{figure}[ht]
\centering
\subfigure[~$e$+jets selection.]{
\includegraphics[width=0.45\linewidth]{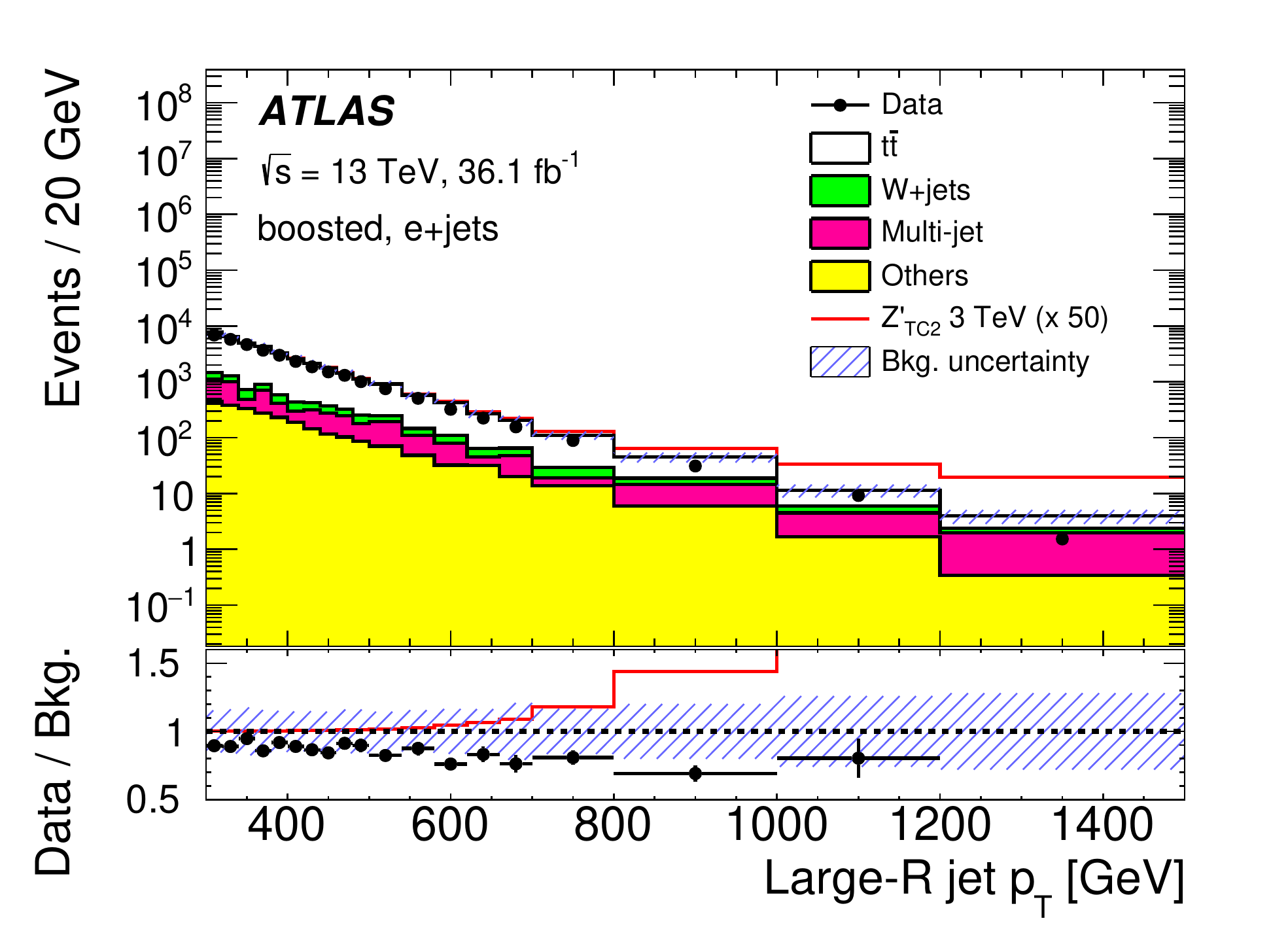}
}
\subfigure[~$\mu$+jets selection.]{
\includegraphics[width=0.45\linewidth]{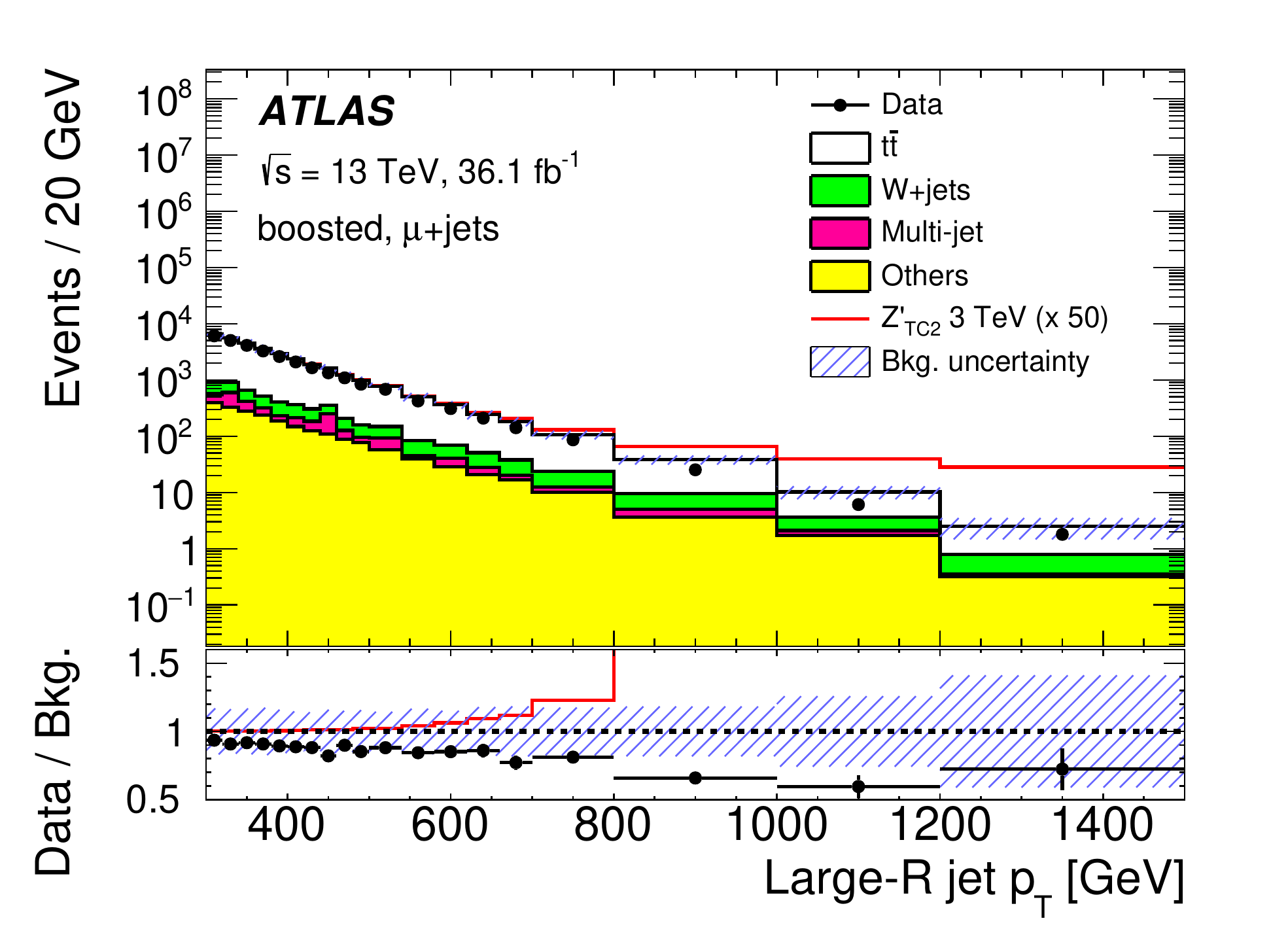}
}
\caption{The distribution of the transverse momentum of the large-$R$ jet in the (a) boosted $e$+jets, and (b) boosted $\mu$+jets selections. The SM background components are shown as stacked histograms. The shaded areas indicate the total systematic uncertainties.
The lower panels in each plot show the ratio of data (points) and a signal example (line) to the background expectation.}
\label{fig:largeJetPt_boosted}
\end{figure}

\begin{figure}[ht]
\centering
\subfigure[~$e$+jets, boosted selection.]{
\includegraphics[width=0.45\linewidth]{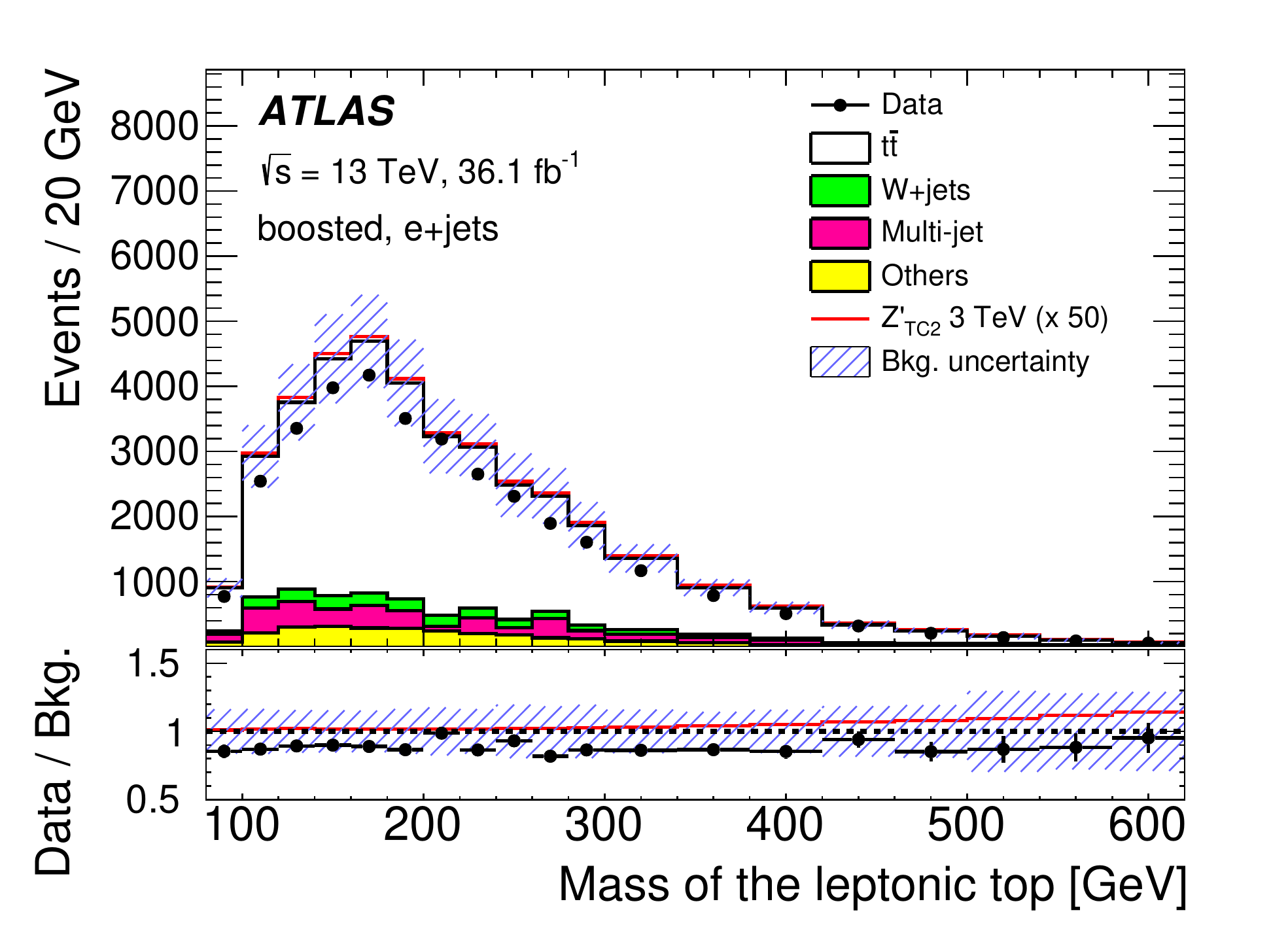}
}
\subfigure[~$\mu$+jets, boosted selection.]{
\includegraphics[width=0.45\linewidth]{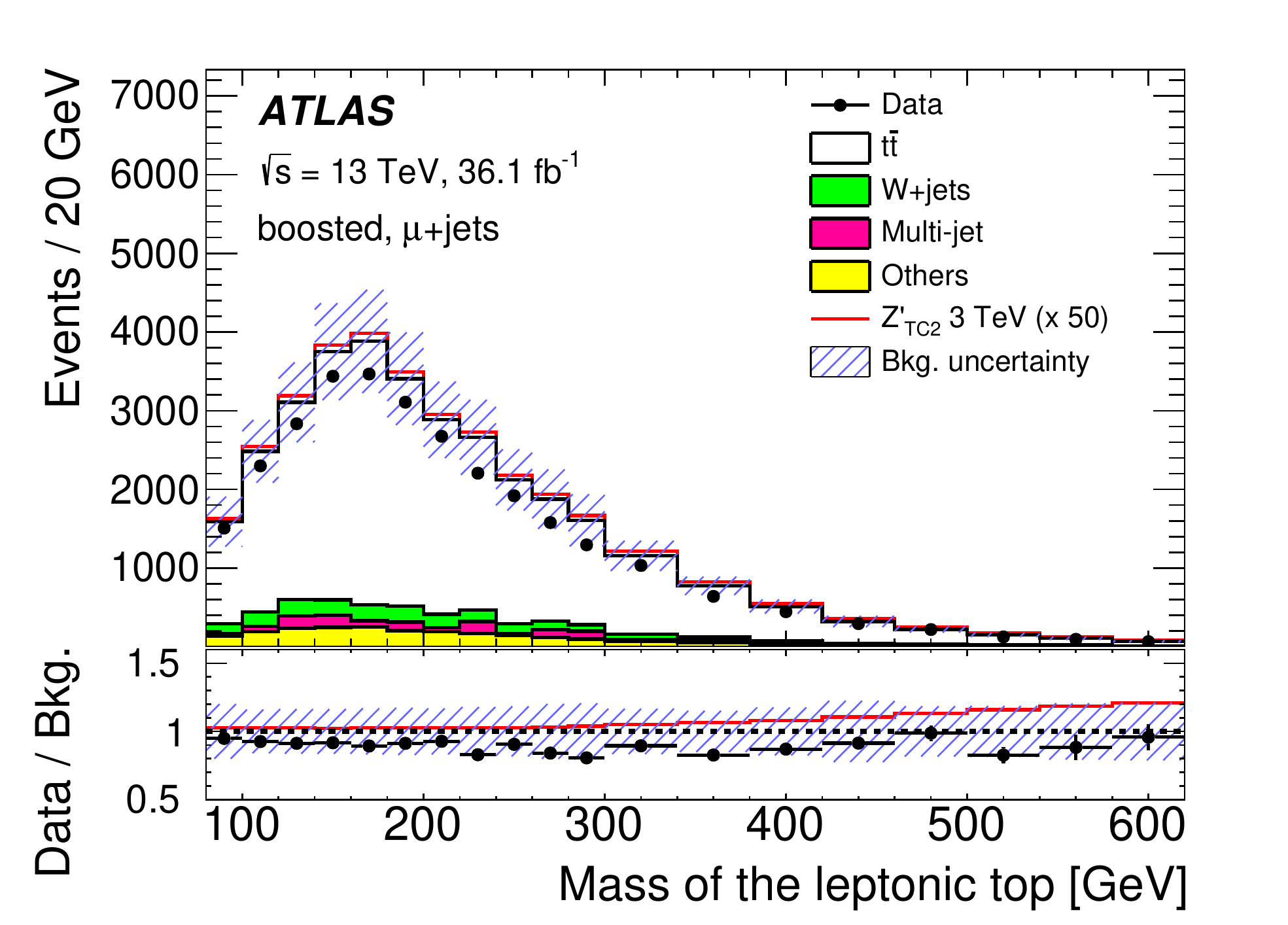}
}
\subfigure[~$e$+jets, resolved selection.]{
\includegraphics[width=0.45\linewidth]{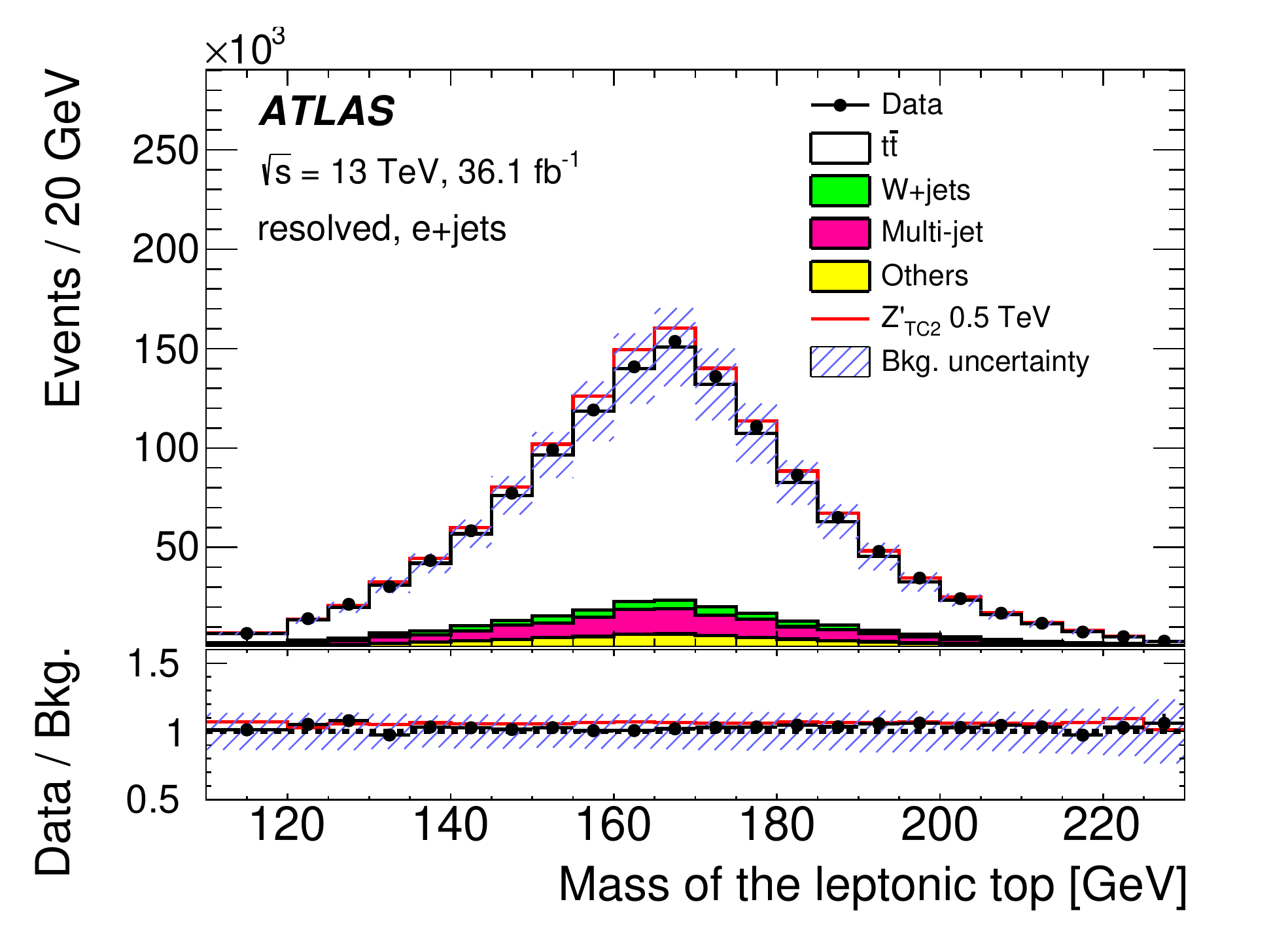}
}
\subfigure[~$\mu$+jets, resolved selection.]{
\includegraphics[width=0.45\linewidth]{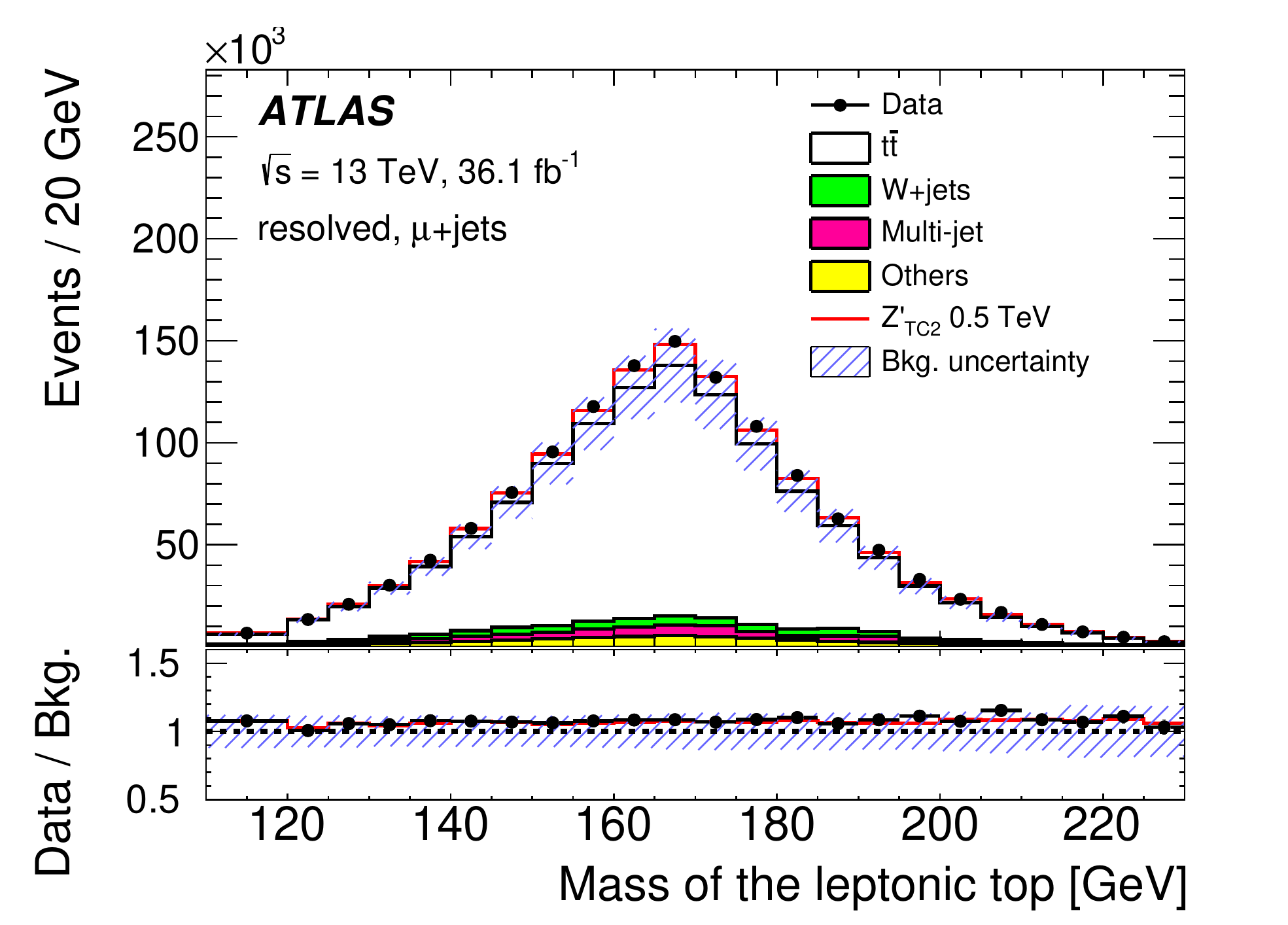}
}

\caption{The distribution of the reconstructed mass of the leptonic-top candidate in the (a) boosted $e$+jets, (b) boosted $\mu$+jets, (c) resolved $e$+jets, and (d) resolved $\mu$+jets selections. The SM background components are shown as stacked histograms. The shaded areas indicate the total systematic uncertainties.
The lower panels in each plot show the ratio of data (points) and a signal example (line) to the background expectation.}
\label{fig:mtlep_boosted}
\end{figure}

\begin{figure}[ht]
\centering
\subfigure[~$e$+jets, boosted selection.]{
\includegraphics[width=0.45\linewidth]{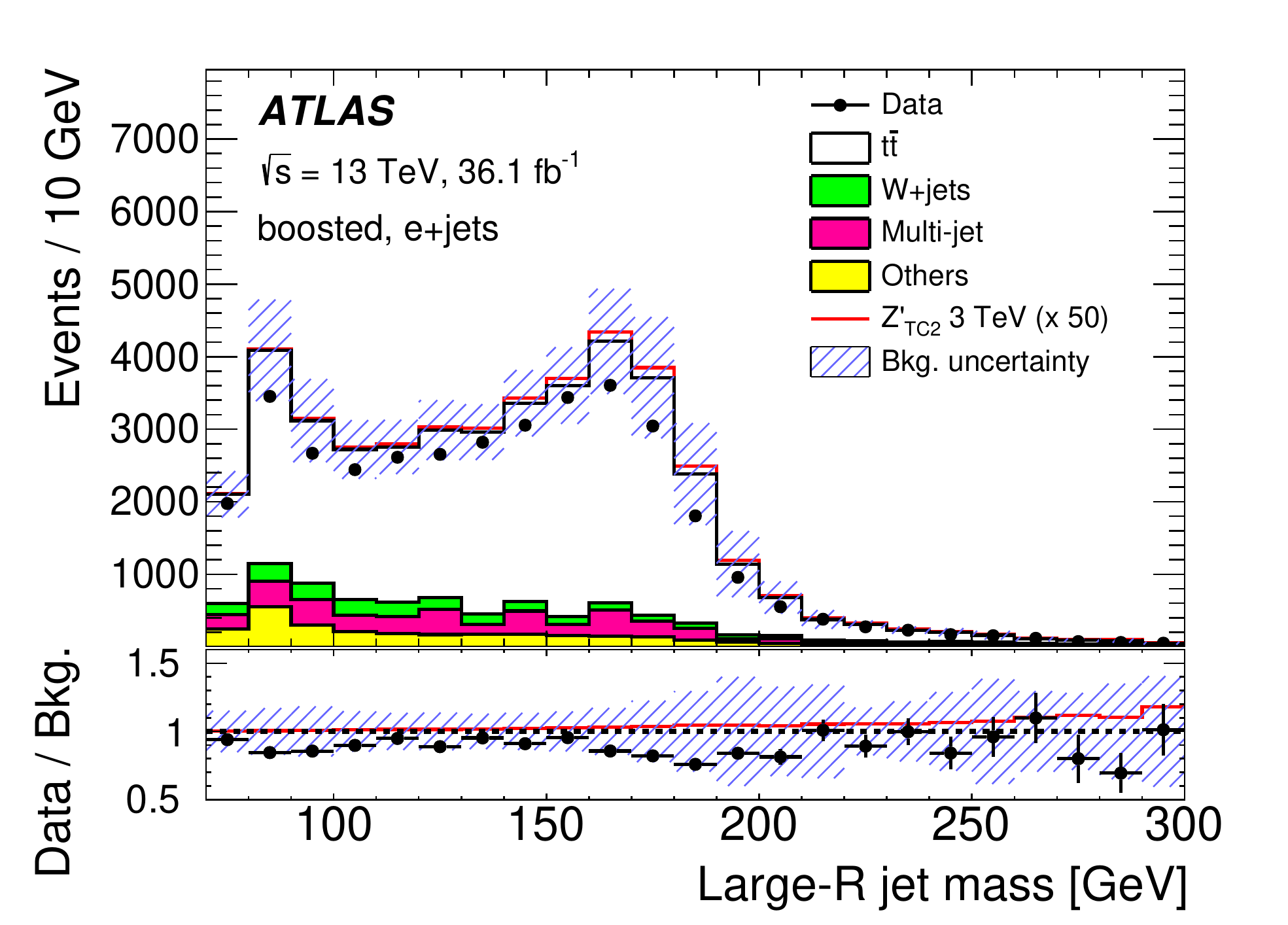}
}
\subfigure[~$\mu$+jets, boosted selection.]{
\includegraphics[width=0.45\linewidth]{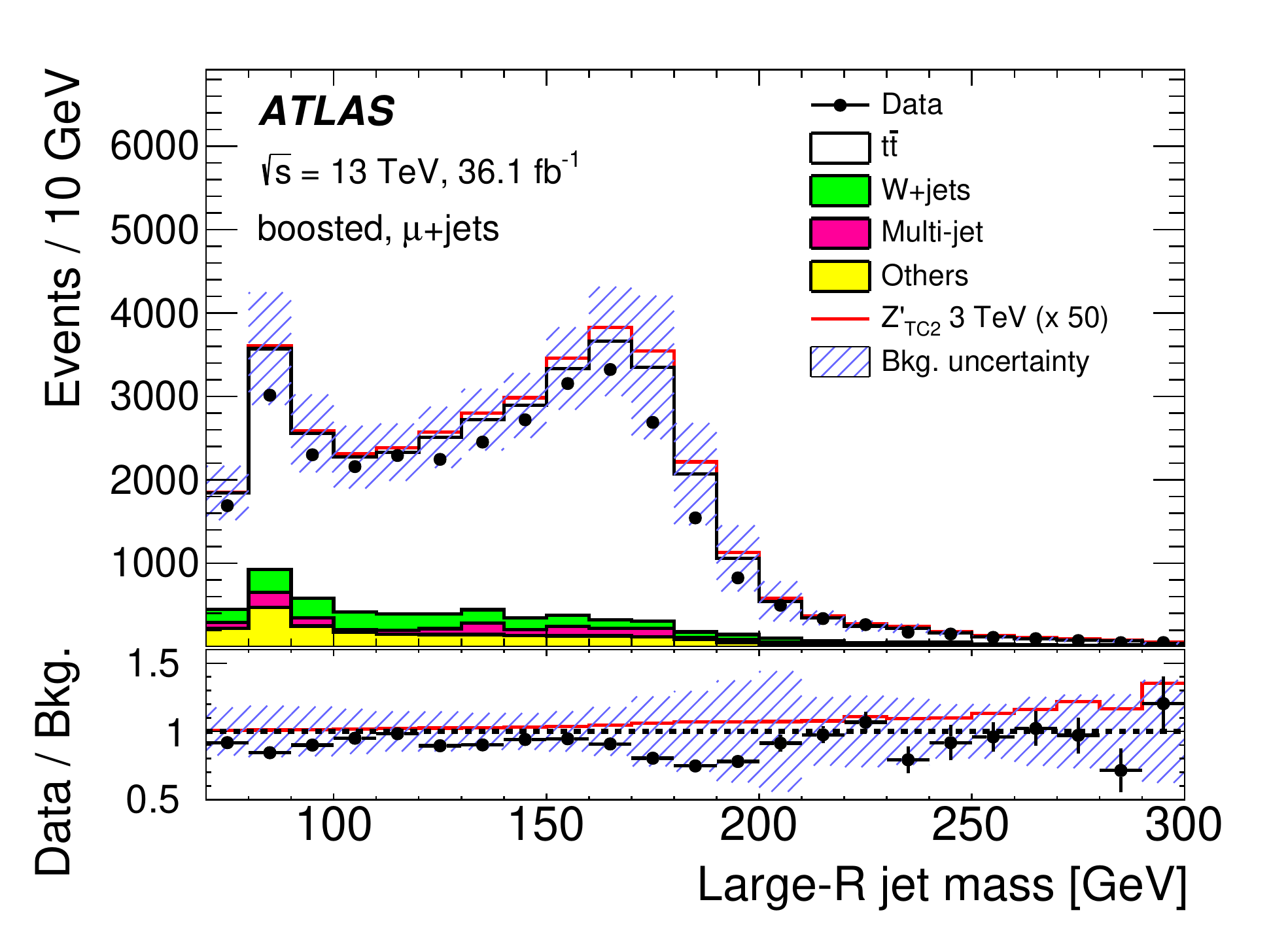}
}
\subfigure[~$e$+jets, resolved selection.]{
\includegraphics[width=0.45\linewidth]{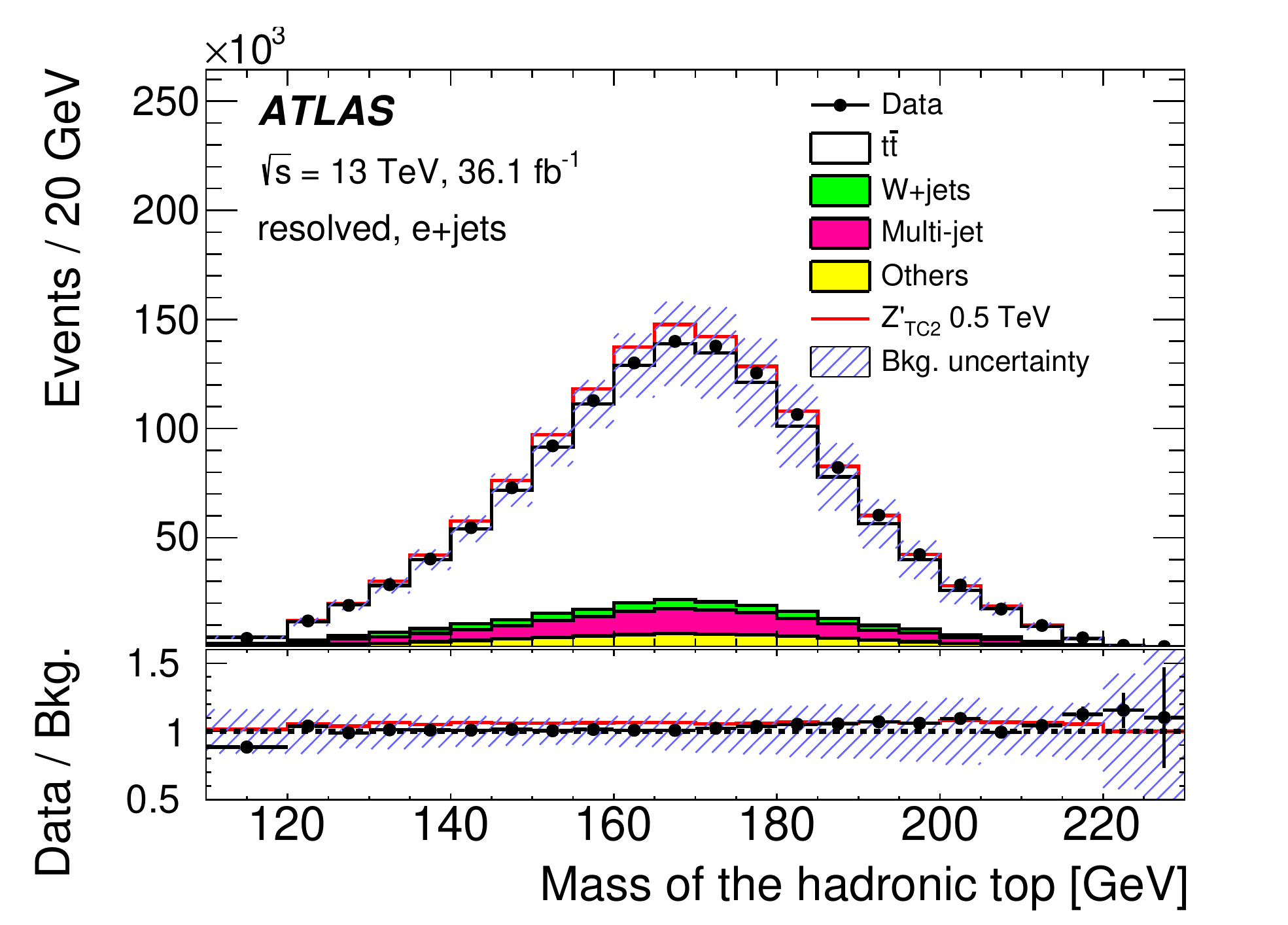}
}
\subfigure[~$\mu$+jets, resolved selection.]{
\includegraphics[width=0.45\linewidth]{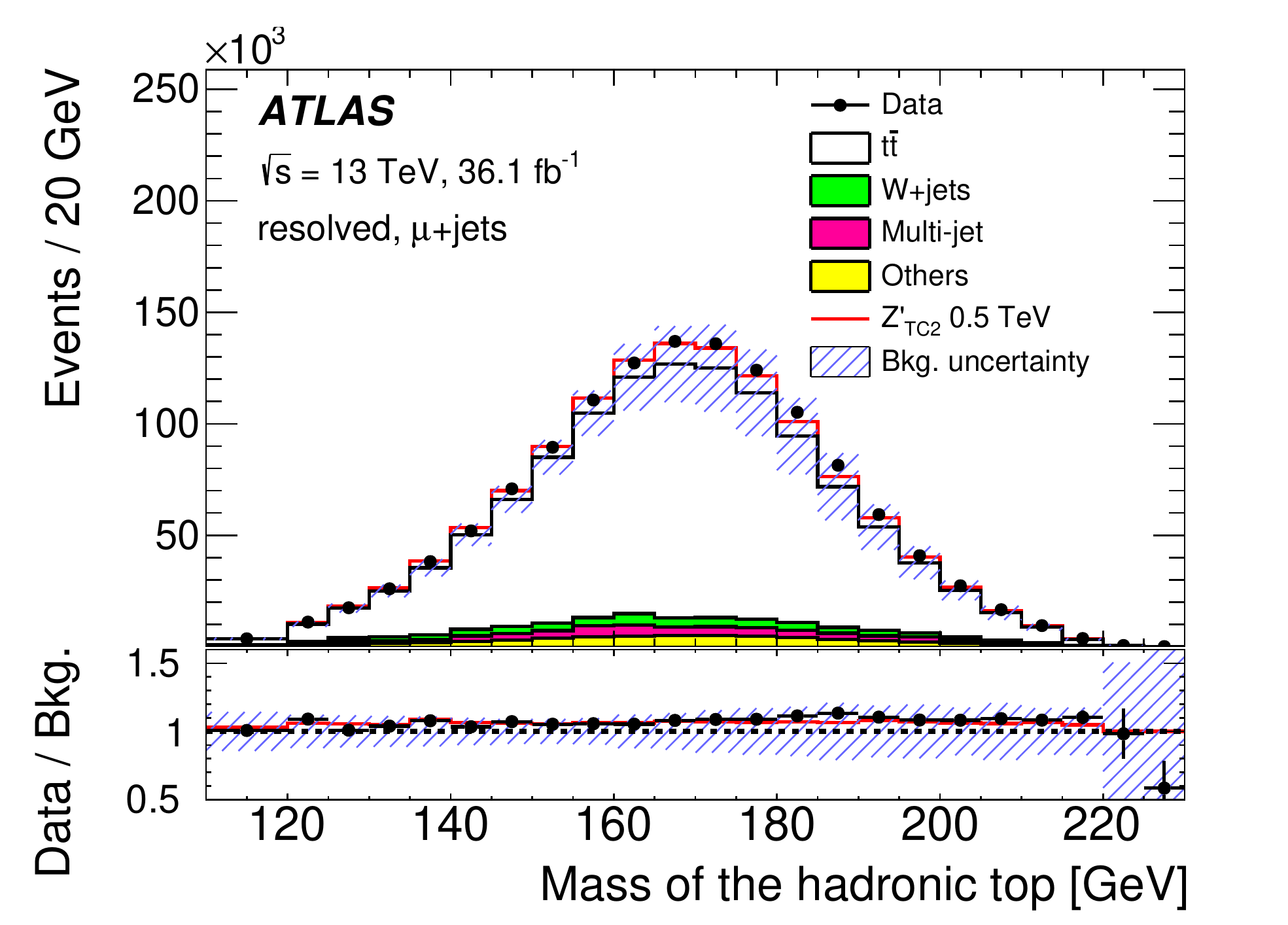}
}
\caption{The distribution of the mass of the large-$R$ jet in the (a) boosted $e$+jets, and (b) boosted $\mu$+jets selections. The mass of the hadronic-top candidate in the (c) resolved $e$+jets, and (d) resolved $\mu$+jets selections. The SM background components are shown as stacked histograms. The shaded areas indicate the total systematic uncertainties.
The lower panels in each plot show the ratio of data (points) and a signal example (line) to the background expectation.}
\label{fig:largeJetM_boosted}
\end{figure}

Figure~\ref{fig:lepPt_boosted} shows the transverse momentum of the charged lepton in the selected events. The $\met{}$ distribution is shown in Figure~\ref{fig:met_boosted}. The transverse momentum of the selected jet and top-tagged jets are shown in Figures~\ref{fig:closeJetPt_boosted} and \ref{fig:largeJetPt_boosted}. Figures \ref{fig:mtlep_boosted} and \ref{fig:largeJetM_boosted} show the reconstructed mass of the leptonic- and hadronic-top candidates. For all of the distributions in the resolved selections, any deviations from expectations are well within the statistical and systematic uncertainties.
As some top-quark decays are not fully contained within the large-$R$ jet,  two peaks in the jet mass distribution are visible  in Figure~\ref{fig:largeJetM_boosted}. One close to the $W$ boson mass for the cases in which only the $W$ boson decay products are reconstructed within the large-$R$ jet, and one close to the top-quark mass. There is a tendency for the expectations in the boosted selections to be 10--20\% below the data while exhibiting a similar shape.

\begin{figure}[ht]
\centering
\subfigure[$e$+jets $b$-tag category 1]{\includegraphics[width=0.3\linewidth]{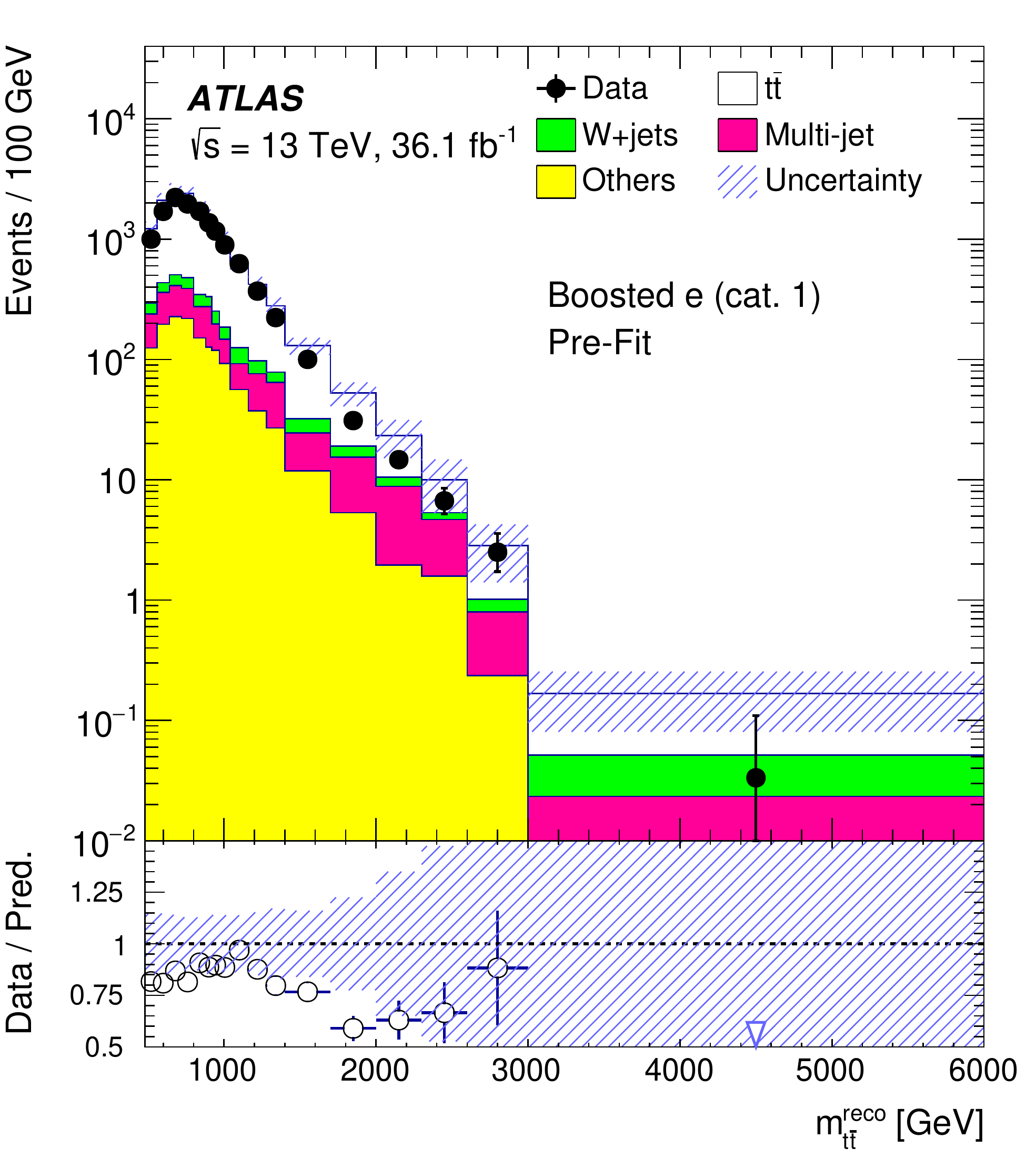}}
\subfigure[$\mu$+jets $b$-tag category 1]{\includegraphics[width=0.3\linewidth]{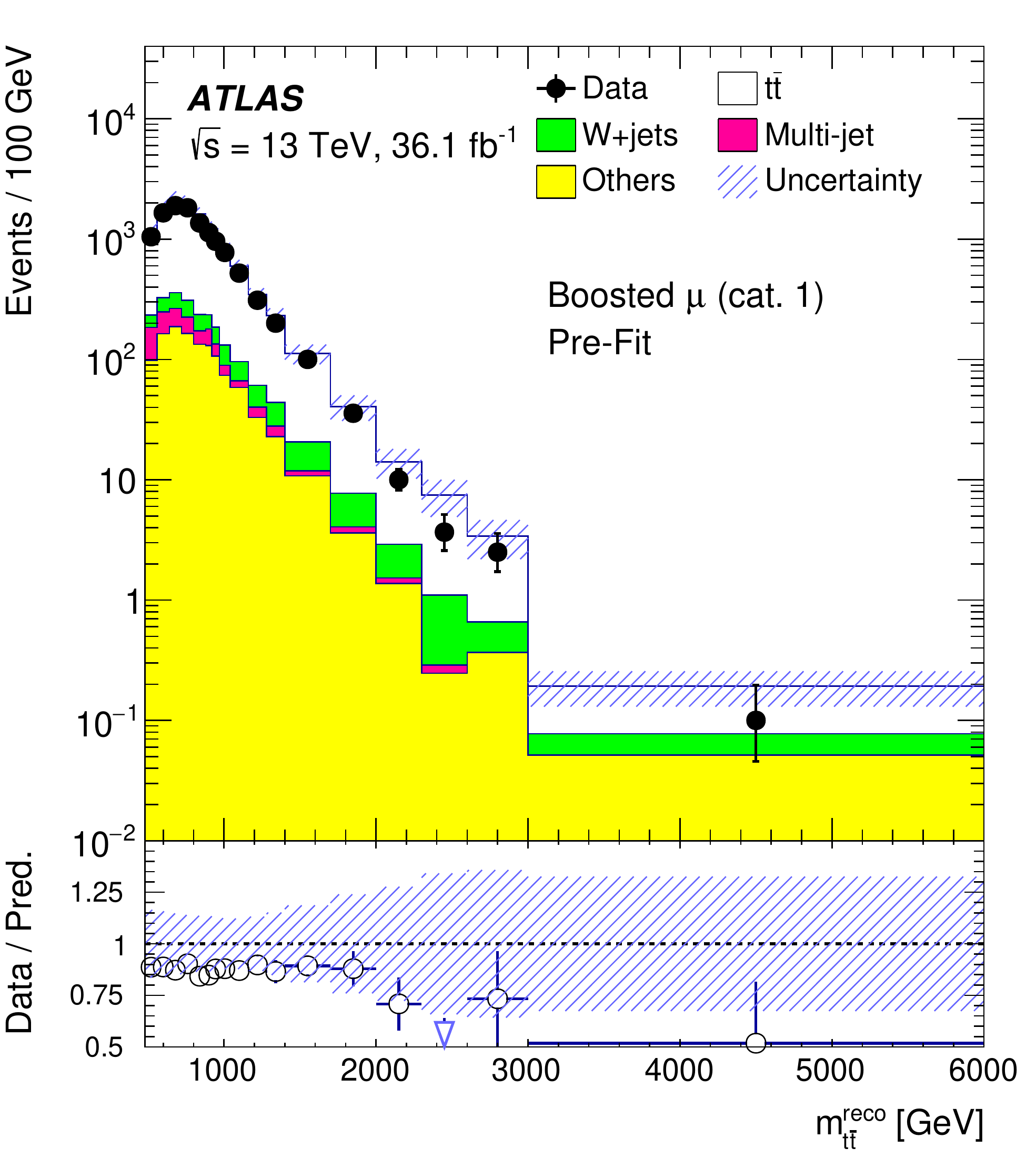}} \\
\subfigure[$e$+jets $b$-tag category 2]{\includegraphics[width=0.3\linewidth]{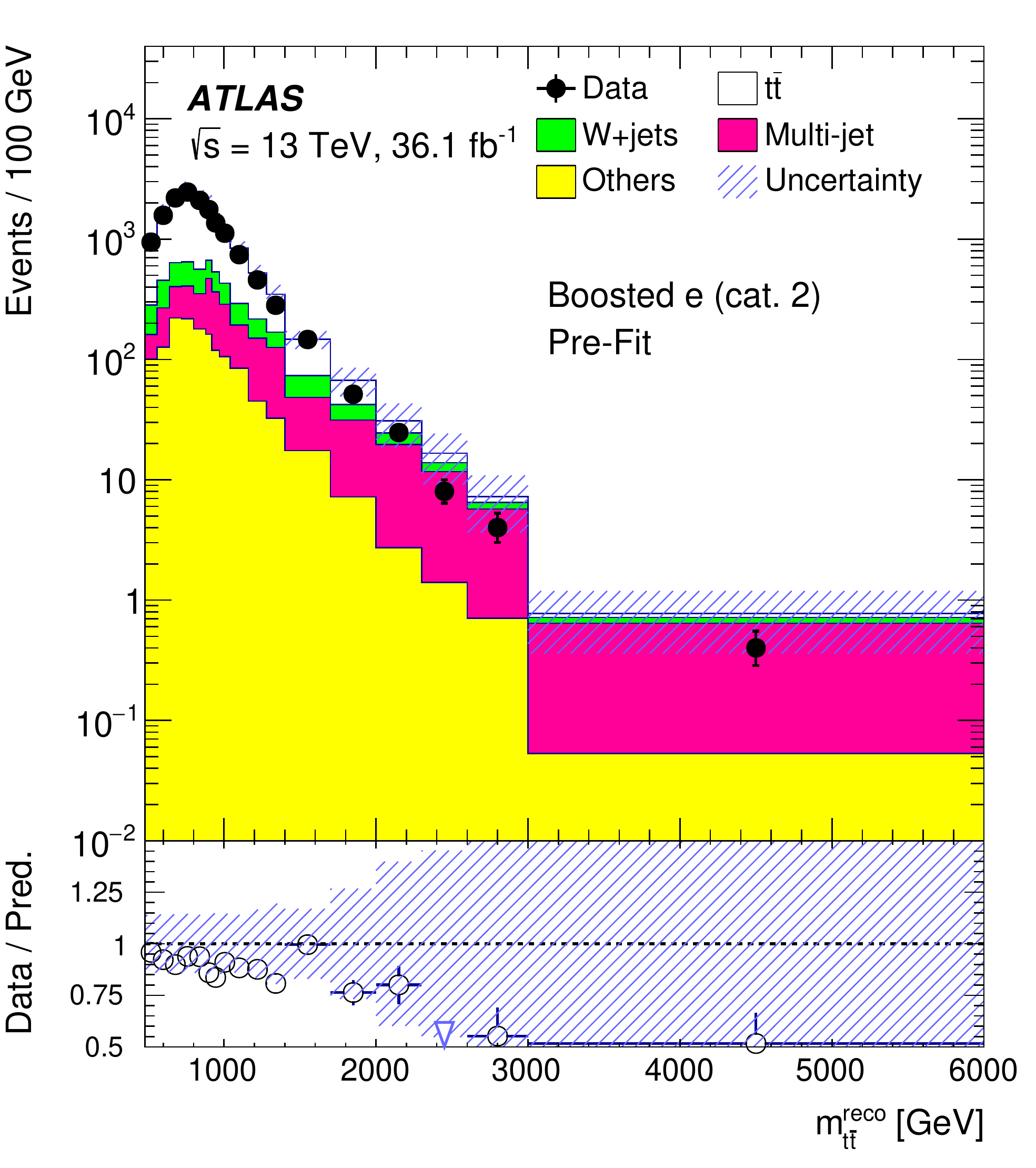}}
\subfigure[$\mu$+jets $b$-tag category 2]{\includegraphics[width=0.3\linewidth]{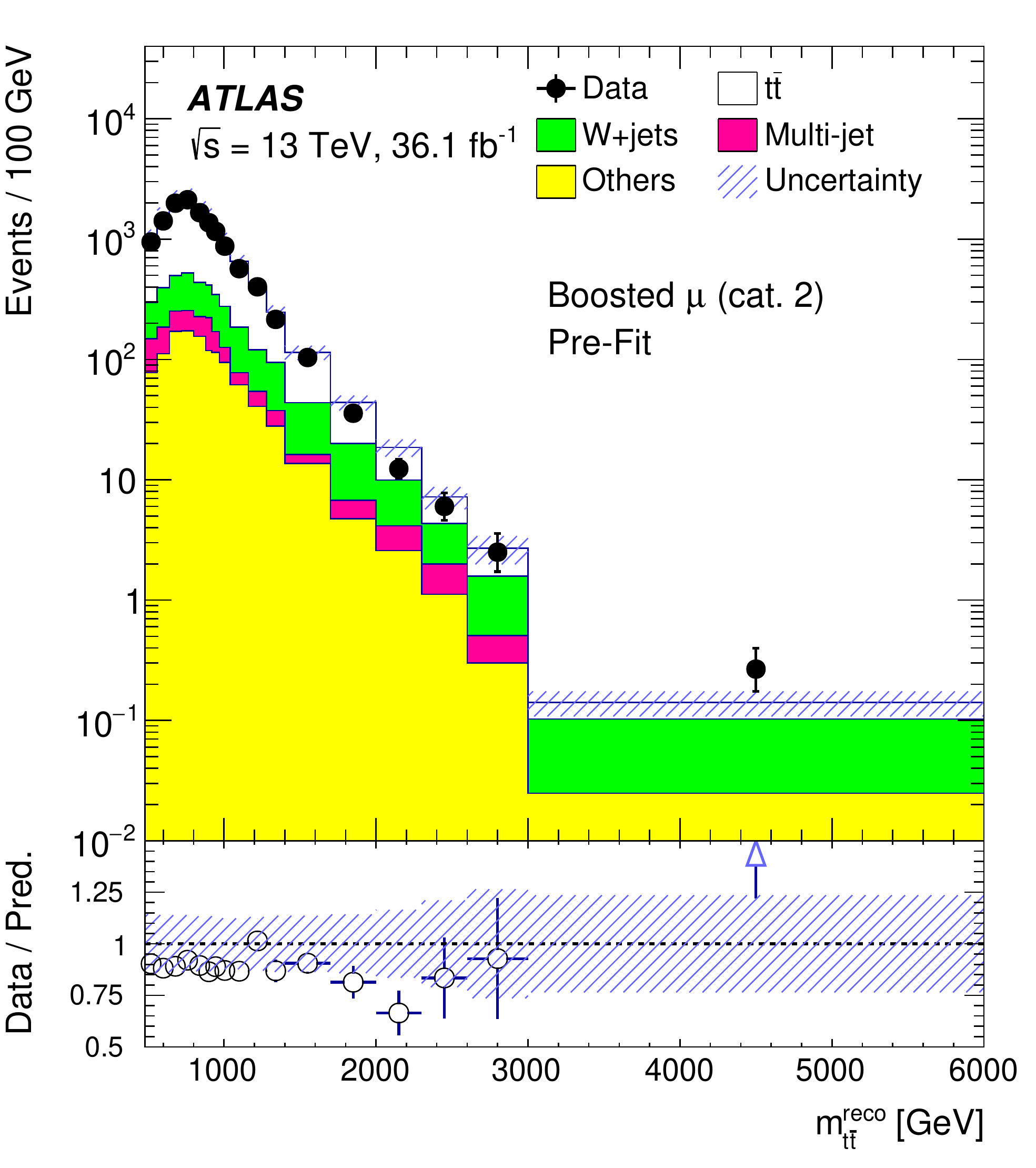}}\\
\subfigure[$e$+jets $b$-tag category 3]{\includegraphics[width=0.3\linewidth]{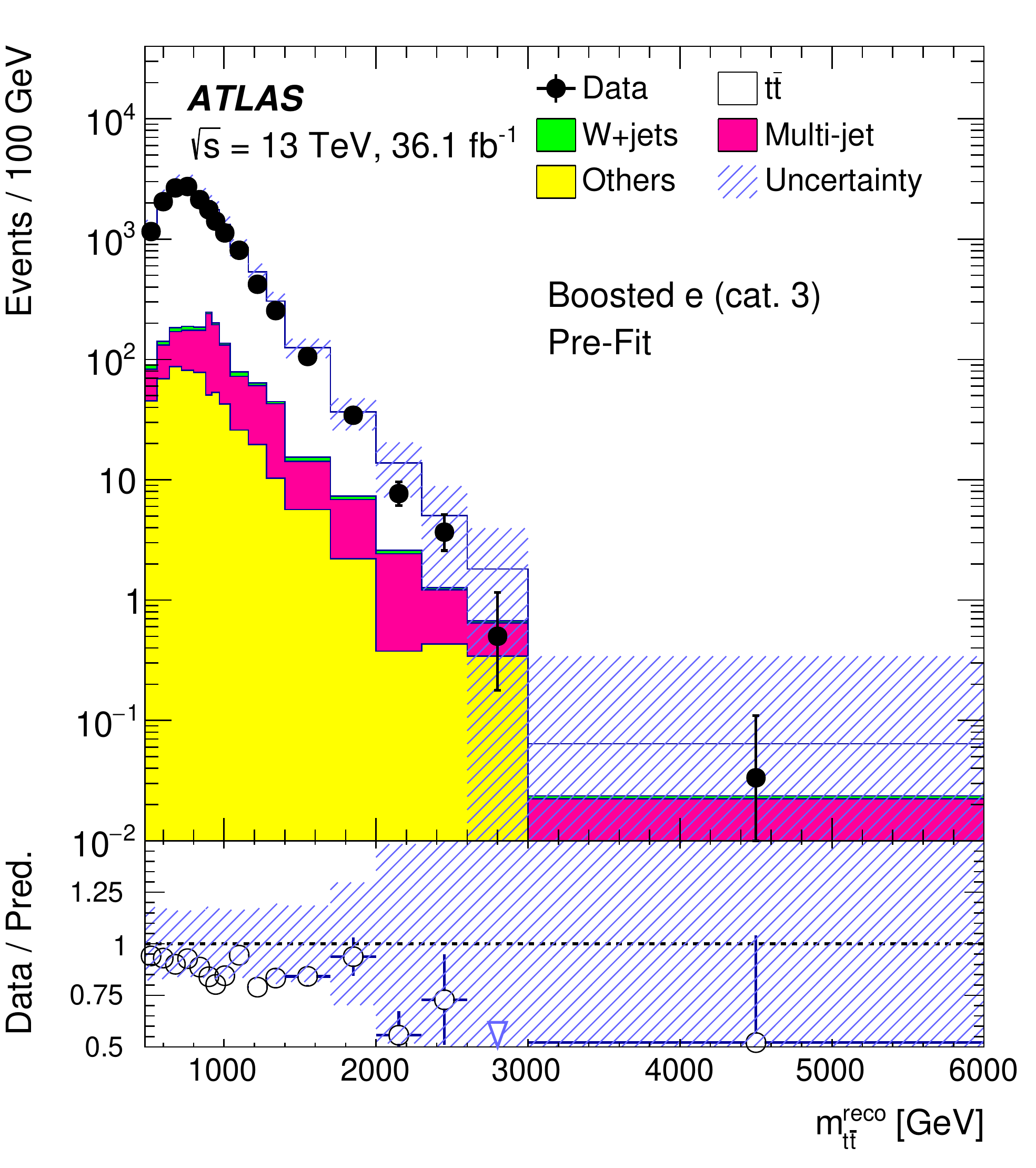}}
\subfigure[$\mu$+jets $b$-tag category 3]{\includegraphics[width=0.3\linewidth]{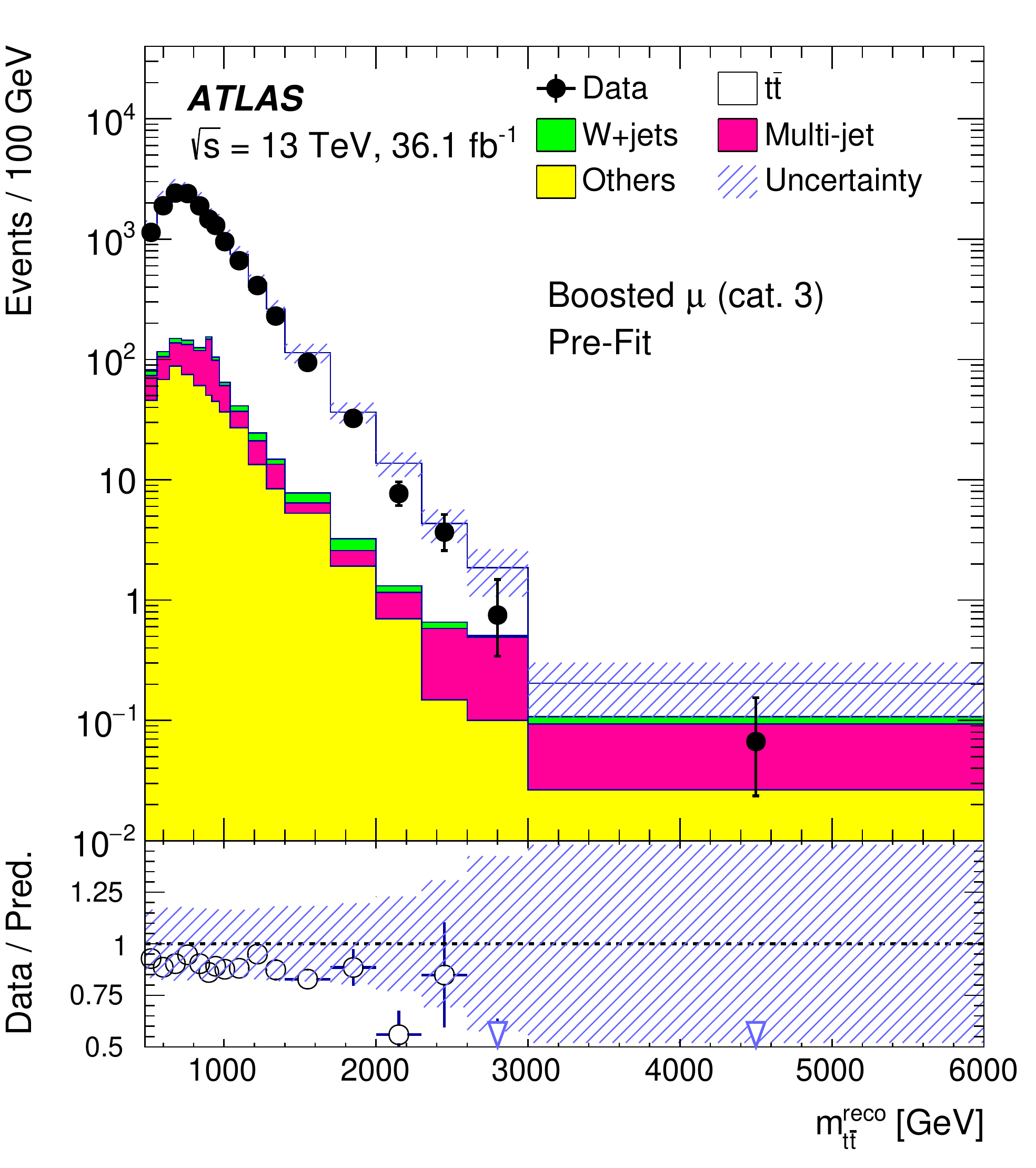}}\\
\caption{The $m^{\text{reco}}_{t\bar{t}}$  distribution before the likelihood fit in the boosted selection. The SM background components are shown as stacked histograms. The shaded areas indicate the total systematic uncertainties. The ratio of the data to the  total expectation from background processes is shown in the lower panel, open triangles indicate that the ratio point would appear outside the panel.}
\label{fig:mtt_boosted}
\end{figure}

\begin{figure}[ht]
\centering
\subfigure[$e$+jets $b$-tag category 1]{\includegraphics[width=0.3\linewidth]{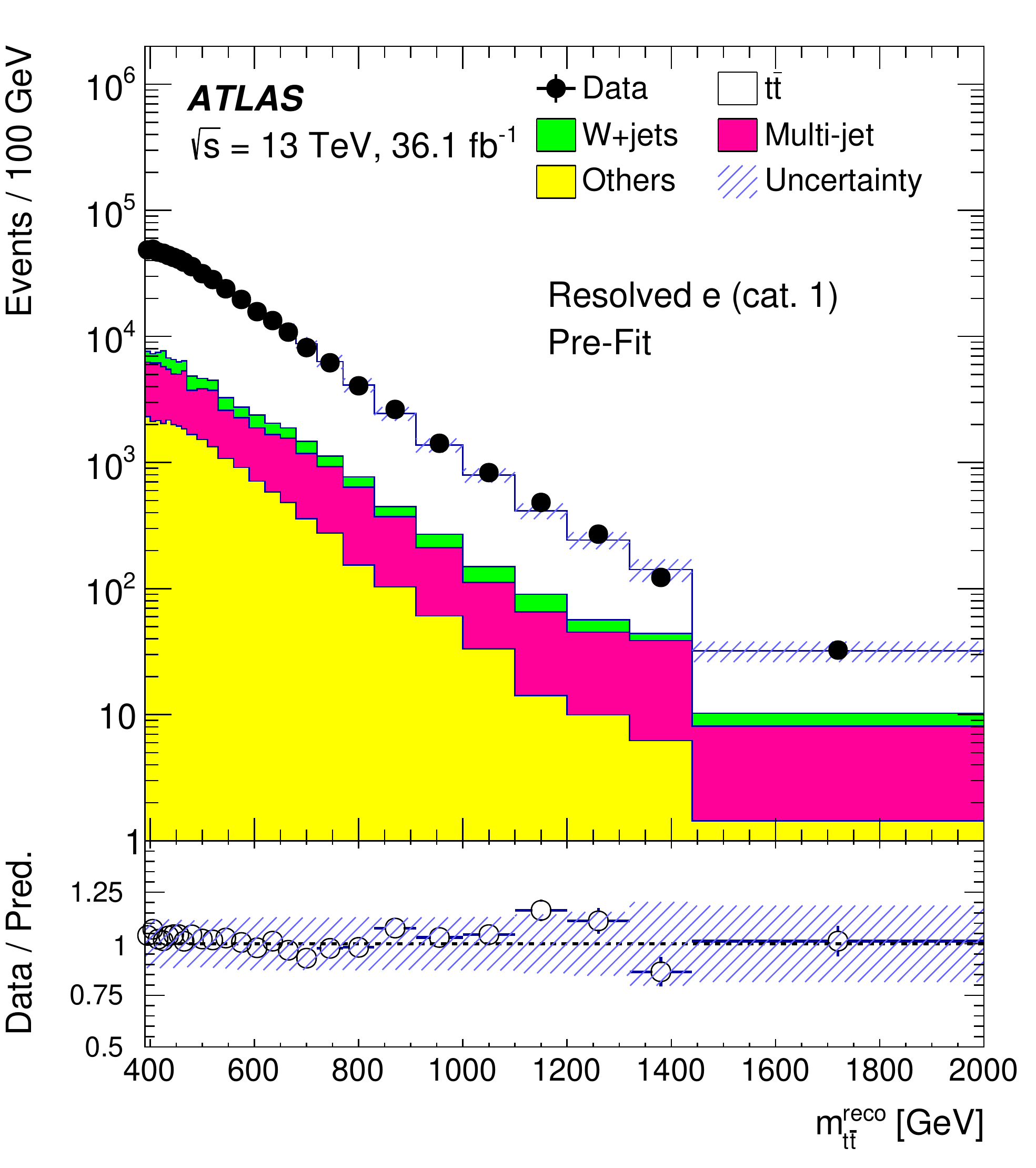}}
\subfigure[$\mu$+jets $b$-tag category 1]{\includegraphics[width=0.3\linewidth]{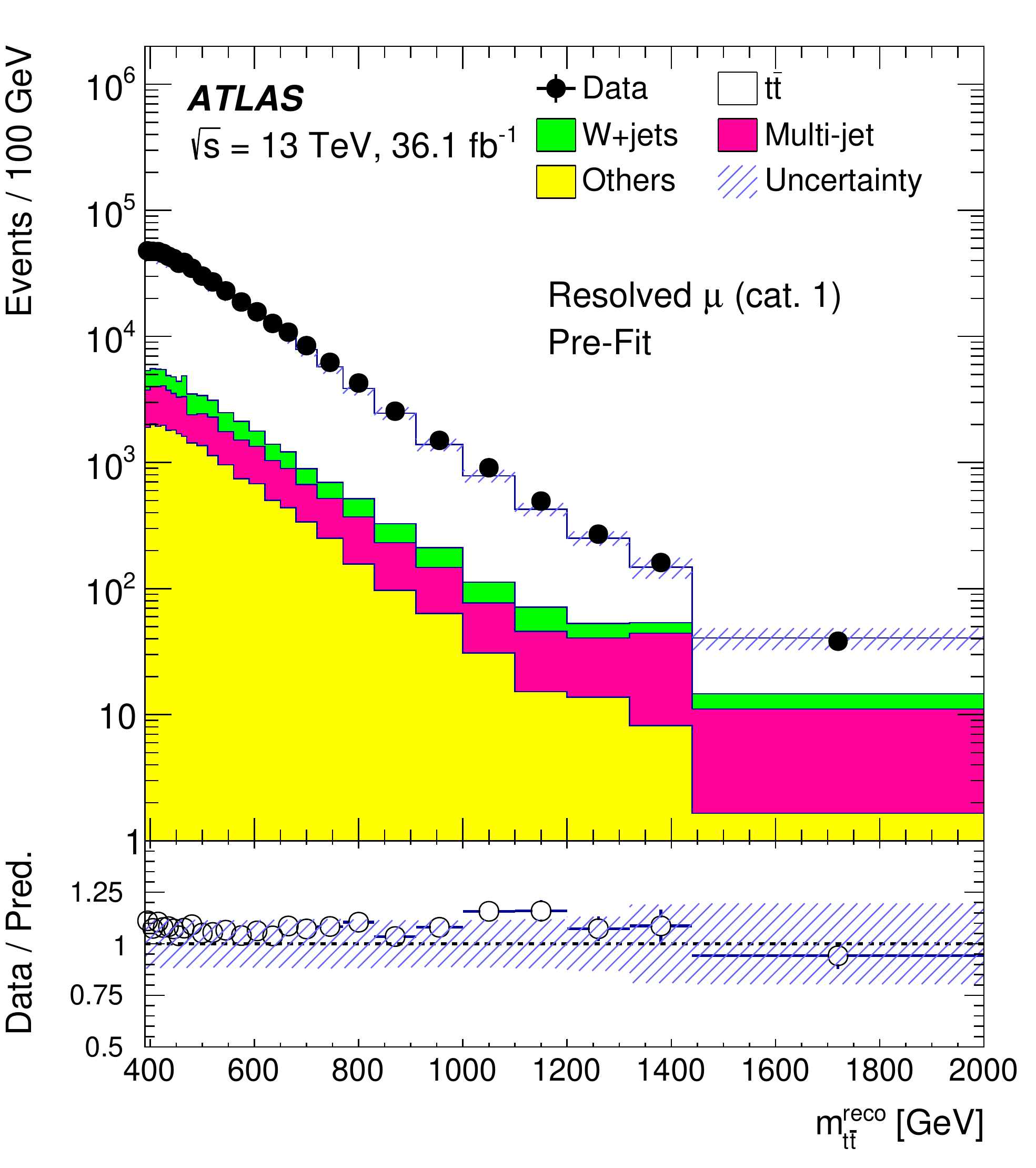}} \\
\subfigure[$e$+jets $b$-tag category 2]{\includegraphics[width=0.3\linewidth]{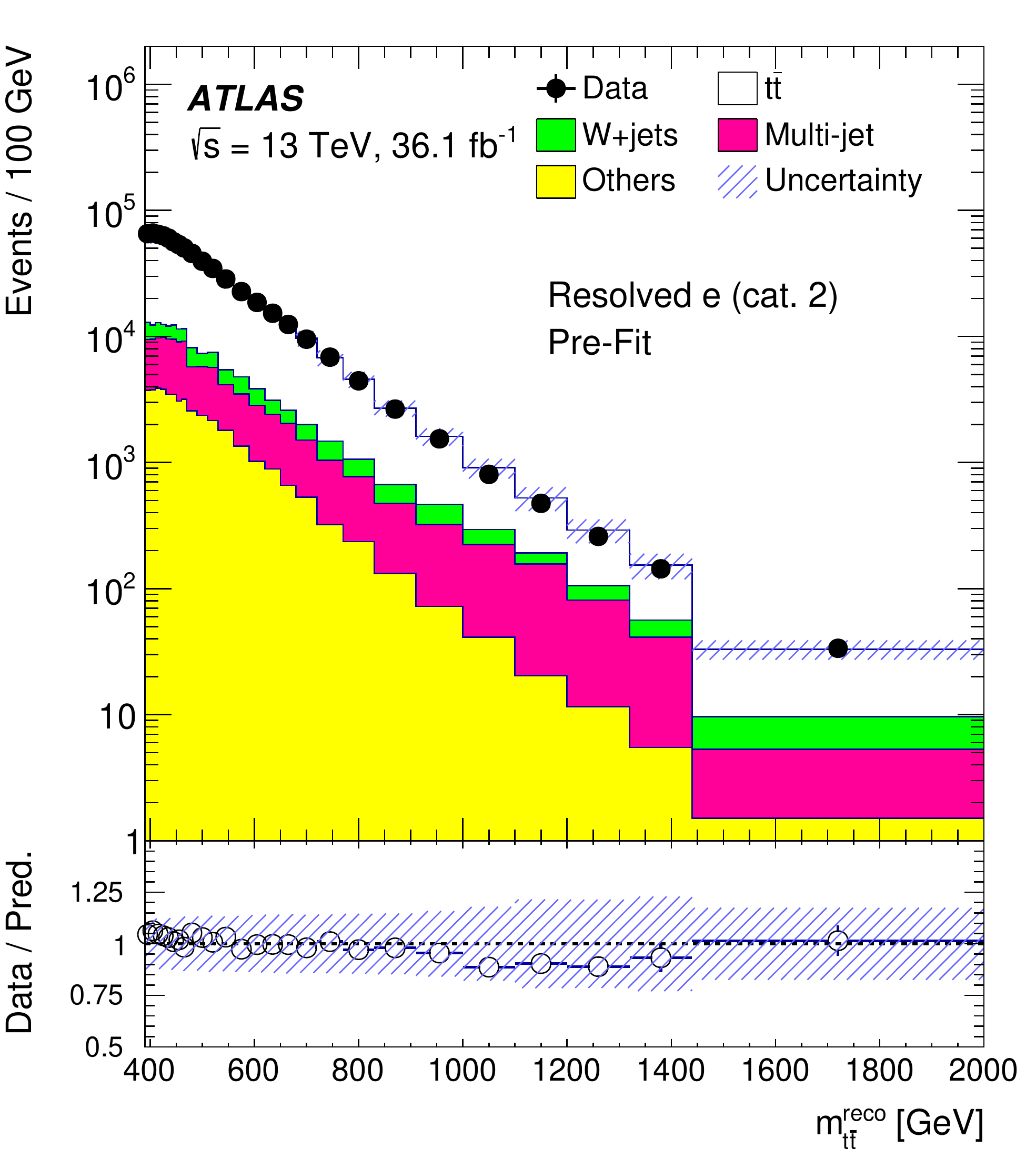}}
\subfigure[$\mu$+jets $b$-tag category 2]{\includegraphics[width=0.3\linewidth]{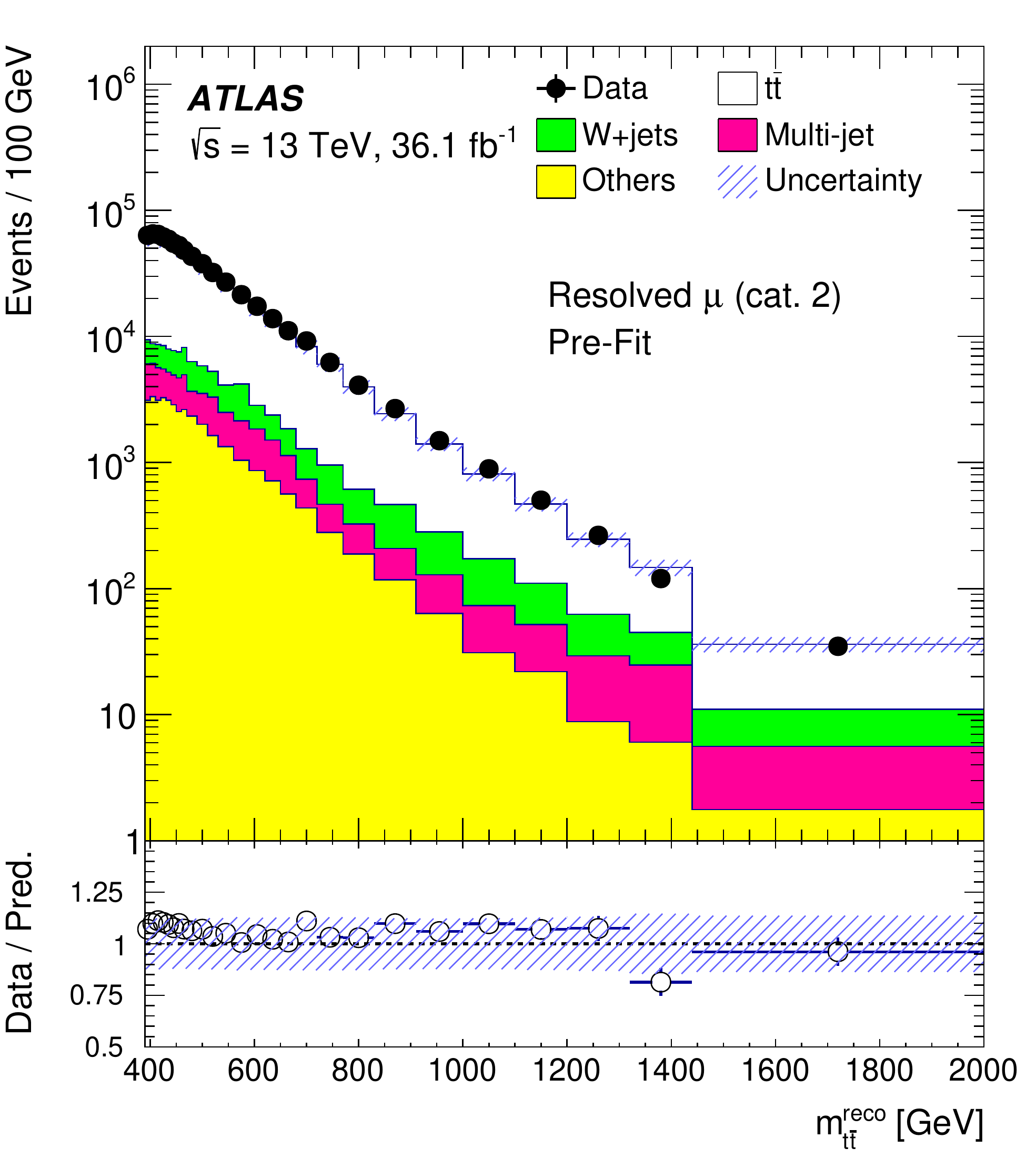}}\\
\subfigure[$e$+jets $b$-tag category 3]{\includegraphics[width=0.3\linewidth]{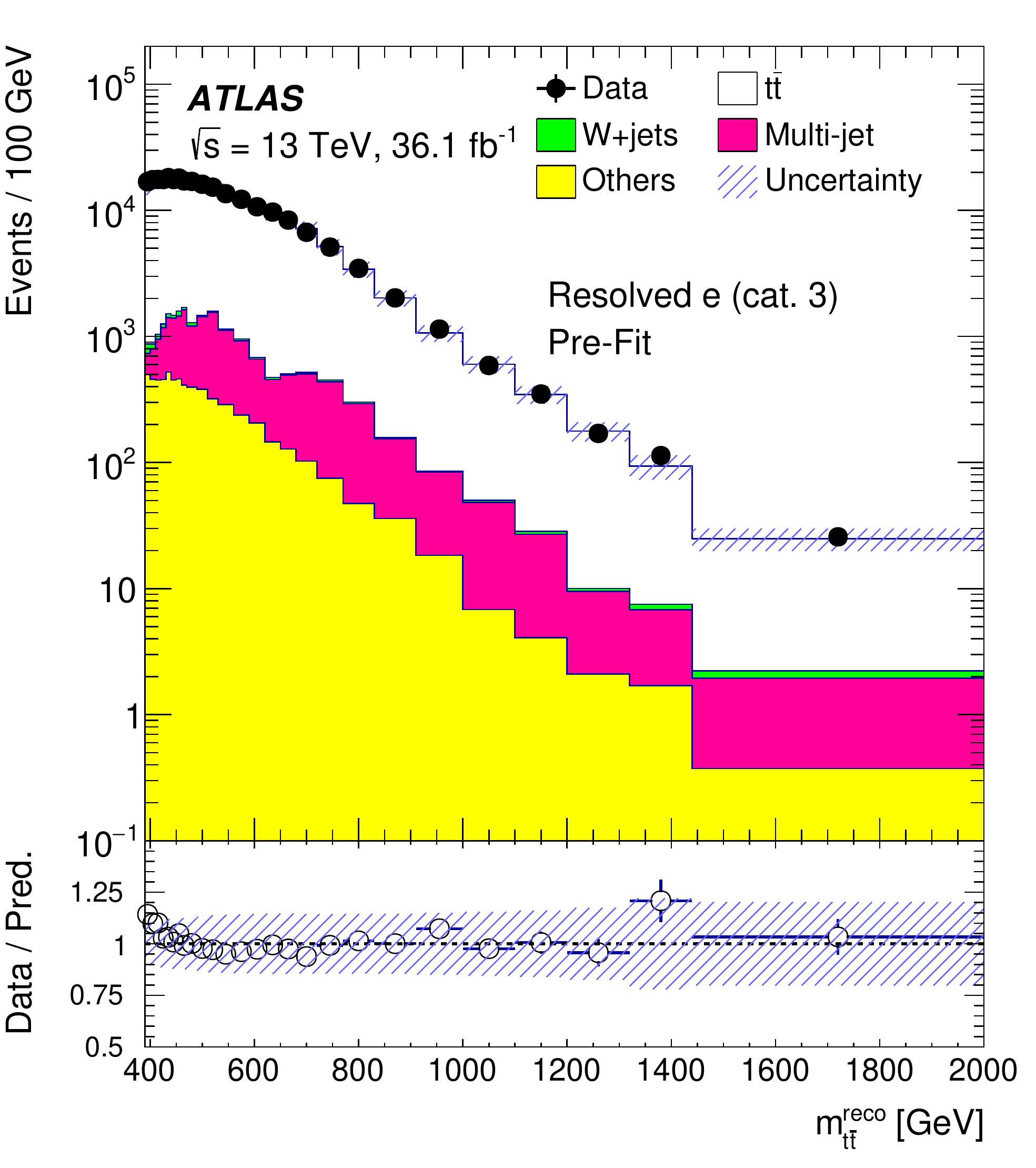}}
\subfigure[$\mu$+jets $b$-tag category 3]{\includegraphics[width=0.3\linewidth]{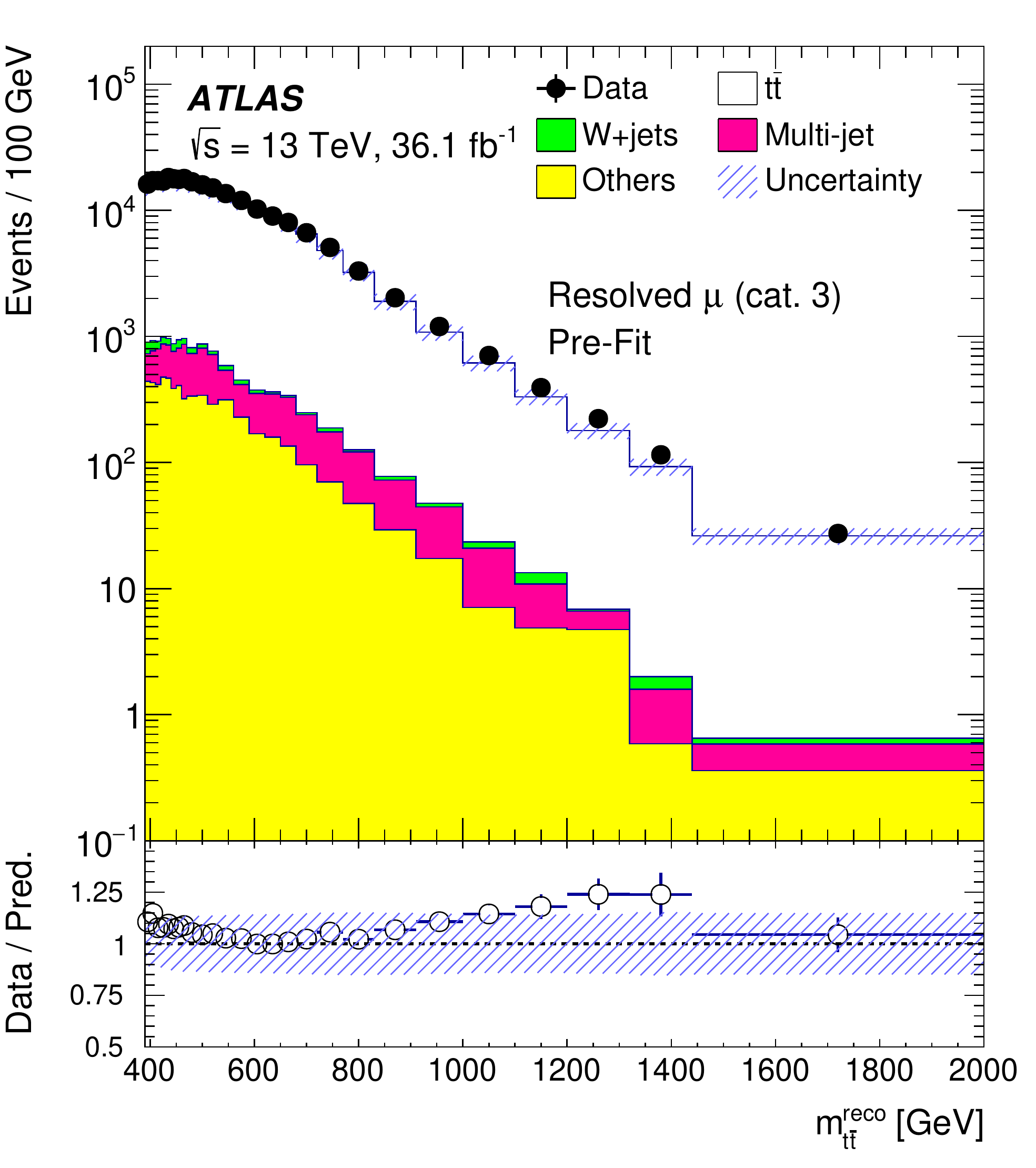}}\\
\caption{The $m^{\text{reco}}_{t\bar{t}}$  distribution before the likelihood fit in the resolved selection. The SM background components are shown as stacked histograms. The shaded areas indicate the total systematic uncertainties.  The ratio of the data to the  total expectation from background processes is shown in the lower panel.}
\label{fig:mtt_resolved}
\end{figure}

The reconstructed \ttbar invariant mass spectra for the electron and muon selections are shown in Figures~\ref{fig:mtt_boosted} and~\ref{fig:mtt_resolved}. The data generally agree with the expected background with slight shape differences seen especially in the high-mass and low-mass regions. 
These deviations are consistent with the nominal predictions within the associated uncertainties. 

The fraction of the SM $W$+jets background increases as a function of $m_{t\bar{t}}^{\text{reco}}$ in the boosted channel, with a higher fraction in the boosted selection in $b$-tag category~2,
where it contributes roughly 50\% of the background for $m_{t\bar{t}}^{\text{reco}} > 3$~\TeV. The fraction in $b$-tag category 3, which is the purest channel, is at most 6\% for $m_{t\bar{t}}^{\text{reco}} > 3$~\TeV.
In the resolved channel, the contribution of the $W$+jets background also grows with $m_{t\bar{t}}^{\text{reco}}$ and it contributes less than 1\%
in the $b$-tag category~3,  while it has up to a 14\% effect in $b$-tag category~2.

\section{Results}
\label{sec:results}

The final discriminating observables used to search for a 
massive resonance are the $m_{t\bar{t}}^{\text{reco}}$ spectra from the two selections.
After the reconstruction of the \ttbar\ mass distribution, the data and expected background distributions are compared  using {\textsc BumpHunter}~\cite{Choudalakis:2011qn}, which is a hypothesis-testing tool that searches the data for local excesses or deficits compared to the expected background, taking the look-elsewhere effect~\cite{Lyons:1900zz} into account over the full mass spectrum in both the boosted (480~\GeV\ to 6~\TeV) and resolved (390~\GeV\ to 2~\TeV) channels.
After accounting for the systematic uncertainties, no significant deviation from the total expected background is found. 
Upper limits are set on the cross-section times branching ratio for each of the signal models using a combined profile likelihood-ratio test build using the 12 categories. 
The CL$_\text{s}$ prescription~\cite{Read:2002hq} is used to derive one-sided  95\% confidence level (CL) limits. 

\begin{figure}[ht]
\centering
\subfigure[$e$+jets $b$-tag category 1]{\includegraphics[width=0.3\linewidth]{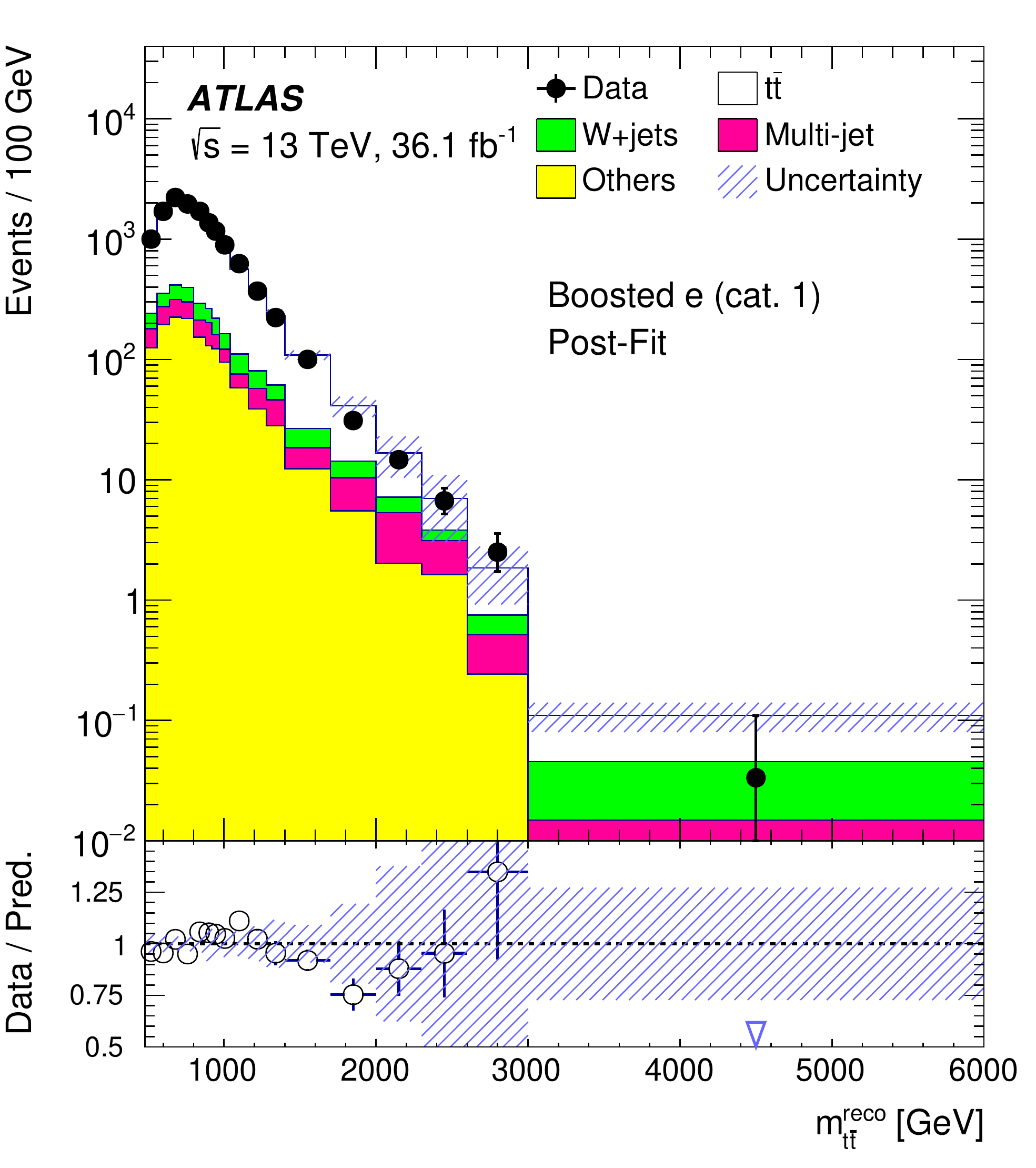}}
\subfigure[$\mu$+jets $b$-tag category 1]{\includegraphics[width=0.3\linewidth]{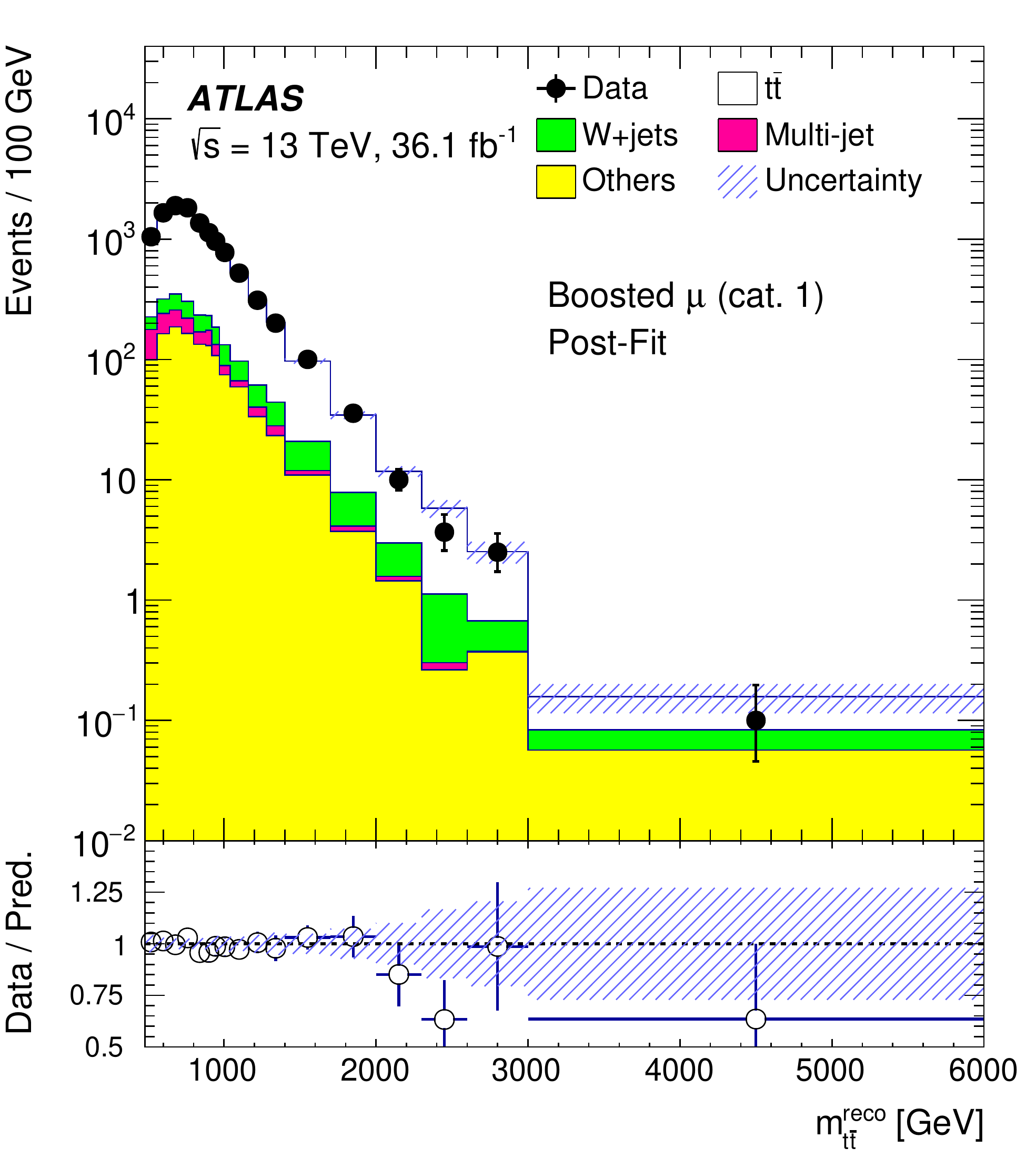}} \\
\subfigure[$e$+jets $b$-tag category 2]{\includegraphics[width=0.3\linewidth]{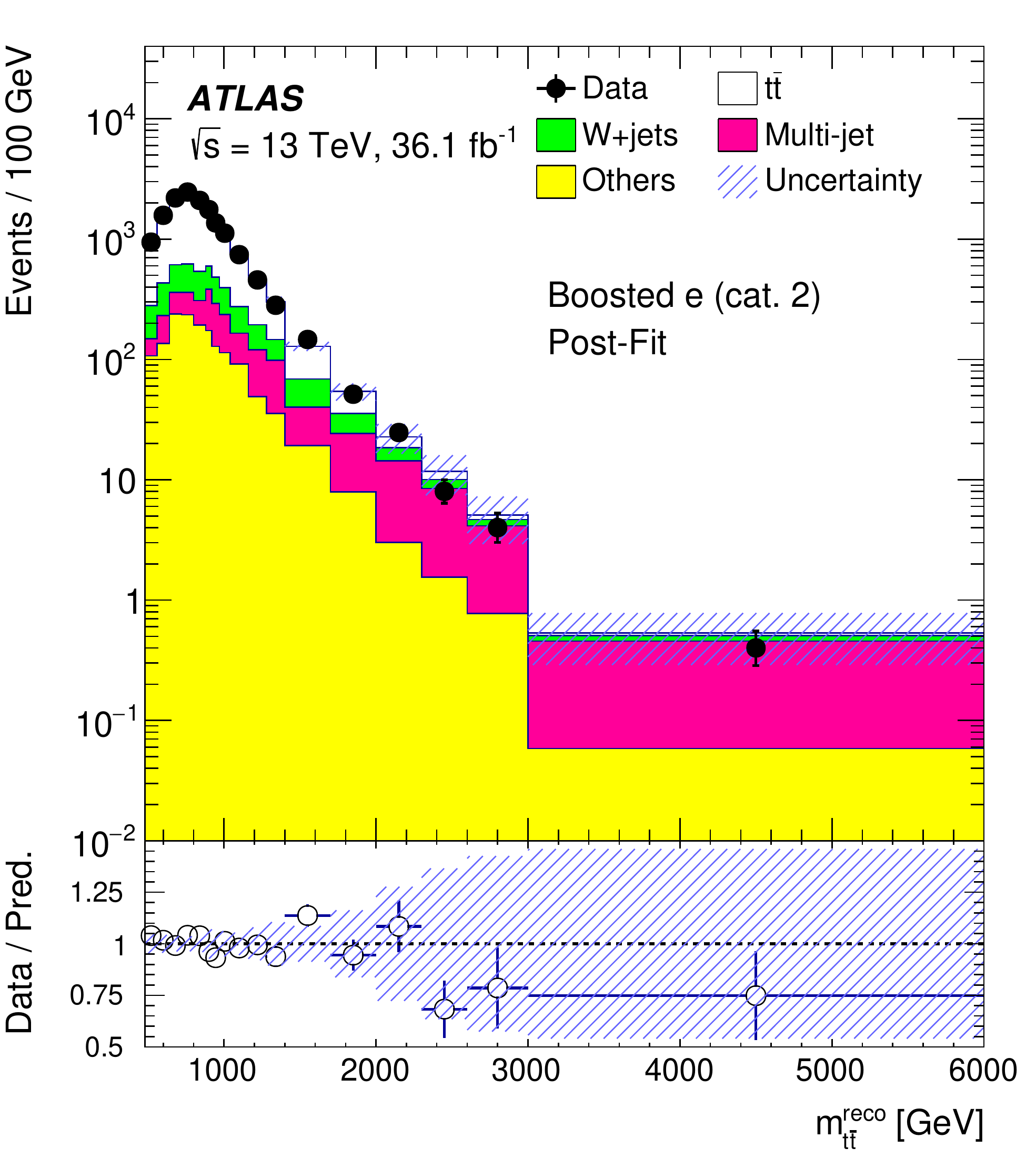}}
\subfigure[$\mu$+jets $b$-tag category 2]{\includegraphics[width=0.3\linewidth]{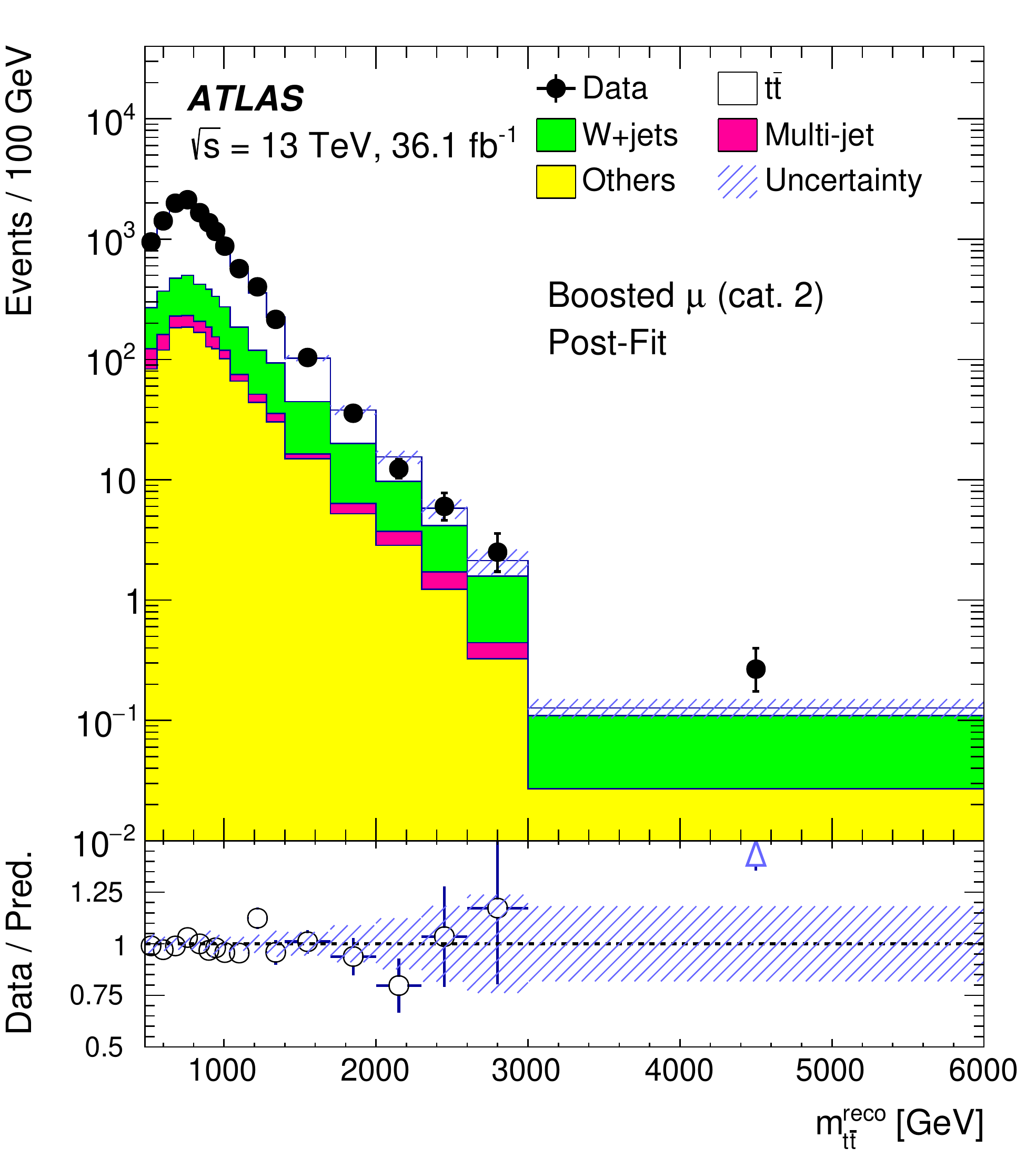}}\\
\subfigure[$e$+jets $b$-tag category 3]{\includegraphics[width=0.3\linewidth]{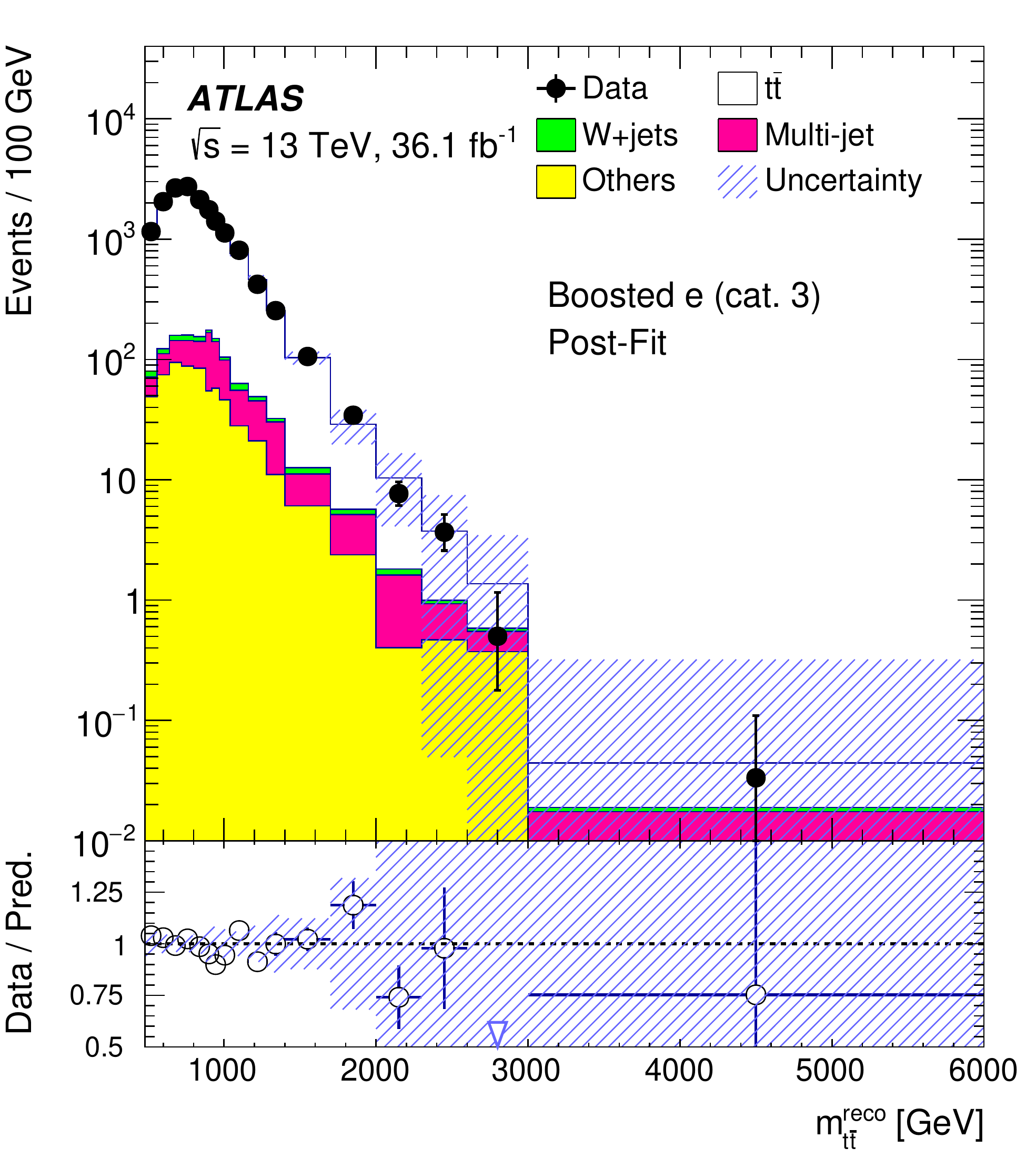}}
\subfigure[$\mu$+jets $b$-tag category 3]{\includegraphics[width=0.3\linewidth]{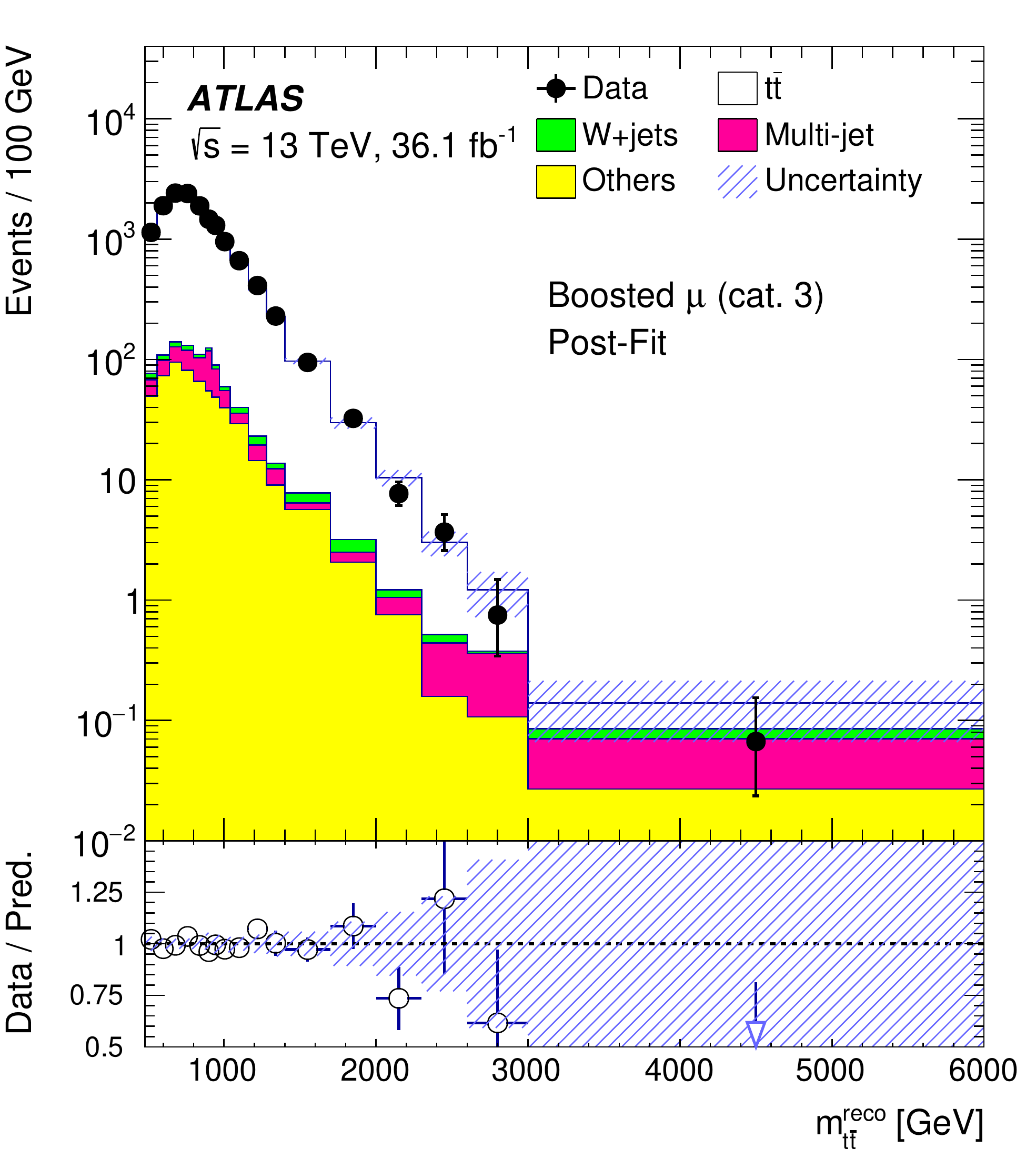}}\\

\caption{The $m^{\text{reco}}_{t\bar{t}}$  distributions, after a likelihood fit under the background-only hypothesis, for the boosted selection. The SM background components are shown as stacked histograms. The shaded areas indicate the total systematic uncertainties. The ratio of the data to the final fitted expectation is shown in the lower panel, open triangles indicate that the ratio point would appear outside the panel.}
\label{fig:mtt_boosted_postfit}
\end{figure}

\begin{figure}[ht]
\centering
\subfigure[$e$+jets $b$-tag category 1]{\includegraphics[width=0.3\linewidth]{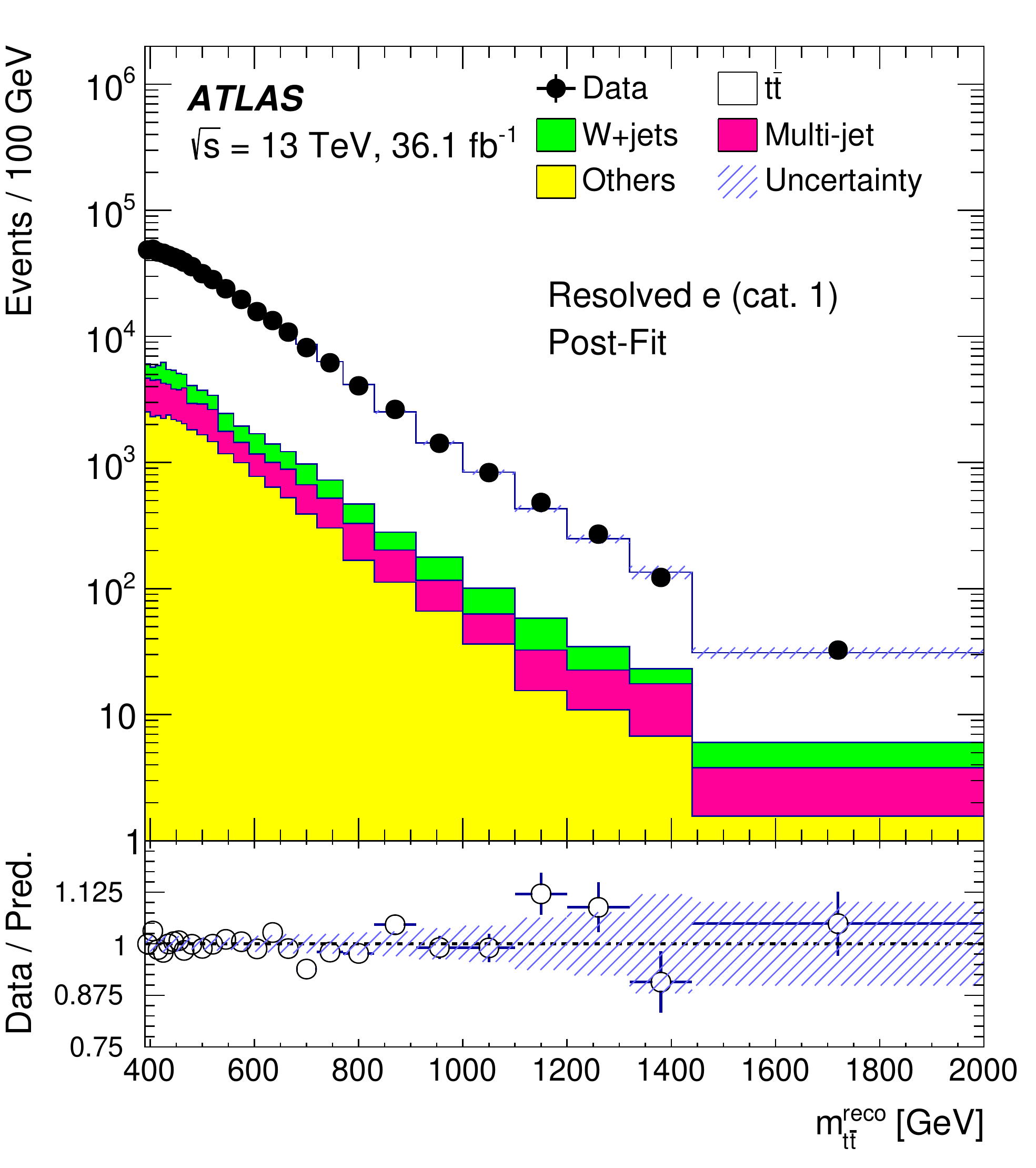}}
\subfigure[$\mu$+jets $b$-tag category 1]{\includegraphics[width=0.3\linewidth]{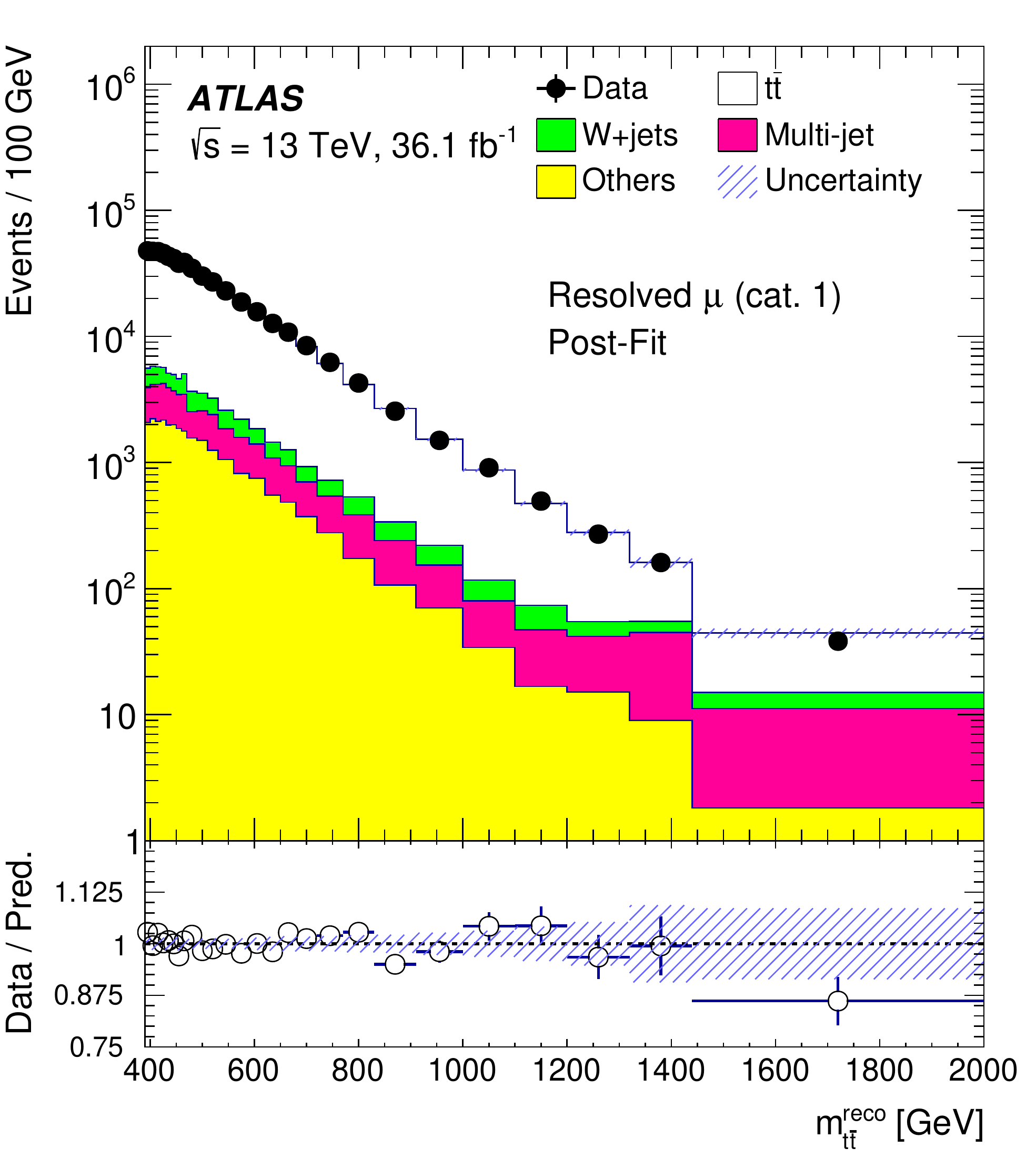}} \\
\subfigure[$e$+jets $b$-tag category 2]{\includegraphics[width=0.3\linewidth]{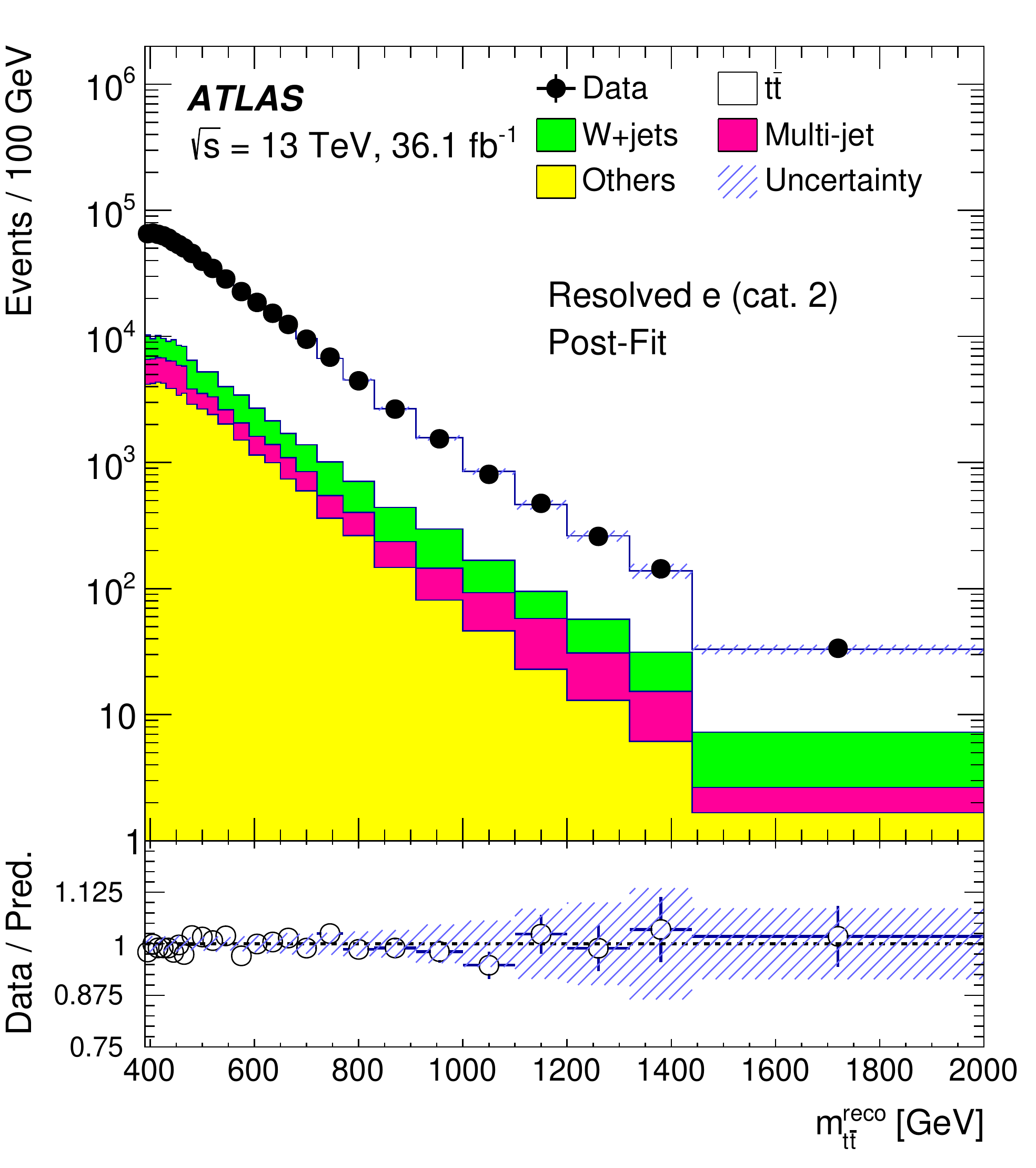}}
\subfigure[$\mu$+jets $b$-tag category 2]{\includegraphics[width=0.3\linewidth]{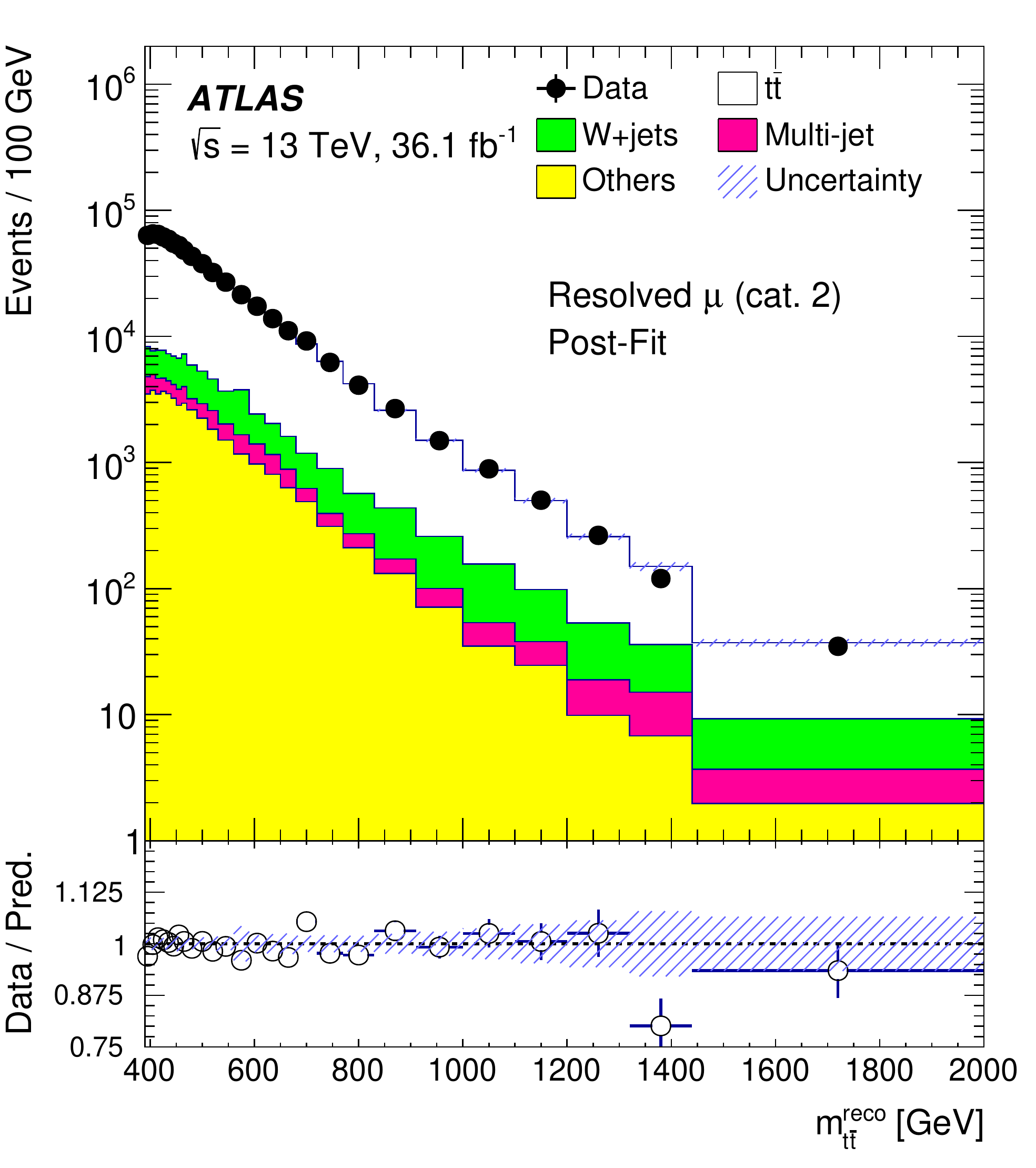}}\\
\subfigure[$e$+jets $b$-tag category 3]{\includegraphics[width=0.3\linewidth]{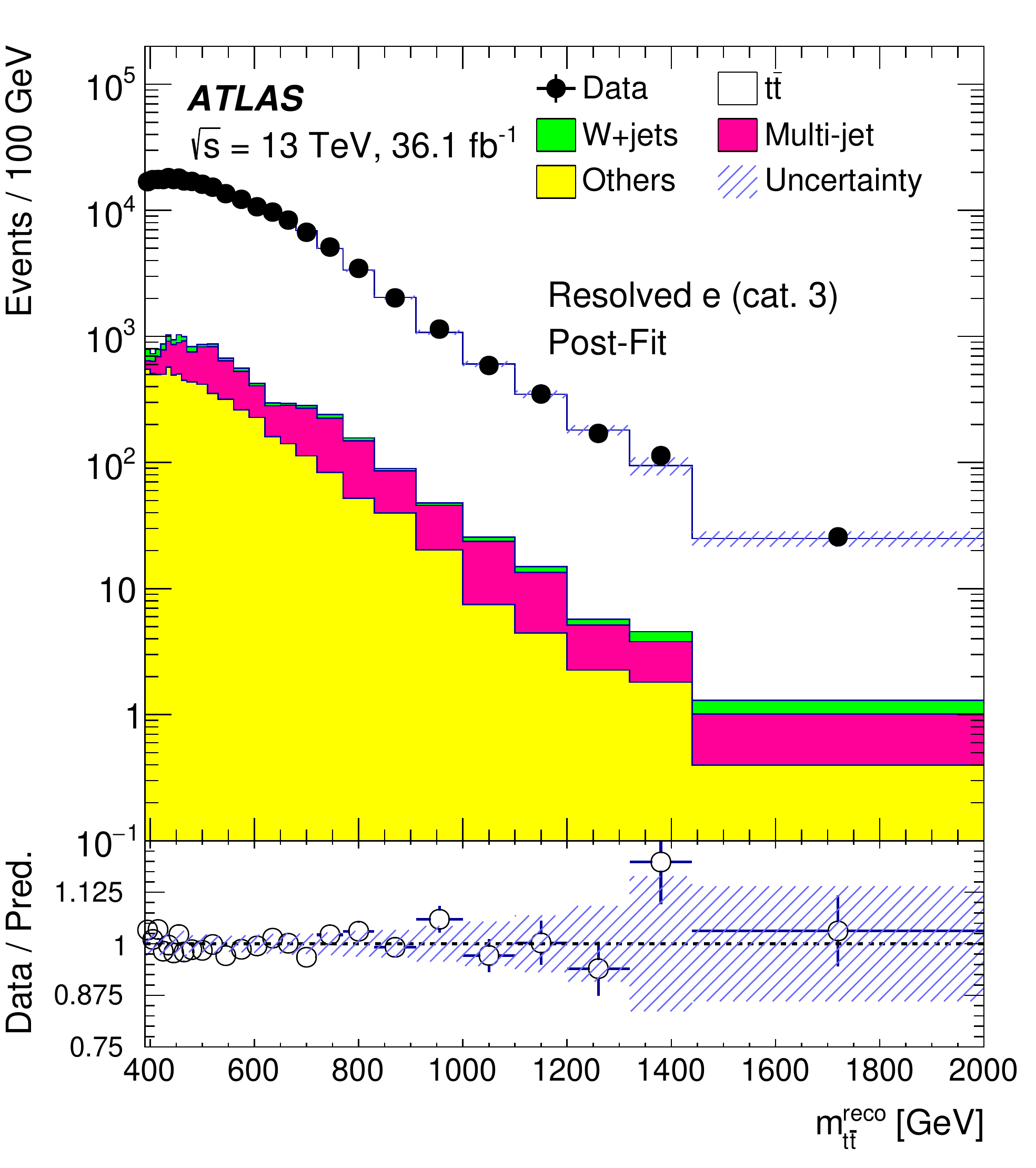}}
\subfigure[$\mu$+jets $b$-tag category 3]{\includegraphics[width=0.3\linewidth]{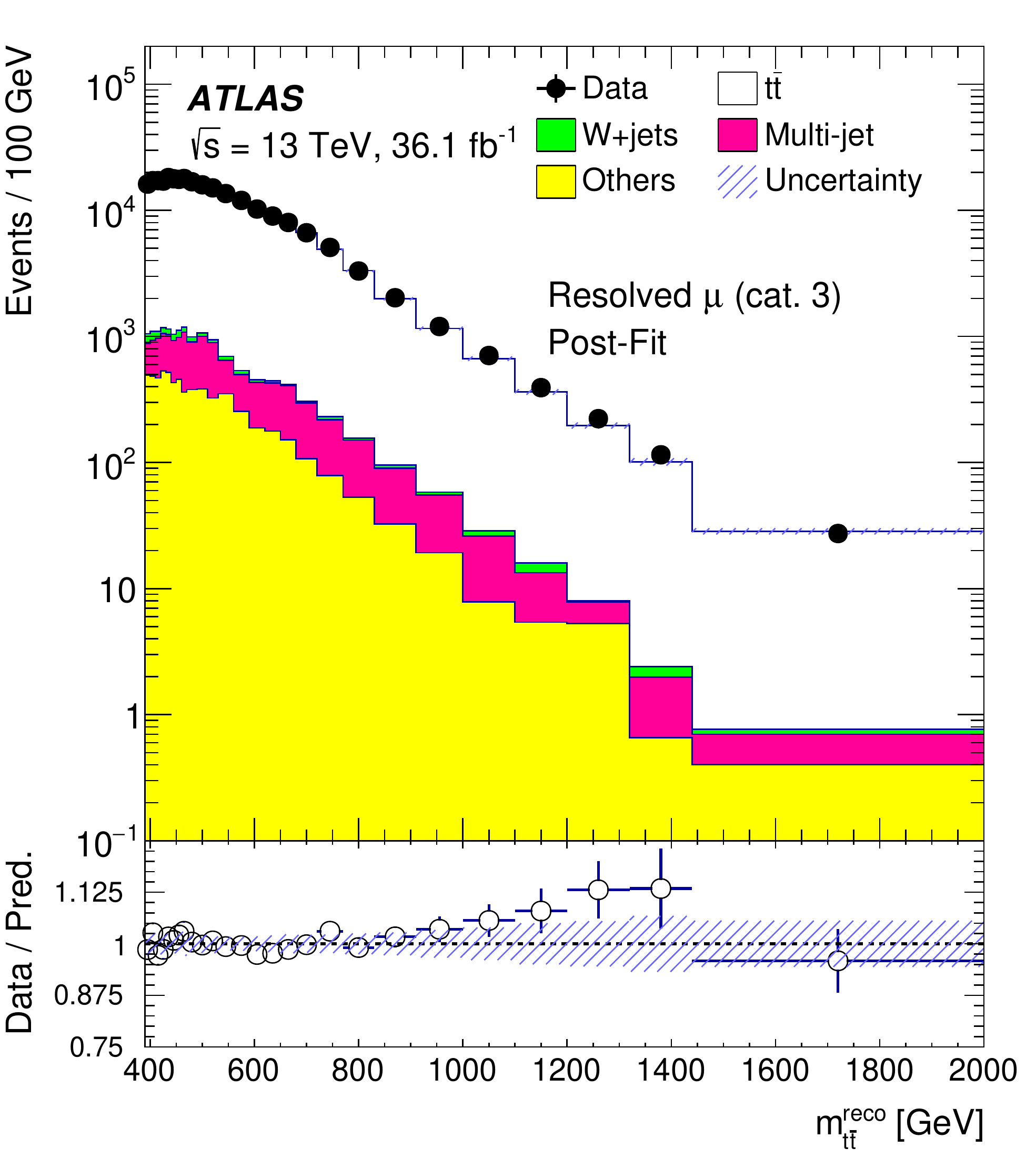}}\\

\caption{The $m^{\text{reco}}_{t\bar{t}}$  distributions, after a likelihood fit under the background-only hypothesis, for the resolved selection. The SM background components are shown as stacked histograms. The shaded areas indicate the total systematic uncertainties. The ratio of the data to the final fitted expectation is shown in the lower panel.} 
\label{fig:mtt_resolved_postfit}
\end{figure}

The statistical and systematic uncertainties in the expected distributions are included in this CL$_\text{s}$ procedure as nuisance parameters in the likelihood fits. 
The nuisance parameters for the systematic uncertainties are constrained by a Gaussian probability density function with a width corresponding to the size of the uncertainty considered. 
Correlations between different channels and bins are taken into account. 
The product of the various probability density functions forms the likelihood function that is 
maximised in the fit by adjusting the free parameter, called the signal strength (a multiplicative factor applied to the signal expected cross-section), and the nuisance parameters. 
The expected  $m_{t\bar{t}}^{\text{reco}}$ distributions are compared to data in Figures~\ref{fig:mtt_boosted_postfit} and~\ref{fig:mtt_resolved_postfit} after a fit of the nuisance parameters under the background-only hypothesis. 
The expected yields after the background-only fit are also shown in Table~\ref{tab:yields_postFit}.
It can be seen that the uncertainties are smaller than in Figures~\ref{fig:mtt_boosted} and~\ref{fig:mtt_resolved}.

Under the background-only hypothesis, a fit to data leads to a constraint of the jet energy resolution and the large-R jet energy scale nuisance parameters amongst the experimental uncertainties. The $t\bar{t}$ generator, radiation and modelling uncertainty nuisance parameters are also constrained, due to the large uncertainty in this background modelling. Amongst the most relevant uncertainties for the  3 TeV $Z^{\prime}_{\text{TC2}}$ model, the $t\bar{t}$ radiation uncertainty nuisance parameter is constrained by a factor of three in the boosted channel and the parton shower uncertainty, by a factor of two.

\begin{table}[tbh!]
\centering
\caption{Data and expected background in all channels after the background-only fit is performed.
The total systematic uncertainty in the expected background yields is also given.
The \ttbar normalisation is extracted from the fit in the boosted channels and its ratio to the pre-fit content is 0.93.
}
\begin{tabular}{c|r@{$\pm$}l|r@{$\pm$}l|r@{$\pm$}l|r@{$\pm$}l}
\hline
\hline
&\multicolumn{8}{c}{Yields} \\
Type       & \multicolumn{2}{c}{Boosted e}  & \multicolumn{2}{c}{Boosted $\mu$} & \multicolumn{2}{c}{Resolved e} & \multicolumn{2}{c}{Resolved $\mu$} \\
\hline
  $t\overline{t}$    & \num[round-mode=figures,round-precision=3]{28484.35} & \num[round-mode=figures,round-precision=1]{472.619399088} & \num[round-mode=figures,round-precision=3]{25954.53} & \num[round-mode=figures,round-precision=1]{393.216983095} & \num[round-mode=figures,round-precision=4]{231137.2} & \num[round-mode=figures,round-precision=2]{1861.11061402} & \num[round-mode=figures,round-precision=4]{225281.6} & \num[round-mode=figures,round-precision=2]{1688.87244019} \\
  W+jets    & \num[round-mode=figures,round-precision=2]{2212.1863} & \num[round-mode=figures,round-precision=2]{235.528077153} & \num[round-mode=figures,round-precision=2]{2153.8962} & \num[round-mode=figures,round-precision=2]{175.772055834} & \num[round-mode=figures,round-precision=2]{9408.273} & \num[round-mode=figures,round-precision=2]{1094.52528303} & \num[round-mode=figures,round-precision=3]{10272.179} & \num[round-mode=figures,round-precision=1]{828.734669723} \\
  Multi-jet    & \num[round-mode=figures,round-precision=2]{1978.554} & \num[round-mode=figures,round-precision=1]{401.286949625} & \num[round-mode=figures,round-precision=2]{780.682} & \num[round-mode=figures,round-precision=2]{203.697748866} & \num[round-mode=figures,round-precision=2]{8199.07} & \num[round-mode=figures,round-precision=2]{1358.9352344} & \num[round-mode=figures,round-precision=2]{7434.88} & \num[round-mode=figures,round-precision=2]{1374.23568453} \\
  Others    & \num[round-mode=figures,round-precision=3]{2879.165} & \num[round-mode=figures,round-precision=2]{233.562327573} & \num[round-mode=figures,round-precision=3]{2421.104} & \num[round-mode=figures,round-precision=2]{178.92204037} & \num[round-mode=figures,round-precision=2]{13407.23} & \num[round-mode=figures,round-precision=1]{649.345668725} & \num[round-mode=figures,round-precision=2]{11623.24} & \num[round-mode=figures,round-precision=1]{549.343712791} \\
  Total    & \num[round-mode=figures,round-precision=3]{35554.3} & \num[round-mode=figures,round-precision=1]{450.801905872} & \num[round-mode=figures,round-precision=3]{31310.18} & \num[round-mode=figures,round-precision=2]{304.574427288} & \num[round-mode=figures,round-precision=4]{262151.5} & \num[round-mode=figures,round-precision=2]{1180.31754031} & \num[round-mode=figures,round-precision=4]{254611.5} & \num[round-mode=figures,round-precision=2]{1077.6881199} \\
\hline
  Data     & \multicolumn{2}{c}{35612}                                     & \multicolumn{2}{c}{31188}                                     & \multicolumn{2}{c}{261554}                                    & \multicolumn{2}{c}{254277}                                    \\
\hline
\hline
\end{tabular}
\label{tab:yields_postFit}
\end{table}

The impact of the fitted nuisance parameters on the fitted signal strength is different at each candidate signal mass.
In order to estimate the impact of a nuisance parameter in the fit of the signal strength, the nuisance parameter is fixed at its central value plus or minus its fit uncertainties, and the variation of the fitted signal strength is tested.
For example, at a $Z'$ mass of 3~\TeV{}, the impact of an uncertainty on the best-fit value is computed by fixing the nuisance parameter $\theta$ to the one-standard-deviation range limits (positive or negative), and repeating the fit for a pseudodata sample with a 1~pb cross-section signal injected. The most significant uncertainties are related to the JES for large-$R$ jets and affect the fitted signal strength by up to 5\%.

\begin{figure}[ht]
\centering
\includegraphics[width=0.8\linewidth]{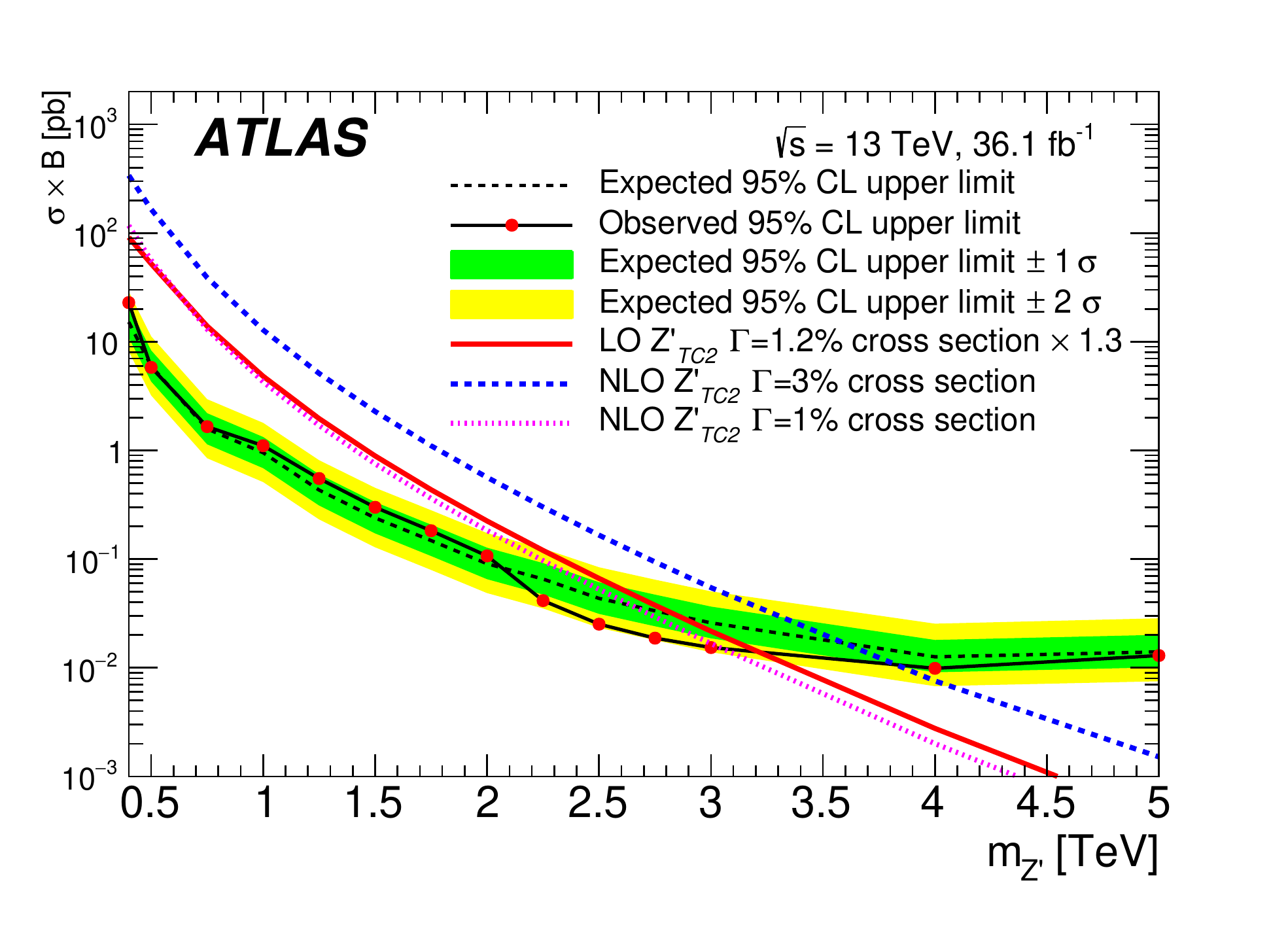}
\caption{The observed and expected cross-section 95\% CL upper limits on the $Z^\prime_{\text{TC2}}$ signal. The theoretical predictions for the production cross-section times branching ratio of $Z^\prime_{\text{TC2}} \rightarrow t\bar{t}$ at the corresponding masses are also shown.}
\label{fig:limitZp}
\end{figure}

\begin{figure}[ht]
\centering
\subfigure[Axial-vector mediator]{\includegraphics[width=0.45\linewidth]{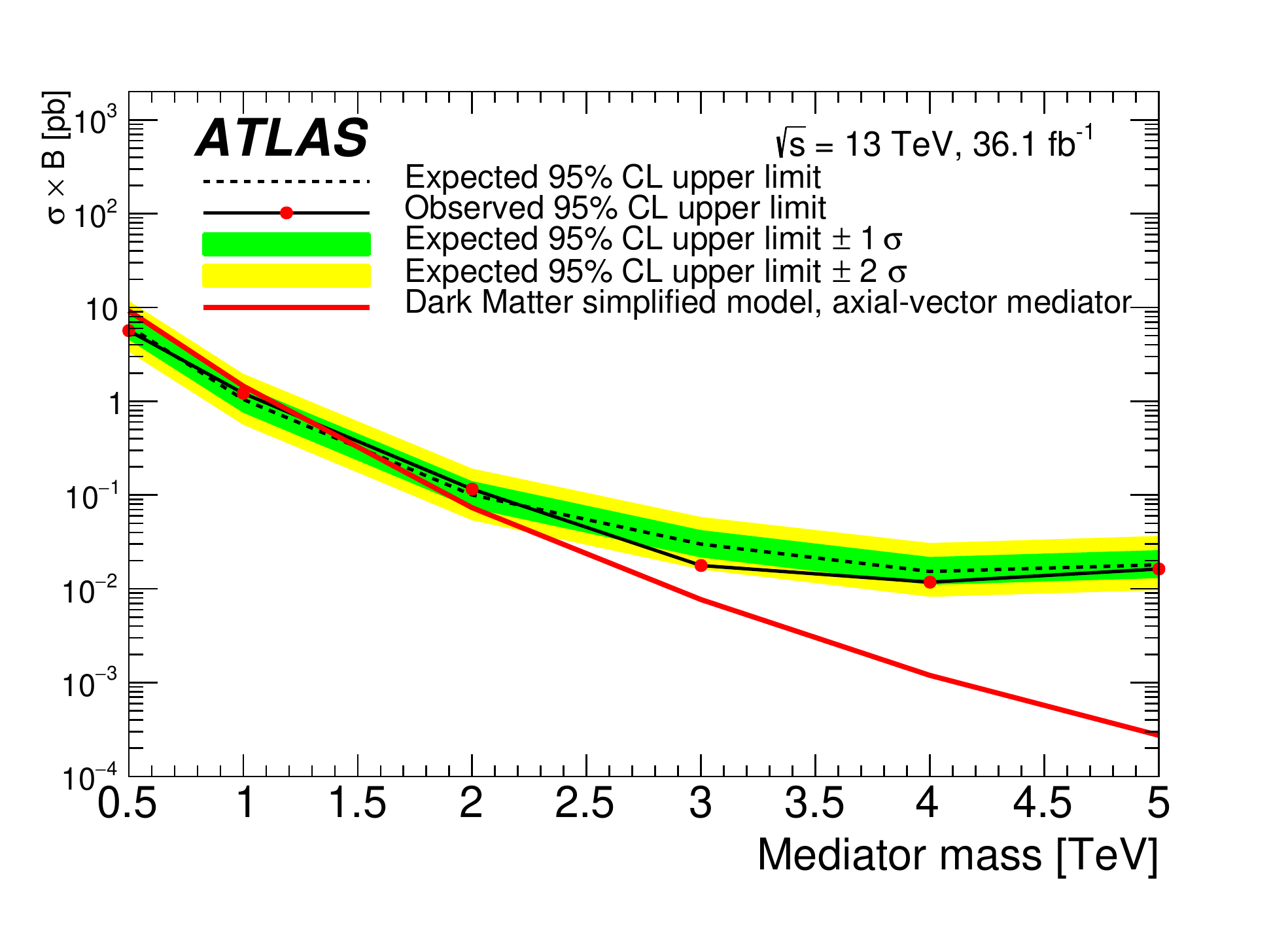}}
\subfigure[Vector mediator]{\includegraphics[width=0.45\linewidth]{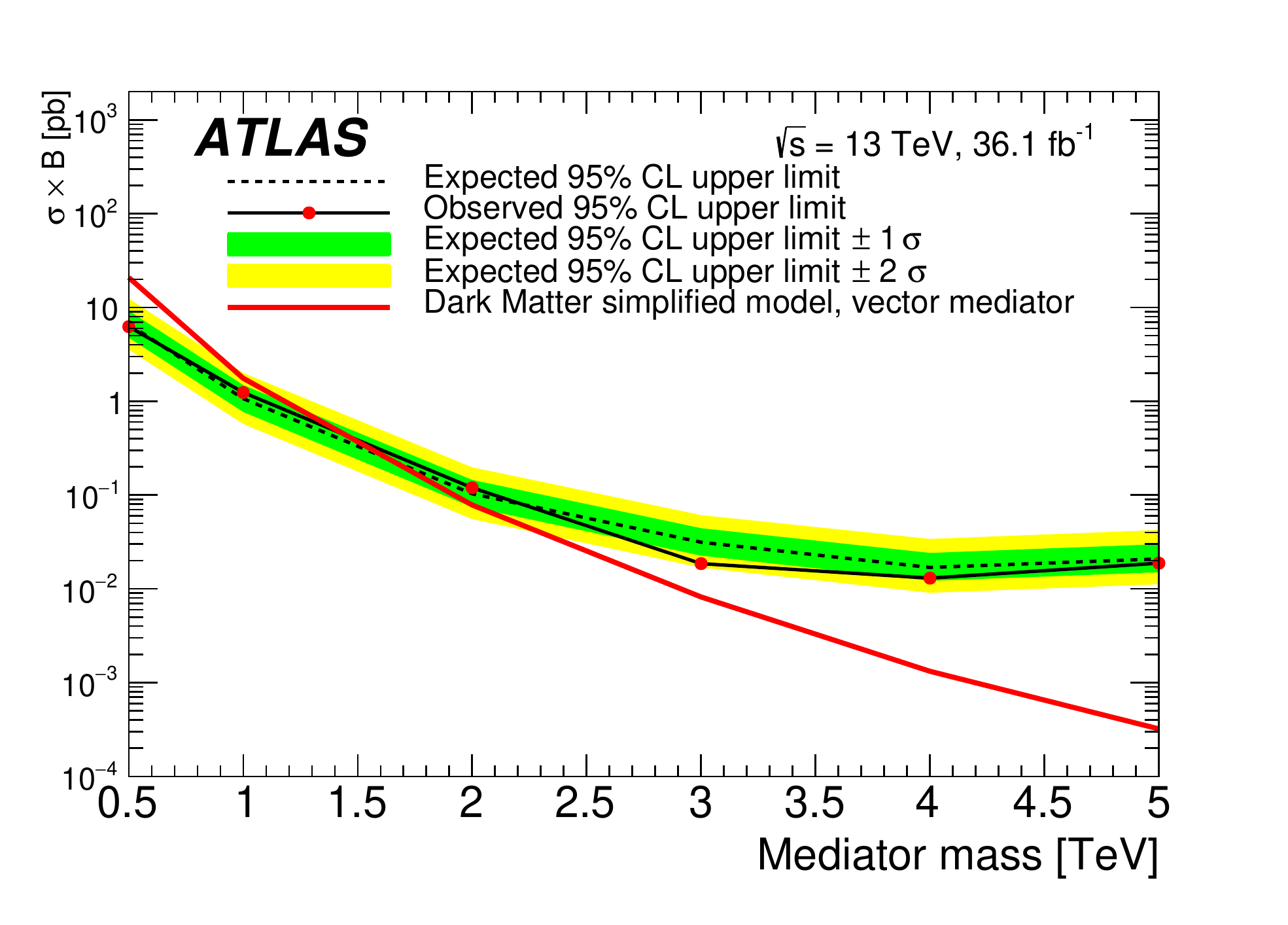}}
\caption{The observed and expected cross-section 95\% CL upper limits on the (a) $Z^\prime_{\text{DM,ax}}$ and (b) $Z^\prime_{\text{DM,vec}}$  signals. The theoretical predictions for the production cross-section times branching ratio of $Z^\prime_{\text{DM}} \rightarrow t\bar{t}$ at the corresponding masses are also shown.}
\label{fig:limitZdm}
\end{figure}

\begin{figure}[ht]
\centering
\includegraphics[width=0.8\linewidth]{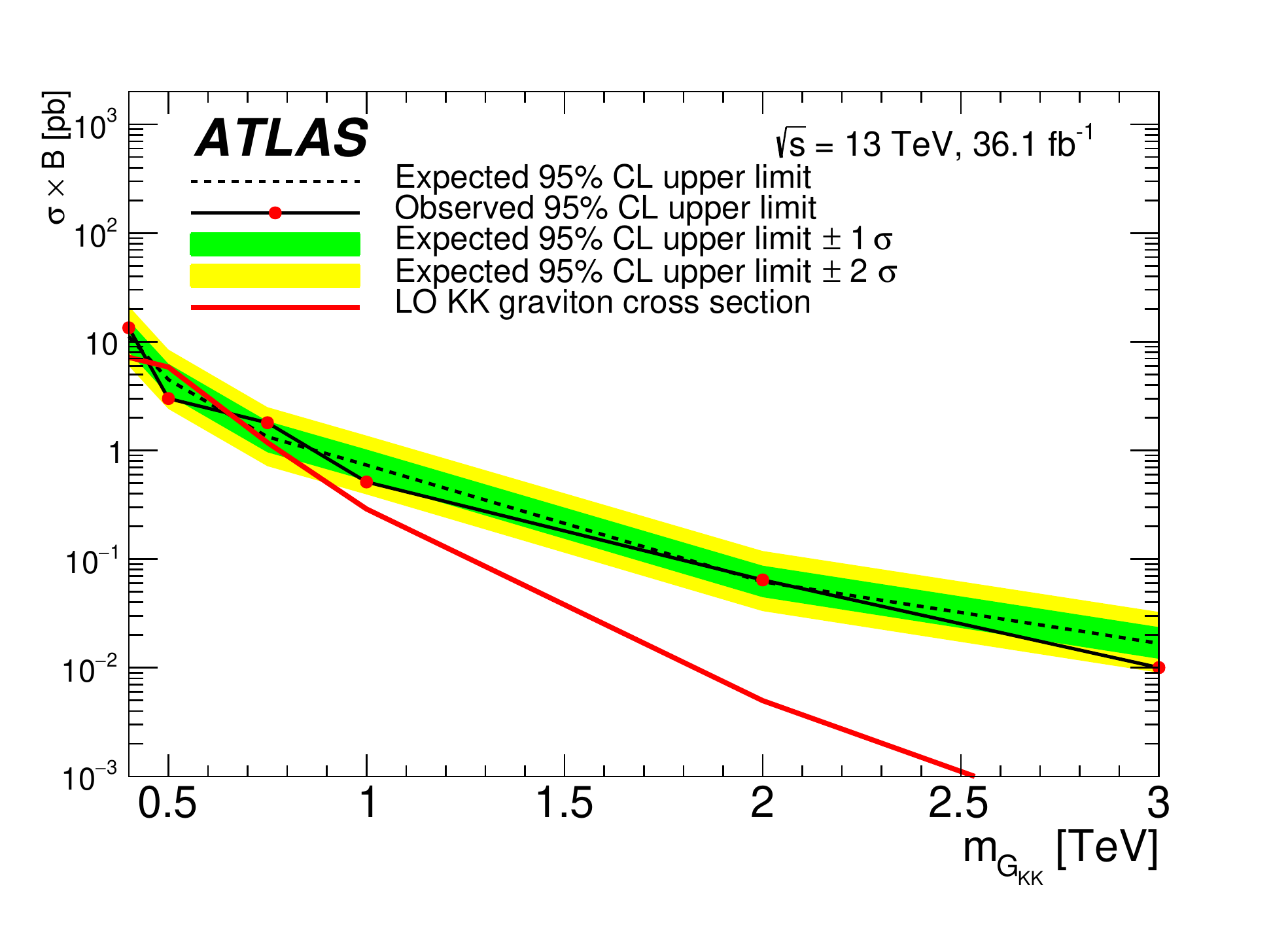}
\caption{The observed and expected cross-section 95\% CL upper limits on the $G_{\text{KK}}$ signal. The theoretical predictions for the production cross-section times branching ratio of $G_{\text{KK}} \rightarrow t\bar{t}$ at the corresponding masses are also shown.}
\label{fig:limitGrav}
\end{figure}

\begin{figure}[ht]
\centering
\subfigure[30\% width]{\includegraphics[width=0.45\linewidth]{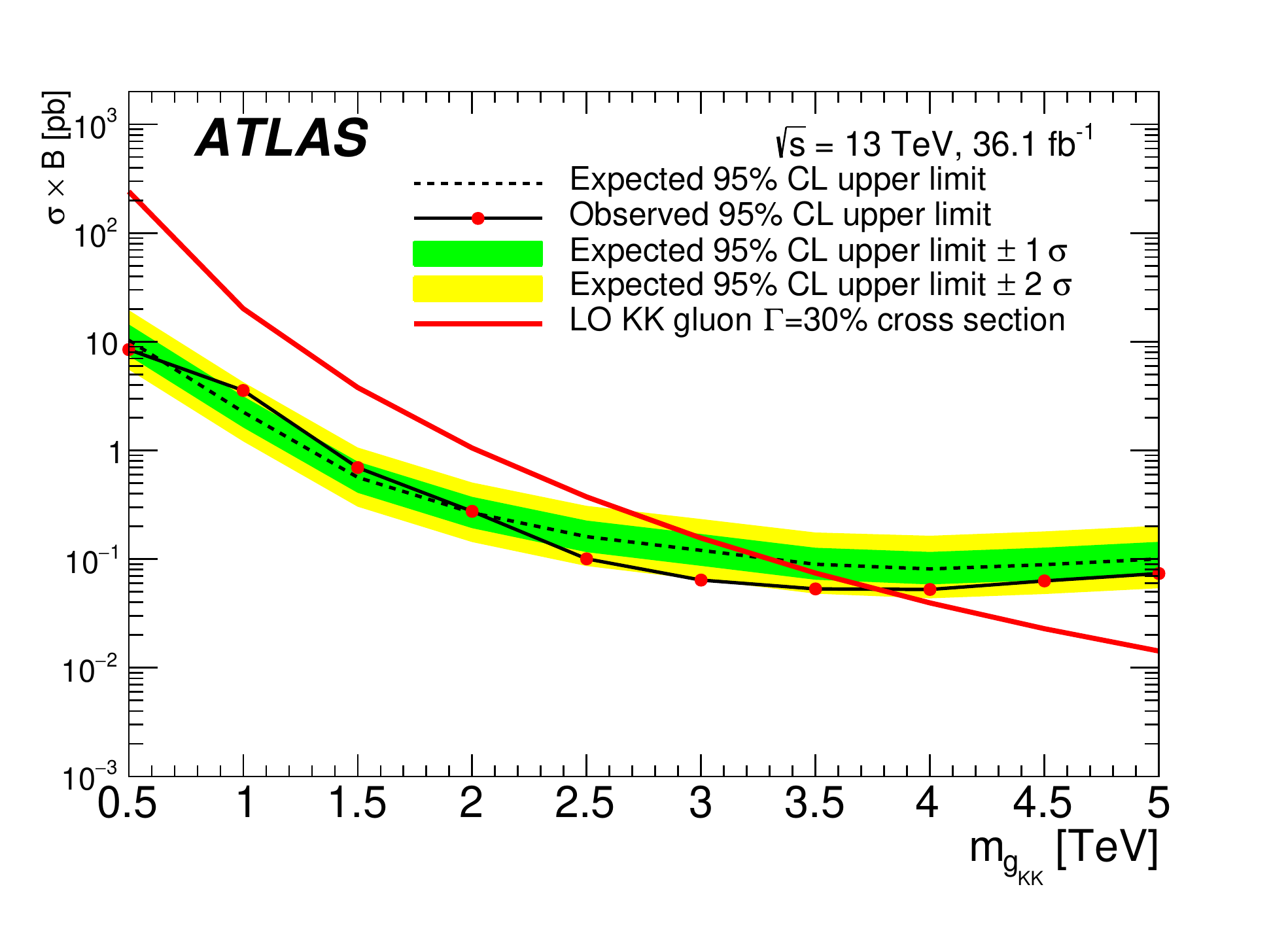}}
\subfigure[15\% width]{\includegraphics[width=0.45\linewidth]{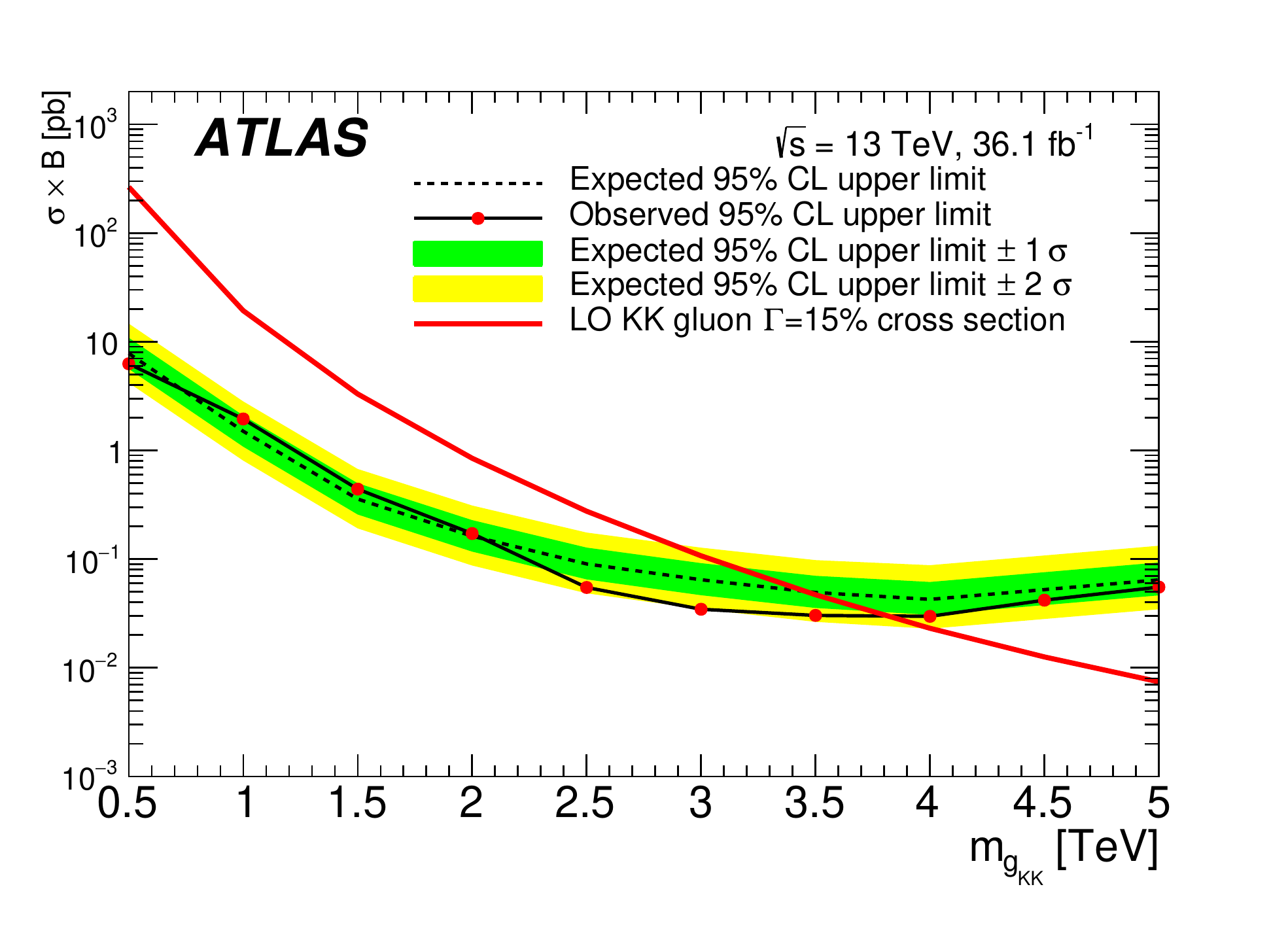}}
\caption{The observed and expected cross-section 95\% CL upper limits on the $g_{\text{KK}}$ signal for resonance widths of (a) 30\% and (b) 15\%. The theoretical predictions for the production cross-section times branching ratio of $g_{\text{KK}} \rightarrow t\bar{t}$ at the corresponding masses are also shown.}
\label{fig:limitGlue}
\end{figure}

The expected and observed limits on the studied signal models versus mass are presented in Figures~\ref{fig:limitZp}--\ref{fig:limitGlue} and summarised in Table~\ref{tab:limits}. 
The cross-section limits are extracted for each mass point, and are interpolated with straight lines in the regions between the mass points.
For the $Z'_{\mathrm{TC2}}$ benchmark, upper limits on the production cross-sections vary from 25~pb  to 0.02~pb for masses from 0.4~\TeV{} to 5~\TeV{}. 
A $Z'_{\text{TC2}}$ of width 1\%  is excluded for masses $m_{Z'_{\text{TC2}}} < 3.0$~\TeV{} while masses in the region $ m_{Z'_{\text{TC2}}} < 2.6$~\TeV{}  are expected to be excluded. 
The $Z'_{\text{DM,ax}}$ considered in this search is excluded for masses in the region  $ m_{Z'_{\text{DM,ax}}} < 1.2$~\TeV{}, while masses in the region $ m_{Z'_{\text{DM,ax}}} < 1.4$~\TeV{}  are expected to be excluded. 
The $Z'_{\text{DM,vec}}$ considered in this search is excluded for masses in the region  $ m_{Z'_{\text{DM,vec}}} < 1.4$~\TeV{} while masses in the region $ m_{Z'_{\text{DM,vec}}} < 1.6$~\TeV{}  are expected to be excluded. The Kaluza--Klein gravitons searched for in this analysis are excluded in the range $0.45~\TeV{}< m_{G_{\text{KK}}} < 0.65$~\TeV{}, which is also the expected exclusion region. A Kaluza--Klein gluon of width $30$\% is excluded for $m_{g_{\text{KK}}} < 3.7$~\TeV{} compared with an expected exclusion for  $m_{g_{\text{KK}}} < 3.2$~\TeV{}.  A Kaluza--Klein gluon of width $15$\% is excluded for $m_{g_{\text{KK}}} < 3.8$~\TeV{} compared with an expected exclusion for  $m_{g_{\text{KK}}} <3.5$~\TeV{}.

\begin{table}[tbh!]
\centering
\caption{Summary of the excluded mass ranges for the signals studied in this analysis. \label{tab:limits}}
\begin{tabular}{c|c|c}
\hline
\hline
\multicolumn{3}{c}{Summary of 95 \% Confidence Level mass exclusion ranges on benchmark models}    \\
\hline
Model                               & Observed excluded mass [TeV] & Expected excluded mass [TeV]   \\
\hline
$Z^\prime_{\text{TC2}}$ (1\% width) & $<$ 3.0                      & $<$ 2.6                        \\
$Z^\prime_{\text{DM,ax}}$           & $<$ 1.2                      & $<$ 1.4                        \\
$Z^\prime_{\text{DM,vec}}$          & $<$ 1.4                      & $<$ 1.6                        \\
$G_{\text{KK}}$                     & [0.45, 0.65]                 & [0.45, 0.65]                   \\
$g_{\text{KK}}$ (15\% width)        & $<$ 3.8                      & $<$ 3.5                        \\
$g_{\text{KK}}$ (30\% width)        & $<$ 3.7                      & $<$ 3.2                        \\
\hline
\hline
\end{tabular}
\end{table}

\begin{figure}[ht]
\centering
\subfigure[$m_{g_{\text{KK}}}=1$~\TeV{}]{\includegraphics[width=0.45\linewidth]{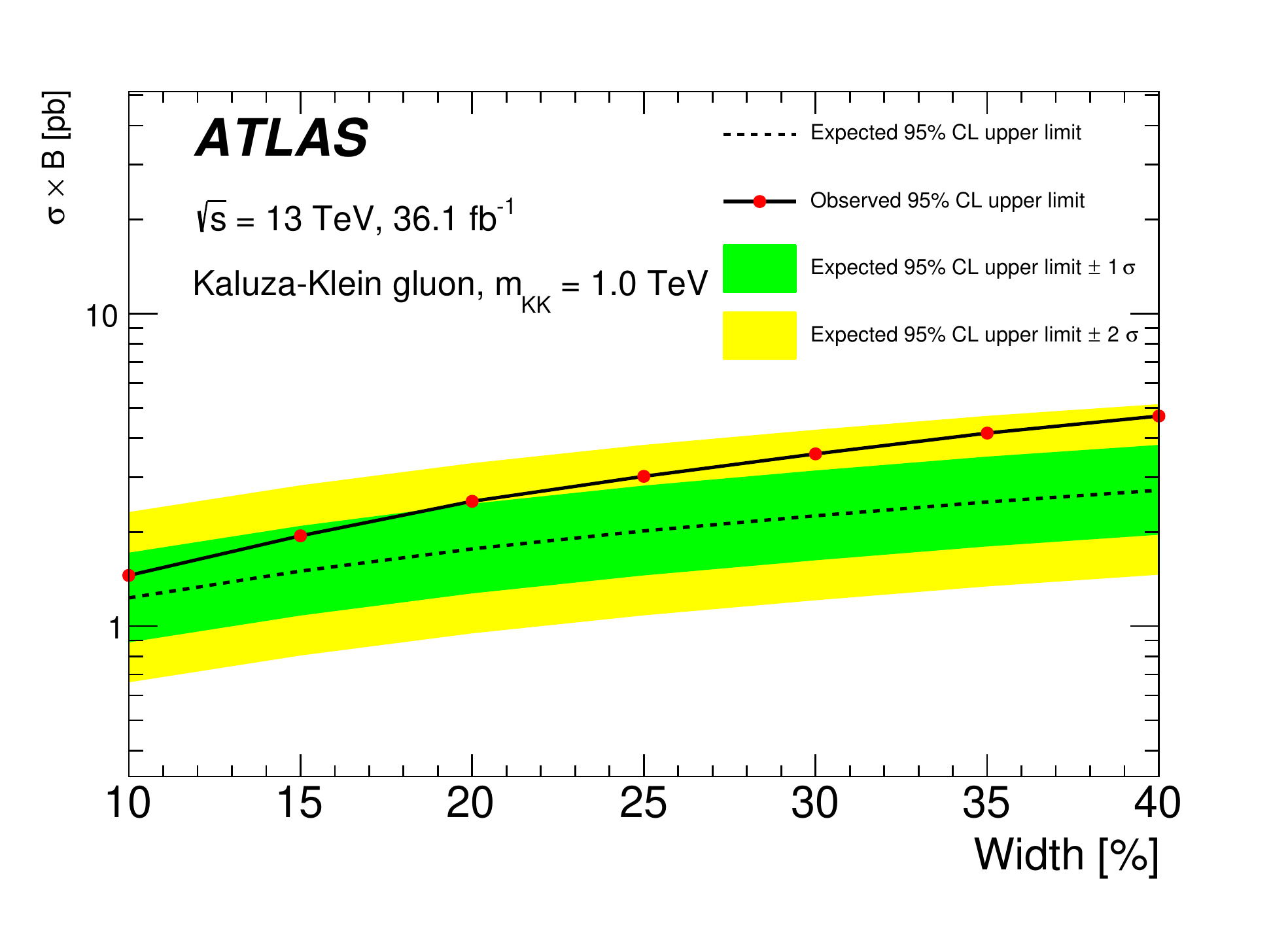}}
\subfigure[$m_{g_{\text{KK}}}=5$~\TeV{}]{\includegraphics[width=0.45\linewidth]{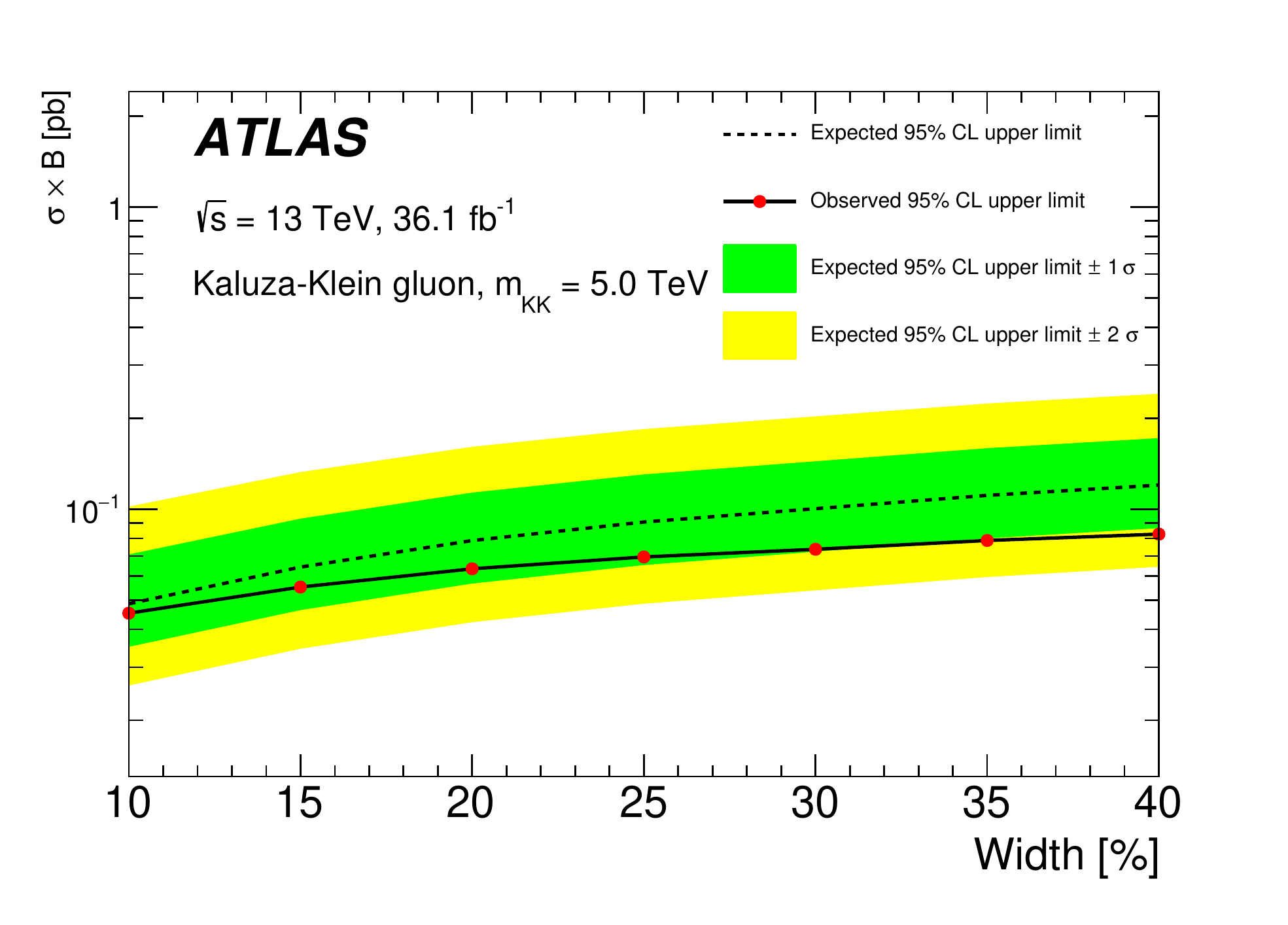}}
\caption{The observed and expected cross-section 95\% CL upper limits on the $g_{\text{KK}}$ signal as a function of the resonance width for masses of (a) 1~\TeV{}, and (b) 5~\TeV{}.}
\label{fig:limitvwidth}
\end{figure}

Furthermore, for the Kaluza--Klein gluons,  the search sensitivity as a function of resonance width was explored. 
Figure~\ref{fig:limitvwidth} shows the excluded cross-sections as a function of width for two different mass points. The cross-section limits deteriorate with increasing resonance width, as the signal peak is smeared out.

\FloatBarrier
\section{Summary}
\label{sec:summary}
A search for heavy particles decaying into $t\bar{t}$  in the lepton-plus-jets decay channel with the ATLAS experiment at the LHC is presented. The search uses data corresponding to an integrated luminosity of 36.1~$\text{fb}^{-1}$ of proton--proton collisions at a centre-of-mass energy of 13~\TeV{}. No excess of events beyond the Standard Model predictions is observed in the $t\bar{t}$ invariant mass spectra.
 Upper limits on the cross-section times branching ratio are set for several heavy resonances in models of new physics.  These results considerably extend the excluded regions for $Z'_{\text{TC2}}$ and $g_{\text{KK}}$ and represent the first mass ranges to be excluded, using the $t\bar{t}$ decay channel, for the dark-matter mediators $Z'_{\text{DM,ax}}$ and $Z'_{\text{DM,vec}}$,  and for $G_{\text{KK}}$.

\section*{Acknowledgements}


We thank CERN for the very successful operation of the LHC, as well as the
support staff from our institutions without whom ATLAS could not be
operated efficiently.

We acknowledge the support of ANPCyT, Argentina; YerPhI, Armenia; ARC, Australia; BMWFW and FWF, Austria; ANAS, Azerbaijan; SSTC, Belarus; CNPq and FAPESP, Brazil; NSERC, NRC and CFI, Canada; CERN; CONICYT, Chile; CAS, MOST and NSFC, China; COLCIENCIAS, Colombia; MSMT CR, MPO CR and VSC CR, Czech Republic; DNRF and DNSRC, Denmark; IN2P3-CNRS, CEA-DRF/IRFU, France; SRNSFG, Georgia; BMBF, HGF, and MPG, Germany; GSRT, Greece; RGC, Hong Kong SAR, China; ISF, I-CORE and Benoziyo Center, Israel; INFN, Italy; MEXT and JSPS, Japan; CNRST, Morocco; NWO, Netherlands; RCN, Norway; MNiSW and NCN, Poland; FCT, Portugal; MNE/IFA, Romania; MES of Russia and NRC KI, Russian Federation; JINR; MESTD, Serbia; MSSR, Slovakia; ARRS and MIZ\v{S}, Slovenia; DST/NRF, South Africa; MINECO, Spain; SRC and Wallenberg Foundation, Sweden; SERI, SNSF and Cantons of Bern and Geneva, Switzerland; MOST, Taiwan; TAEK, Turkey; STFC, United Kingdom; DOE and NSF, United States of America. In addition, individual groups and members have received support from BCKDF, the Canada Council, CANARIE, CRC, Compute Canada, FQRNT, and the Ontario Innovation Trust, Canada; EPLANET, ERC, ERDF, FP7, Horizon 2020 and Marie Sk{\l}odowska-Curie Actions, European Union; Investissements d'Avenir Labex and Idex, ANR, R{\'e}gion Auvergne and Fondation Partager le Savoir, France; DFG and AvH Foundation, Germany; Herakleitos, Thales and Aristeia programmes co-financed by EU-ESF and the Greek NSRF; BSF, GIF and Minerva, Israel; BRF, Norway; CERCA Programme Generalitat de Catalunya, Generalitat Valenciana, Spain; the Royal Society and Leverhulme Trust, United Kingdom.

The crucial computing support from all WLCG partners is acknowledged gratefully, in particular from CERN, the ATLAS Tier-1 facilities at TRIUMF (Canada), NDGF (Denmark, Norway, Sweden), CC-IN2P3 (France), KIT/GridKA (Germany), INFN-CNAF (Italy), NL-T1 (Netherlands), PIC (Spain), ASGC (Taiwan), RAL (UK) and BNL (USA), the Tier-2 facilities worldwide and large non-WLCG resource providers. Major contributors of computing resources are listed in Ref.~\cite{ATL-GEN-PUB-2016-002}.

\clearpage
\appendix

\printbibliography

\clearpage

 
\begin{flushleft}
{\Large The ATLAS Collaboration}

\bigskip

M.~Aaboud$^\textrm{\scriptsize 34d}$,    
G.~Aad$^\textrm{\scriptsize 99}$,    
B.~Abbott$^\textrm{\scriptsize 125}$,    
O.~Abdinov$^\textrm{\scriptsize 13,*}$,    
B.~Abeloos$^\textrm{\scriptsize 129}$,    
S.H.~Abidi$^\textrm{\scriptsize 165}$,    
O.S.~AbouZeid$^\textrm{\scriptsize 143}$,    
N.L.~Abraham$^\textrm{\scriptsize 153}$,    
H.~Abramowicz$^\textrm{\scriptsize 159}$,    
H.~Abreu$^\textrm{\scriptsize 158}$
% ...
\bigskip
\\

% ...
$^{13}$Institute of Physics, Azerbaijan Academy of Sciences, Baku; Azerbaijan.\\
% ...
$^{34}$$^{(a)}$Facult\'e des Sciences Ain Chock, R\'eseau Universitaire de Physique des Hautes Energies - Universit\'e Hassan II, Casablanca;$^{(b)}$Centre National de l'Energie des Sciences Techniques Nucleaires (CNESTEN), Rabat;$^{(c)}$Facult\'e des Sciences Semlalia, Universit\'e Cadi Ayyad, LPHEA-Marrakech;$^{(d)}$Facult\'e des Sciences, Universit\'e Mohamed Premier and LPTPM, Oujda;$^{(e)}$Facult\'e des sciences, Universit\'e Mohammed V, Rabat; Morocco.\\
% ...
$^{99}$CPPM, Aix-Marseille Universit\'e, CNRS/IN2P3, Marseille; France.\\
% ...
$^{125}$Homer L. Dodge Department of Physics and Astronomy, University of Oklahoma, Norman OK; United States of America.\\
% ...
$^{129}$LAL, Universit\'e Paris-Sud, CNRS/IN2P3, Universit\'e Paris-Saclay, Orsay; France.\\
% ...
$^{143}$Santa Cruz Institute for Particle Physics, University of California Santa Cruz, Santa Cruz CA; United States of America.\\
% ...
$^{153}$Department of Physics and Astronomy, University of Sussex, Brighton; United Kingdom.\\
% ...
$^{158}$Department of Physics, Technion, Israel Institute of Technology, Haifa; Israel.\\
$^{159}$Raymond and Beverly Sackler School of Physics and Astronomy, Tel Aviv University, Tel Aviv; Israel.\\
% ...
$^{165}$Department of Physics, University of Toronto, Toronto ON; Canada.\\
% ...
$^{a}$ Also at  Department of Physics, University of Malaya, Kuala Lumpur; Malaysia.\\
$^{b}$ Also at Borough of Manhattan Community College, City University of New York, NY; United States of America.\\
$^{c}$ Also at California State University, East Bay; United States of America.\\
$^{d}$ Also at Centre for High Performance Computing, CSIR Campus, Rosebank, Cape Town; South Africa.\\
$^{e}$ Also at CERN, Geneva; Switzerland.\\
% ...
$^{*}$ Deceased

\end{flushleft}




\end{document}